\documentclass{jfm}
\usepackage{graphicx}        % For eps figures, newer & more powerfull
\usepackage{amssymb}        % useful mathemati cal symbols
\usepackage{amsmath}
\usepackage{threeparttable}
\usepackage{xcolor}
\usepackage[normalem]{ulem}
\usepackage{MR_AA}
\usepackage{interval}
\usepackage{soul}
\usepackage[utf8]{inputenc}
\usepackage{verbatim}
\usepackage[overload]{empheq}
\usepackage{commath}

\providecommand\bnabla{\boldsymbol{\nabla}}

\newcommand\buu{{\boldsymbol{u}}}
\newcommand\bv{{\boldsymbol{v}}}

\newcommand\bC{{\boldsymbol{C}}}

\newcommand\be{{\boldsymbol{e}}}
\newcommand\bff{{\boldsymbol{f}}}
\newcommand\br{{\boldsymbol{r}}}
\newcommand\bq{{\boldsymbol{q}}}

\newcommand\bg{{\boldsymbol{g}}}

\newcommand\bF{{\boldsymbol{F}}}

\DeclareMathOperator{\cotan}{cotan}

\newcommand\Alm{A_{l-1}^{l}}
\newcommand\Alp{A_{l+1}^{l}}

\newcommand\Blm{B_{l-1}^{l}}
\newcommand\Blp{B_{l+1}^{l}}

 \def\corr#1{{\color{red} \sf #1}}

\title{Stress-driven spin-down of a viscous fluid within a spherical shell}

\author{D. Gagnier\aff{1} \aff{2}
  \corresp{\email{dgagnier@irap.omp.eu}},
 M. Rieutord\aff{1}}

\affiliation{\aff{1}IRAP, Universit\'e de Toulouse, CNRS, UPS, CNES,
14, avenue \'{E}douard Belin, F-31400 Toulouse, France
\aff{2}Department of Astronomy, University of Geneva, Chemin des Maillettes 51,
1290, Versoix, Switzerland }

\begin{document}

\maketitle

\begin{abstract}

We investigate the linear properties of the steady and axisymmetric stress-driven spin-down flow of a viscous fluid inside a spherical shell, both within the incompressible and anelastic approximations, and in the asymptotic limit of small viscosities. 
From boundary layer analysis, we derive an analytical geostrophic solution for the 3D incompressible steady flow, inside and outside the cylinder $\mathcal{C}$ that is tangent to the inner shell. The Stewartson layer that lies on $\mathcal{C}$ is composed of two nested shear layers of thickness $O(E^{2/7})$ and $O(E^{1/3})$. We derive the lowest order solution for the $E^{2/7}$-layer. A simple analysis of the $E^{1/3}$-layer laying along the tangent cylinder, reveals it to be the site of an upwelling flow of amplitude $O(E^{1/3})$. Despite its narrowness, this shear layer concentrates most of the global meridional kinetic energy of the spin-down flow. %The meridional circulation associated with the spin-down flow in these two regions scales with the Ekman number $E$. It is essentially parallel to the rotation axis inside $\mathcal{C}$ and perpendicular to the rotation axis outside $\mathcal{C}$. \corr{...}
%A simple analysis of the inner Stewartson shear layer of thickness $O(E^{1/3})$ and laying along the tangent cylinder, reveals it to be the site of an upwelling flow of amplitude $O(E^{1/3})$. Despite its narrowness, this shear layer concentrates most of the global meridional kinetic energy of the system. 
Furthermore, a stable stratification does not perturb the spin-down flow provided the Prandtl number is small enough. If this is not the case, the Stewartson layer disappears and meridional circulation is confined within the thermal layers.
The scalings for the amplitude of the anelastic secondary flow have been found to be the same as for the incompressible flow in all three regions, at the lowest order. However, because the velocity no longer conforms the Taylor-Proudman theorem, its shape differs outside the tangent cylinder $\mathcal{C}$, that is, where differential rotation takes place. Finally, we find the settling of the steady-state to be reached on a viscous time for the weakly, strongly and thermally unstratified incompressible flows. Large density variations relevant to astro- and geophysical systems, tend to slightly shorten the transient.

%We investigate the linear properties of the steady stress-driven spin-down flow of a viscous fluid inside spherical shell, both within the incompressible and anelastic approximations, and in the asymptotic limit of small viscosities. From boundary layer analysis, we derive an analytical geostrophic solution for the 3D incompressible flow, inside and outside a cylinder $\mathcal{C}$ that is tangent to the inner shell. The meridional circulation associated with the spin-down flow in these two regions scales with the Ekman number $E$. It is essentially parallel to the rotation axis inside $\mathcal{C}$ and perpendicular to the rotation axis outside $\mathcal{C}$. A power expansion of the fields with the small parameter $E^{1/3}$ corresponding to the width scaling of the vertical shear layer sitting along the tangent cylinder wall, reveals it to be the site of an upwelling flow of amplitude $O(E^{1/3})$. Despite its narrowness, the motion within this shear layer is of crucial importance regarding the strength of the generated secondary flow, as it concentrates most of the global meridional kinetic energy of the system. [\textbf{Thermal stratification}] The same scalings for the amplitude of the anelastic secondary flow have been found in all three regions, at the lowest order. However, because the velocity no longer conforms the Taylor-Proudman theorem, its shape differs outside the tangent cylinder $\mathcal{C}$, that is, where differential rotation takes place. 

\end{abstract}


\section{Introduction}

%It is also shown that three nested vertical boundary layer form tangentially to the inner shell and consist in a free shear layer (or boundary layer) of thickness $O(E^{1/3})$ that is located inside two other vertical boundary layers of thickness $O(E^{1/4})$ and $O(E^{2/7})$. The role of the boundary layer of thickness $O(E^{1/4})$ is to smooth out the $O(1)$ azimuthal primary flow discontinuity while the $O(E^{2/7})$ layer removes a singularity in the gradient of the azimuthal velocity. The inner layer of thickness $O(E^{1/3})$ is often the means by which the fluid is returned from the inner Ekman layer to the outer one (or the opposite in case of spin-up).
One of the long-lasting problem in stellar astrophysics is the nature of the  mechanisms responsible for the angular momentum and chemicals transport within the stably stratified radiative envelope of rotating massive stars.
%Our concern with this problem is mainly motivated by an interest in the so-called rotation-induced mixing (of angular momentum and chemicals) inside the radiative envelope of rotating massive stars. 
Such transport can result from various physical processes such as internal gravity waves \citep[e.g.,][]{rogers_etal13, Lee2014}, turbulence from shear instabilities due to differential rotation  \citep[e.g.,][]{zahn74, zahn92,Prat2013,Prat2014,Prat2016,Garaud2017,Gagnier2018,Kulenthirarajah2018}, and centrifugal driving of meridional circulations. The latter is usually associated with the steady baroclinic state of the radiative envelope of rotating massive stars \citep[e.g.,][]{garaud02,R06,ELR13,RB14,HR14,Hypolite2018}. In such a state, isobar and isopycnic lines (or equipotentials) are different and result in a baroclinic torque $(\bnabla p \times \bnabla \rho)/\rho^2$. Because of the strength of the pressure and density gradients in stellar interiors, a slight misalignment of the two vectors is sufficient to drive a sizeable baroclinic
flow. In fact, the misalignment remains less than a degree, even for the most rapidly rotating stars \citep{ELR11}. The effects of baroclinicity can however be supplemented by the meridional circulation induced by spin-up or spin-down flows resulting from stellar contraction or expansion \citep[][]{HR14}. Furthermore, in massive stars, rapid rotation usually combines with strong radiation-driven winds that may lead to significant latitude-dependent outward mass flow  and angular momentum flux \citep[e.g.,][]{ Pelupessy2000, MaederMeynet2000, Georgy2011,Gagnier2019a, Gagnier2019b}. When this is the case, typically for stars more massive than seven solar masses, the baroclinic flow is supplemented by a meridional circulation resulting from the surface stress induced by the mass-loss
\citep{zahn92}. The transport of products of nuclear reactions in the stellar core can then enrich the surface layers in chemical elements, locally increasing opacity and thus enhancing the radiation-driven outward mass and angular momentum fluxes \citep[][]{Kudritzki1987, Puls2000, Puls2008}. The physical process of wind-driven spin-down flow is therefore of crucial importance for the understanding of the secular evolution of rotating massive stars.

So far, however, there is still no consensus on the appropriate way to model the mixing induced by such fluid flows in stellar evolution codes. Indeed, while the transport of chemicals is always accounted for as a diffusive process \citep[as justified by][]{chaboyer_Z92}, the angular momentum transport is either treated as an advection-diffusion process following  \citet{zahn92}, \citet{MeynetMaeder1997}, and \citet{MZ98} or as a purely diffusive process \citep{paxton_etal11}. Moreover, when facing fast rotation, developments beyond the current 1D-model approximations become necessary. In this context,
the achievement of the first self-consistent 2D models of rapidly-rotating early-type  stars, worked out by Espinosa Lara and Rieutord \cite[e.g.][]{ELR13, Rieutordal2016} and in which the differential rotation as well as the meridional circulation arising from baroclinicity are computed self-consistently, opens the door to the exploration of the evolution of fast stellar rotators. Because the sources of rotation-induced mixing are multiple, a meticulous study of each transport mechanism appears to be necessary for a better understanding of stellar evolution.  In this paper we address this issue and investigate the properties of the primary and secondary flows driven by radiation-driven outward mass and angular
momentum fluxes at the surface of rotating massive stars.

%\textbf{To this extent, this paper aims at investigating the properties of the primary and secondary flows generated from a prescribed surface-stress, mimicking a weak angular momentum outward flux at the surface of rotating massive stars.}
%In particular, we wish to determine if the amplitude of the induced secondary flow can compare to the baroclinic meridional circulation, and thus establish the importance of the rotation-induced mixing resulting from radiation-driven winds in the radiative envelope of rotating massive stars.

However, the complete modelling of astrophysical rotation-induced mixing is a complex problem. Hence, to understand its different facets, it is useful to study simplified set-ups which incorporate, step by step, the various physical phenomena that contribute to the whole realistic model. For instance, we shall be particularly interested in the scaling laws which control the viscous effects.
%that would require relaxing the spherical symmetry and computing the spin-down and baroclinic flows, as well as the turbulent transport from shear instabilities altogether.
%To get a detailed understanding of this problem we investigate in this paper a simplified set-up which will enlighten us on the scaling laws governing the flows, in particular with respect to viscosity.

When a star loses mass, the associated wind extracts angular momentum. At some place inside the star, but close to the surface, this extraction generates a radial differential rotation that further extracts angular momentum from the deeper layers. The upper layers therefore impose a torque to the interior of the star, which slowly spins down.
As is well-known \cite[see][]{Green69}, the spin-down of an incompressible fluid inside a rigid container with no-slip boundaries occurs on a time scale P$_{\rm rot}/\sqrt{E}$, where P$_{\rm rot}$ is the rotation period of the fluid and $E$ is the Ekman number (see below for its definition). Since in most situations $E\ll1$, the spin-down time scale is much shorter that the viscous diffusion time scale P$_{\rm rot}/E$. In stars however, boundary conditions are not rigid. Rather, the wind imposes a global angular momentum loss which amounts to a torque applied to the inner layers of the star. Hence, at some depth, the fluid is spun-down by a (turbulent) viscous stress. As in the no-slip case, secondary meridional flows arise as shown by \citet{friedlander76}.

The set-up is further complicated by the convective core of the massive stars at hands. Compared to the envelope, the core may be viewed as a very viscous region, since thermal convection is highly turbulent there. To simplify, we shall represent it with the limiting case of a solid ball. A solid boundary or an important jump in viscosity both lead to the formation of Stewartson shear layers staying along the tangent cylinder of the core, parallel to the rotation axis \cite[e.g.][]{stewar66,R06}. Hence, the secondary flows are certainly more complex that those arising in a full sphere. Moreover, we wish to know how long it takes for chemical elements generated in the core by nucleosynthesis to reach the stellar surface, where they can be observed. Hence, it is important to know the meridional circulation turnover time and its dependence with viscosity. In addition thermal stratification and density variations between the core and the surface also influence the spin-down flow.

Our simplified model takes into account all these features of the problem so as to examine their particular role. Let us summarise this model: We consider a spherical shell containing a viscous fluid where the rotation of the core is imposed and where an imposed large-scale stress is applied on the outer surface. We first consider a fluid of constant density and temperature, then discuss the case with a stable thermal stratification (as requested by radiative envelopes), and finally, we allow for radial density variations using a polytropic envelope within the anelastic approximation \cite[e.g.][]{jones_etal09}. 
We further assume that the angular momentum extraction and the associated surface stress is weak enough that linearised equations can be used for the axisymmetric flow determinations.

The paper is organised as follows: in Sect.~\ref{sec:Formulation} we introduce our model and describe the numerical method used for solving the equations of motion. We then consider the case of an incompressible flow and we discuss the time scales governing the transient phase and the asymptotic properties of the stationary primary (differential rotation) and secondary (meridional circulation) flows in the limit of small Ekman numbers in Sect.~\ref{sec:incompr}. In Sect.~\ref{sec:thermal}, we discuss the role of thermal stratification using the Boussinesq approximation, and we finally include density variations of the background using the anelastic approximation 
in Sect.~\ref{sec:anelastic}.
Discussions and conclusions follow in Sect.~\ref{sec:conclusions}.

\section{Formulation of the problem}\label{sec:Formulation}

\begin{figure}
\centering
      \includegraphics[width=0.6\textwidth]{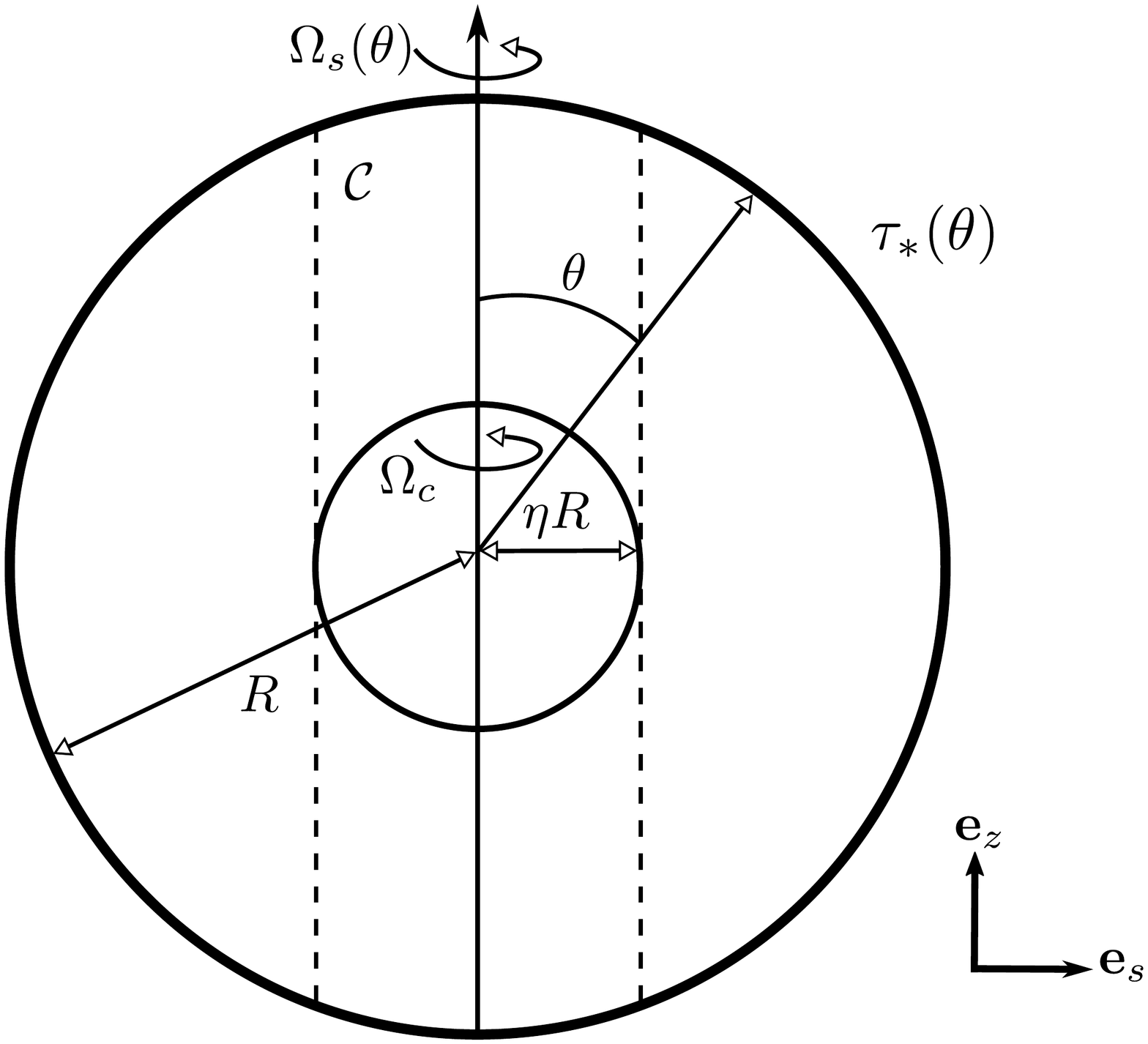}
   \caption{Schematic view of the system: the inner shell of radius $\eta R$ rotates with an angular velocity $\Omega_c$ while the outer one  of radius $R$ rotates differentially at $\Omega_s(\theta)$ by means of a prescribed tangential surface stress $\tau_{*}(\theta)$. The dashed lines correspond to the edges of the tangent cylinder $\mathcal{C}$, circumscribing the inner sphere. }
\label{fig:config}
\end{figure}

\subsection{Description}

A fluid of constant kinematic viscosity $\nu$ is enclosed between two spheres (see below). The inner one is rigidly rotating with a constant angular velocity $\Omega_c$ while the outer shell supports a prescribed tangential surface-stress $\tau_{*}(\theta)$. $R$ and $\eta R$ are the radii of the outer and inner shells, respectively, and $\theta$ is the colatitude. We sketch out this model in Fig.~\ref{fig:config}. We consider the driving stress to be sufficiently weak that the rotation period of the system is much shorter than the typical turn-over times associated with the flow in the rotating frame of reference. We thus consider the non-linear terms to be negligible, which we justify \textit{a posteriori} in Appendix \ref{sec:A2}. The dimensionless equation of vorticity and mass conservation equation in the rotating frame with angular velocity $\Omega_c$, and in the context of an ``anelastic" flow, read

\begin{equation}\label{eq:nucst}
\begin{cases}
\begin{aligned}
&\frac{\partial}{\partial t} (\bnabla \times \rho \buu)+ \bnabla \times (\be_z \times  \rho \buu) = E \bnabla \times \bff_{\rm visc} \\ \\ &\bnabla \cdot \rho \buu = 0 \ ,
\end{aligned}
\end{cases}
\end{equation}
where we have used $R$ as the length scale, $(2\Omega_c)^{-1}$ as the time scale, and $\rho_c$, the density at the inner shell, as the density scale. $E$ is the Ekman number defined as 

\begin{equation}
    E=\frac{\nu}{2 \Omega_c R^2} \ , 
\end{equation}
and 
\begin{equation}
    \bff_{\rm visc}= \rho \left(\Delta \buu + \frac{1}{3}\bnabla {\rm div} \buu \right) + 2(\bnabla \rho \cdot \bnabla)\buu  + \bnabla \rho \times (\bnabla \times \buu) - \frac{2}{3} \bnabla (\rho {\rm div} \buu) \ ,
\label{eq:visc}
\end{equation}
is the dimensionless viscous force. This expression of the viscous force is meant to represent the case where the envelope is pervaded by some small-scale turbulence with constant diffusive properties as often used in stellar physics \citep{BJNRST96,KMB12}. Centrifugal effects are neglected altogether, and the flows are considered axisymmetric.

This system is then completed by boundary conditions. On the outer shell, we impose a specified stress, which represents the torque resulting from the stellar wind. The simplest dimensionless expression of such a stress, which we take equatorially symmetric, is

\[\tau(\theta)=\sigma_{r \phi}/\rho = -A\sqrt{\frac{3}{4\pi}} \sin \theta, \]
where $[\sigma]$ is the dimensionless stress tensor and $A$ 
%\corr{\[K=A \sqrt{3/4\pi} \]}
is a positive constant that sets the amplitude of the prescribed (braking) stress. Although the wind imposes a small radial flow at the surface, we shall ignore it in this first approach, and impose a vanishing normal velocity on the outer shell. Finally,  we impose no-slip boundary conditions at the inner boundary, as justified by the important viscosity jump at the stellar envelope/core interface. Boundary conditions thus read

\begin{subequations}\label{eq:BC}
\begin{equation}
    u_r = u_\theta = u_\phi = 0  \qquad  {\rm at \quad } r=\eta 
\end{equation}
\begin{equation}\left\{ \begin{array}{l}
\displaystyle{u_r = r\frac{\partial}{\partial r}\left( \frac{u_\theta}{r}\right) = 0} \ ,  \\
\\
\displaystyle{\frac{\sigma_{r\phi}}{\rho}=r \frac{\partial}{\partial r}\left( \frac{ u_\phi}{r}\right) = -A\sqrt{\frac{3}{4\pi}} \sin \theta  }
\end{array}\right. \qquad {\rm at \quad } r=1 \ .
\end{equation}
\end{subequations}

\subsection{Numerical method}

We use a spectral decomposition and expand the fields in spherical harmonics for the angular part and Chebyshev polynomials for the radial part \citep[see][]{RV97}. %This allows to express all partial derivatives analytically. 
We write

\begin{equation}
\bq=\rho \buu = \sum^{+\infty}_{l=0} \sum^{+l}_{m=-l} u_{m}^{l}(r) \boldsymbol{R}^{m}_l + v_{m}^{l}(r) \boldsymbol{S}^{m}_l + w_{m}^{l}(r) \boldsymbol{T}^{m}_l \ ,
\end{equation}
where
\begin{equation}
  \boldsymbol{R}^{m}_l =  Y^{m}_l \be_r, \quad \boldsymbol{S}^{m}_l =  \bnabla Y^{m}_l, \quad \boldsymbol{T}^{m}_l = \bnabla \times  \boldsymbol{R}^{m}_l \ ,
\end{equation}
with $Y^{m}_l$ being the usual normalised scalar spherical harmonic function. Because the $\bq$-flow is divergenceless both in the incompressible and anelastic models, $v_{m}^{l}$ can be expressed as a function of $u_m^l$ only \citep{R87}. Projecting the vorticity equation (\ref{eq:nucst}) on $\boldsymbol{R}^{m}_l$ and $\boldsymbol{T}^{m}_l$ for an axisymmetric flow ($m=0$), we obtain the following system of equations for the radial parts

\begin{equation}\label{eq:syst}
\begin{cases}
\begin{aligned}
&\frac{\partial w^l}{\partial t} - \left[\Alm r^{l-1}\frac{\partial}{\partial r}\left(\frac{r \ulmm}{r^{l-1}}\right) + \Alp r^{-l-2}\frac{\partial}{\partial r}(r^{l+3} \ulp)\right]   =  E  \hll \ , \\
&\frac{\partial}{\partial t}\left( \Delta_l r u^l \right) + \Blm r^{l-1}\frac{\partial}{\partial r}\left(\frac{w^{l-1}}{r^{l-1}}\right) + \Blp r^{-l-2}\frac{\partial}{\partial r}(r^{l+2} \wlp) =-\frac{E}{r} \left(f^l - \frac{\partial r g^l}{\partial r} \right) \ , \\
&\vll=\frac{1}{r \Lambda} \frac{\partial r^2 \ull}{\partial r} \ ,
\end{aligned}
\end{cases}
\end{equation}
where $\Lambda=l(l+1)$. $\Alm$, $\Alp$, $\Blm$, and $\Blp$ are the coupling coefficients defined as

\begin{equation}
\Alm = A_l^{l-1}=\frac{1}{l}\sqrt{\frac{1}{(2l-1)(2l+1)}} \ ,  \quad \Blm = B_l^{l-1}=\sqrt{\frac{l(l^2-1)}{(2l-1)(2l+1)}} \ .
\end{equation}
The viscous force is projected on the spherical harmonics basis as
well, namely
\begin{equation}
\bff_{\rm visc} = \sum^{+\infty}_{l=0} \sum^{+l}_{m=-l} f_{m}^{l}(r) \boldsymbol{R}^{m}_l + g_{m}^{l}(r) \boldsymbol{S}^{m}_l + h_{m}^{l}(r) \boldsymbol{T}^{m}_l \ ,
\end{equation}
where

\begin{equation}
\begin{aligned}
&f^l = 2 \frac{\partial \rho}{\partial r} \frac{\Lambda b \vll - 2 b\ull}{r} + \frac{\Lambda \rho}{r} z^l + \frac{\partial }{\partial r} \left( \frac{4 \rho}{3} d^l \right) \ , \\ 
&g^l=\frac{1}{r}\frac{\partial}{\partial r} (r \rho z^l) + \frac{2}{r}\frac{\partial \rho}{\partial r} (b\ull - b\vll) + \frac{4 \rho}{3 r} d^l \ , \\ 
&h^l= \rho \Delta_l b \wl + r\frac{\partial \rho}{\partial r} \frac{\partial}{\partial r} \left( \frac{b \wl}{r}\right) \ ,  
\end{aligned}
\end{equation}
with 

\begin{equation}
z^l= \frac{1}{r}\frac{\partial r b \vll}{\partial r} - \frac{b \ull}{r}, \quad d^l = \frac{1}{r^2}\dr{r^2bu^l} - \frac{\Lambda b v^l}{r} \ ,
\end{equation}
and where $b(r)=1/\rho(r)$ is the inverse density function.

Finally, the boundary conditions in the rotating frame of reference are projected  on the spherical harmonics and read 

\begin{equation}\label{eq:BC_spect}
\begin{cases}
\begin{aligned}
&\ull = \vll = \wl = 0  \ , && {\rm at \ } r=\eta \\
\\
&\ull = r\frac{\partial }{\partial r}\left( \frac{b\vll}{r}\right) = 0 \ , \quad
r \frac{\partial }{\partial r}\left( \frac{ b\wl}{r}\right) = - A\delta_{l1}  \ ,\qquad && {\rm at \ } r=1 \ ,
\end{aligned}
\end{cases}
\end{equation}
where $\delta_{ij}$ is the Kronecker symbol.

%To validate our numerical method we first compute the spherical Taylor-Couette flow and we recover the results of \citet{Proudman1956}, \citet{stewar66}, and \citet{Dormy1998}, which we briefly expose in Appendix \ref{sec:ProudStew}. 

\section{The incompressible flow}\label{sec:incompr}

%\subsection{Description}
In this section, and as a first step, we solve equations (\ref{eq:nucst}) for constant density throughout the domain, namely for $\rho(r)= 1$, with the boundary conditions (\ref{eq:BC}), and with the initial condition $\buu= \boldsymbol{0}$. Of course, before the steady-state is reached, the flow is in a transient state during which the driving surface stress has not yet been fully communicated to the interior of the fluid. We now briefly analyse this transitional stage so as to estimate the transient time scale.
%Note that one should not confuse steady-state time scale and spin-down time scale, the latter can be seen as the time required for the angular velocity of every fluid particle to be modified by the applied boundary condition and is determined by the scaling of the radial velocity of the flow. We then estimate the time scale on which the Stewartson layer forms.

%We note that, as expected, the azimuthal velocity $u_\phi$ is quasi-geostrophic outside the vertical shear layers, thus motivating the inquiry into an analytical geostrophic solution. We also see that the meridional circulation is essentially concentrated in the Stewartson vertical nested shear layer, and we note that, contrary to the standard spherical Taylor-Couette flow where the angular velocity inside $\mathcal{C}$ is the average of that of the inner and outer shells, it is now almost zero in the rotating frame. That is because of the near-inviscid Taylor-Proudman theorem applied to the no-slip boundary condition at the inner container wall.
%Finally, we find the system to be divided in three distinct regions: inside and outside the tangent cylinder $\mathcal{C}$ circumscribing the inner sphere, and the separating narrow vertical shear layers.

\subsection{The transient phase}

As to assess the relevance of the steady-state approximation for geophysical and astrophysical applications, it is important to determine the time scales the flow characteristics are governed by. To do so,  we measure the time at which the evolution of the total angular momentum in the rotating frame may be considered to its end. We note that the change in angular momentum of the fluid is due to the difference between the torque exerted on the fluid by the outer boundary condition and the torque that the fluid exerts on the steady inner sphere. Therefore, a stationary state is reached when

\begin{equation}\label{eq:torque}
    \Delta \Gamma \equiv \Gamma(1) - \Gamma(\eta) = 0 \ ,
\end{equation}
where $\Gamma(1)$ and $\Gamma(\eta)$ are the torques about the $z$-axis exerted on the outer and inner boundary surface, respectively. The torque about the $z$-axis exerted on a layer located at some radius $r$ can be written 

\begin{equation}\label{eq:torque1}
\Gamma(r)=\int_{\partial r} \boldsymbol{\sigma} \cdot \be_z  dS \ ,
\end{equation}
where $\boldsymbol{\sigma}$ is the local stress applied on the sphere of radius $r$. We expand $\boldsymbol{\sigma}$ on the spherical harmonics basis, hence the torque about the $z$-axis reads

\begin{equation}\label{eq:torque2}
\Gamma(r)= - \int_{\partial r} r \sin \theta \rho(r) \sum_{l} r^2 t^l  \frac{\partial Y_l}{\partial\theta}  d\Omega  \ ,   
\end{equation}
where 

\begin{equation}
    t^l=r \frac{\partial}{\partial r} \left(\frac{w^l}{r} \right) \ .
\end{equation}
Using Legendre polynomials recurrence relations, (\ref{eq:torque2}) actually reads

     \begin{equation}
\Gamma(r)= 4 \sqrt{\frac{\pi}{3}} r^3  \rho(r)  t^1(r) \ .
 \end{equation}
 Applying boundary conditions (\ref{eq:BC_spect}), (\ref{eq:torque}) can finally be written
 
 \begin{equation}\label{eq:torque_diff}
   \Delta \Gamma  =  -4 \sqrt{\frac{\pi}{3}}  \left( A \rho_s + \eta^3 \frac{\partial w^1}{\partial r}\Bigg|_{r=\eta}\right) \ ,
 \end{equation}
 where the adimensional surface density $\rho_s=\rho(1)=1$ for the considered incompressible model.
 
\begin{figure}
\centering
     \includegraphics[width=0.6\textwidth]{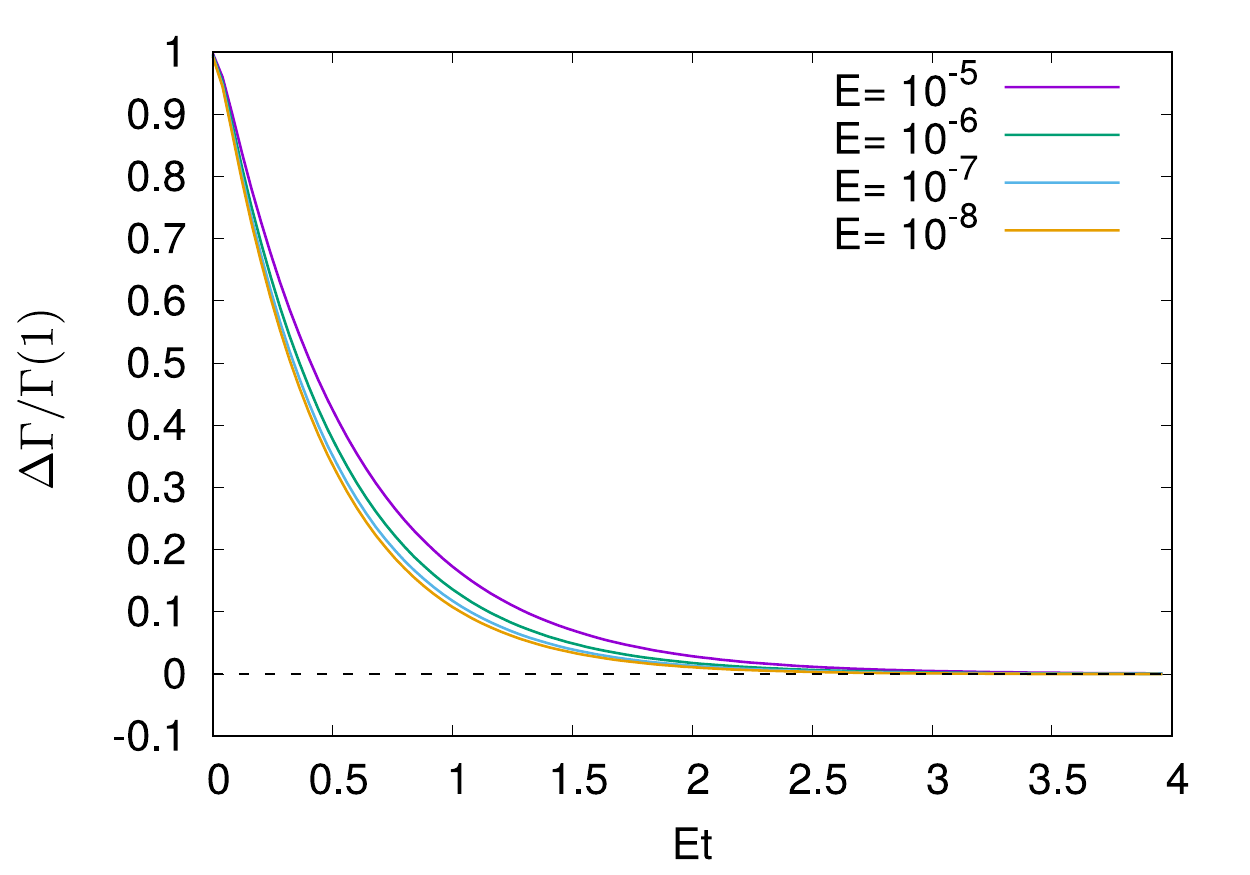}
   \caption{Relative difference between the torque applied on the outer sphere and the torque exerted by the fluid on the stationary inner sphere, as a function of the reduced time $E t$. We find the steady-state to be reached on a $O(E^{-1})$ time scale, that is on a viscous time scale, in the asymptotic regime of small Ekman numbers.}
\label{fig:torquediff}
\end{figure}

 We monitor the evolution of the relative difference between the torques $\Delta \Gamma/\Gamma(1)$ for various Ekman numbers and show the result in Fig.~\ref{fig:torquediff}. It shows that on a time scale $O(E^{-1})$ the stress is completely communicated to the interior flow, and a steady-state  is reached. %, and we arbitrarily consider the stationary state to be reached when the relative difference between the torques is of less than a percent.
 
 Of course in the astrophysical or geophysical context, it may become necessary to relax the constant rotation of the inner sphere. The torque exerted by the fluid on this boundary can make the inner sphere spin-down. According to the angular momentum theorem
 
 \begin{equation}
    \Gamma_*(\eta R)=\frac{d L^{\rm core}_z}{dt_*} = I_c \frac{d \Omega_c}{dt_*} \ ,
\end{equation}
where $\Gamma_*$ is the dimensional torque, $I_c$
%=(2/5)M\eta^2$
is the moment of inertia of the core assumed to be in solid-body rotation, and $L^{\rm core}_z$ is its total angular momentum. Starred quantities are dimensional.

The core spin-down then engenders an additional Euler force $\dot{\Omega_c}\be_z \times \br$ since our frame is attached to the core. This force can however be neglected if the timescale associated with the core spin-down is larger than the (viscous) timescale during which the angular momentum is redistributed. In that case, the flow can reach a quasi-steady state. In the present work, for simplicity, we shall use this assumption ($\dot{\Omega_c}=0$). Astrophysically, we justify its adoption by the Roche approximation, often used for massive stars, where the whole mass of the star is assumed to be in the core.

\begin{figure}
\centering
     \includegraphics[width=0.6\textwidth]{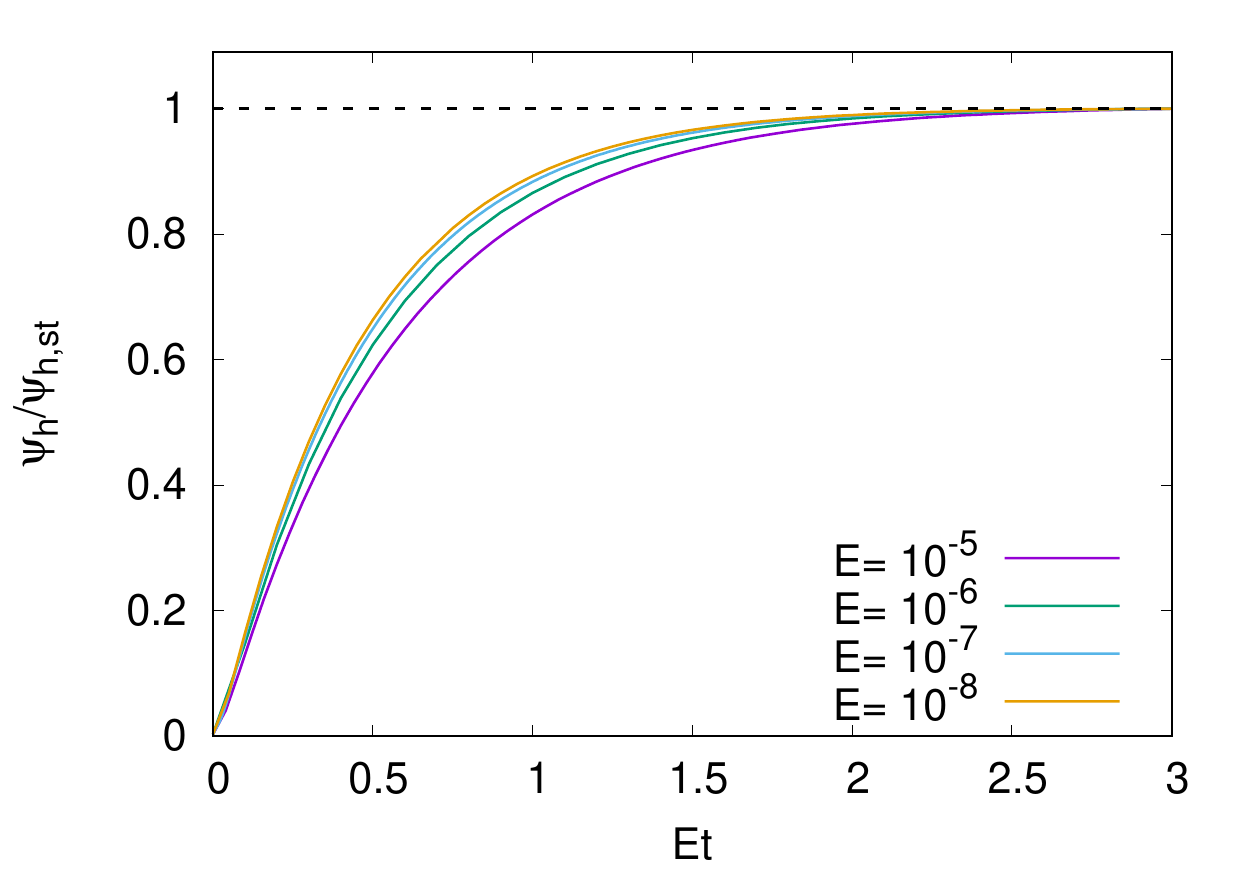}
   \caption{Amplitude of the stream function taken at $s=\eta$ and $z=1/2$ normalised by its value at the steady-state $\psi_h/\psi_{h,st}$, as a function of the reduced time $Et$ for various Ekman numbers. We find the Stewartson to appear within a few inner sphere revolutions and to evolve in a viscous time scale.}
\label{fig:psicentr}
\end{figure}

Another important time scale is that of the rise of the Stewartson layer. This layer appears as the result of the equatorial singularity of the inner Ekman boundary layer. It is well known that within an equatorial band of latitude $O(E^{1/5})$, the thickness of the Ekman boundary layer is $\delta_E = O(E^{2/5})$ \citep{RS63}. Hence, this singular equatorial viscous boundary layer is expected to be fully developed, and to initiate the development of the Stewartson layer, on the  $O(\delta_E^2 /E)=O(E^{-1/5})$ time scale, that is much shorter than the $O(E^{-1})$ viscous time scale. Likewise, the central Stewartson layer of thickness $O(E^{1/3})$ is fully developed on the $O(E^{-1/3})$ time scale, again, much shorter than the viscous time scale. We therefore expect the Stewartson layer to start developing within a few revolutions of the inner shell and to be fully developed on a time scale that is much shorter than the viscous one on which we expect the shear flow inside it to evolve towards the steady-state. We verify this in Fig.~\ref{fig:psicentr} showing the amplitude of the stream function $\psi_h/\psi_{h,st}$, taken at cylindrical radius $s=\eta$ and $z=1/2$, and normalised by its value at the steady-state, as a function of the reduced time $Et$ for various Ekman numbers.

%Another important time scale is that of the rise of the Stewartson layer. The main role of such layer is to remove a singularity in the gradient of azimuthal velocity resulting from the no-slip inner boundary condition.  %This singularity appears within a few revolutions thus on the $O(1)$ time scale. We therefore expect the Stewartson layer to appear on a $O(1)$ time scale and then to evolve on a $O(E^{-1})$ time scale towards the steady-state. Fig.~\ref{fig:psicentr} shows the amplitude of the stream function $\psi_h/\psi_{h,st}$, taken at cylindrical radius $s=\eta$ and $z=1/2$ normalised by its value at the steady-state, as a function of the reduced time $Et$ for various Ekman numbers. We find that the Stewartson layer does indeed appear almost immediately after the surface-stress is applied, and then evolves on a $O(E^{-1})$ time scale towards the steady-state.

The analysis of the transient phase preceding the settling of a steady-state thus underlines that the entire meridional flow evolves on a viscous time scale of order $E^{-1}$. In the astrophysical context characterised by extremely small Ekman numbers (typically $E<10^{-10}$), such a scaling implies that a steady-state of the radiative envelope of massive stars is not reached during their lifetime.

%and that appears as soon as the effects of rotation have made them felt, that is within a few revolutions, one could expect the Stewartson layer to  form on a  $O(1)$ time scale. The Ekman layers also form within a few revolutions thus on the $O(1)$ time scale  \citep{Green69}, the existence of the Stewartson to transport mass from one Ekman layer to the other is thus necessary on this time scale.

%Of course, it is well known that the Ekman boundary layers form as soon as the effects of rotation have made them felt, that is within a few revolutions \citep{Green69}. One could therefore expect the Stewartson layer to also form on this $O(1)$ time scale, as its roles are to remove a singularity in the gradient of azimuthal velocity resulting from the no-slip boundary condition on the inner sphere and to transport mass from one Ekman layer to the other.

%We now focus on the steady-flow. First, we derive the analytical expression of the geostrophic flow in these two regions, starting with the outer domain.

\subsection{The steady flow}

Even if it is not reached during the lifetime of the star, the steady state is worth studying since it owns many simple features and its structure is very similar to that of the transient flow.

%For completeness we nevertheless investigate the resulting steady-flow.
It is well known that the linear and stationary vorticity equation of an inviscid incompressible flow verifies  Taylor-Proudman theorem \citep[][]{Proudman1916,Taylor1921}, namely

\begin{equation}
(\be_z \cdot \bnabla ) \buu =  \boldsymbol{0}    \ .
\end{equation}

\begin{figure}
\centering
     \includegraphics[width=0.6\textwidth]{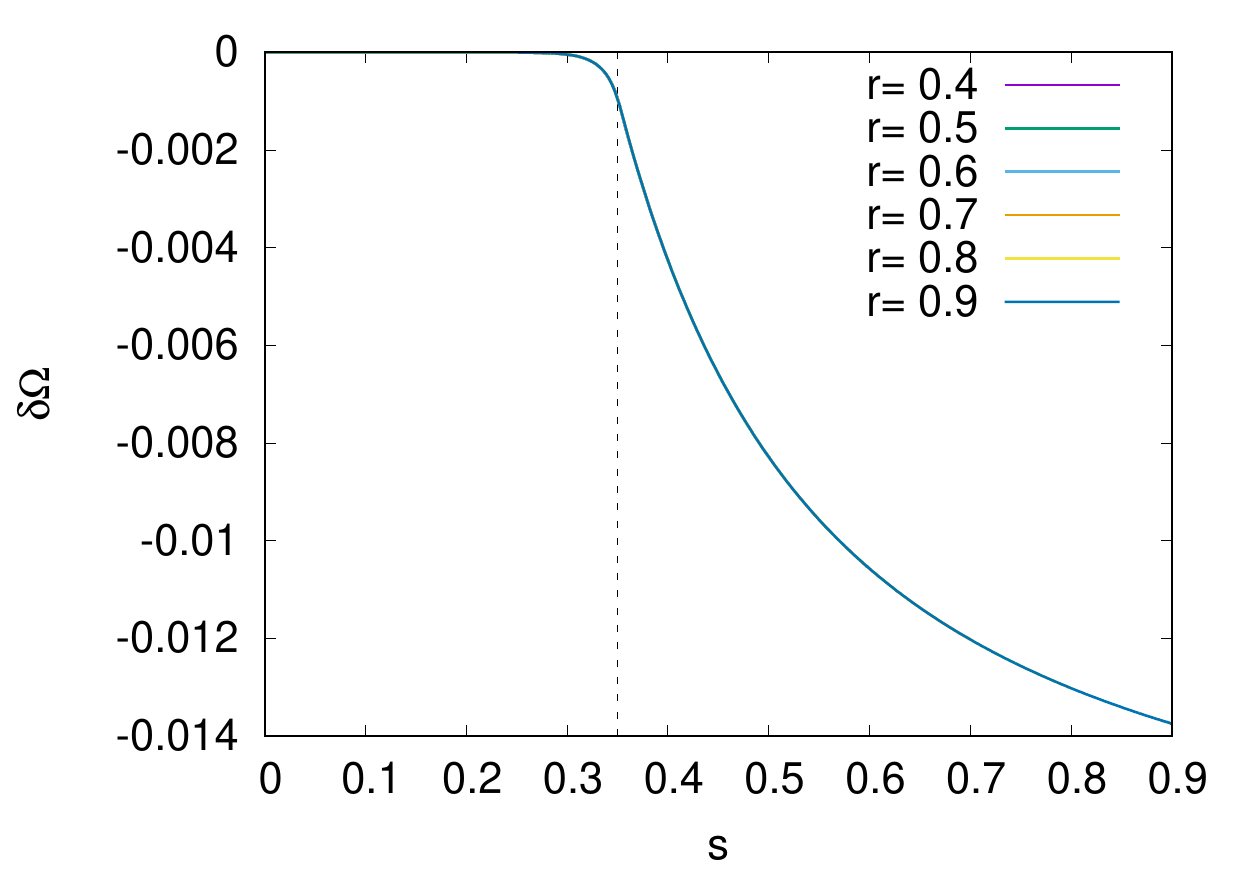}
   \caption{Normalised angular velocity  as a function of the cylindrical radial coordinate $s$ for $E=10^{-7}$, $\eta=0.35$, $A=0.01$, and for various $r$.}
\label{fig:uphi_stress}
\end{figure}

\begin{figure}
\centering
     \includegraphics[width=0.5\textwidth]{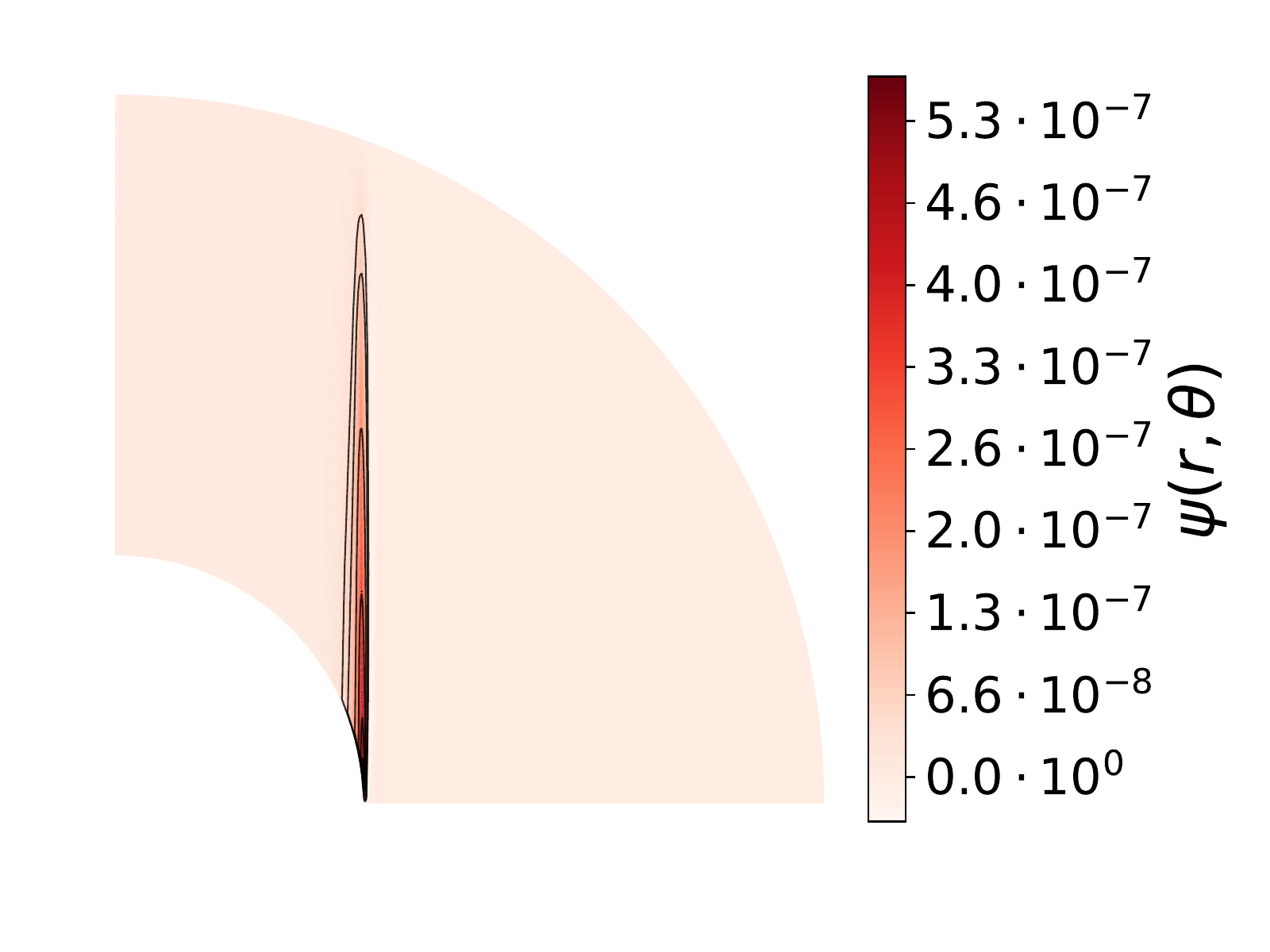}\hfill
     \includegraphics[width=0.5\textwidth]{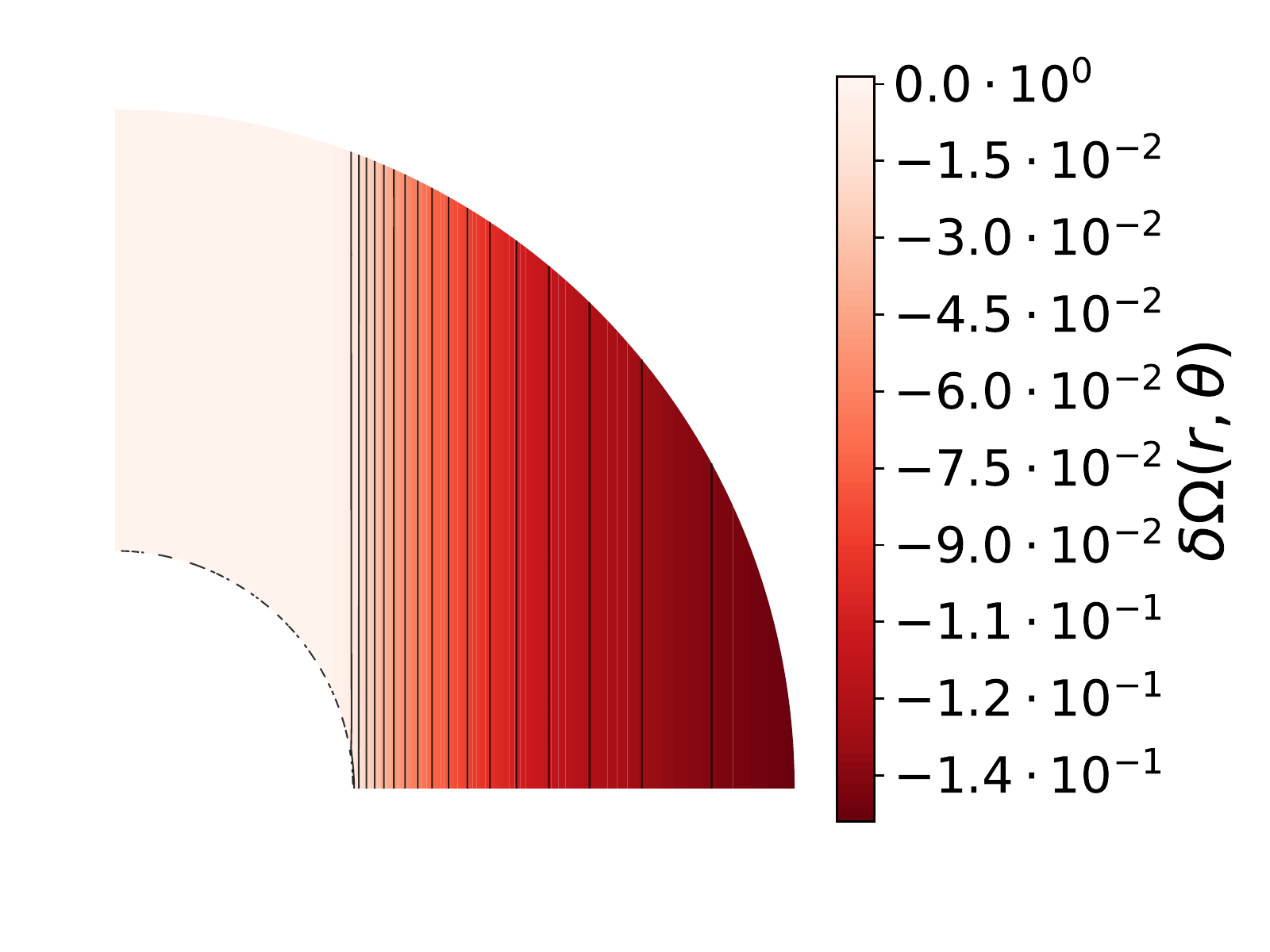}\hfill     
   \caption{Meridional view of the stream function $\psi$ (left) and of the differential rotation in the rotating frame of reference $\delta \Omega = (\Omega-\Omega_c)/2\Omega_c$ (right) for $E=10^{-7}$, $\eta=0.35$, and $A=0.01$. Three regions can be distinguished: a Stewartson layer on the outer boundary of a tangent cylinder $\mathcal{C}$ of radius $\eta$ separating two regions of weak amplitude circulation. The region inside the cylinder $\mathcal{C}$ corotates with the inner sphere and the region outside $\mathcal{C}$ rotates differentially with a columnar profile.}
\label{fig:incompr}
\end{figure}

For quasi-inviscid interior flows, that is for sufficiently small Ekman numbers, we thus expect a quasi-geostrophic solution for the velocity field. Fig.~\ref{fig:uphi_stress} shows the normalised angular velocity  as a function of the cylindrical radial coordinate $s$ for $E=10^{-7}$, $\eta=0.35$, and for various radial distances $r$. Fig.~\ref{fig:incompr} shows a 2D-view of the meridional circulation as well as the differential rotation in the rotating frame for the same model. 
The resulting flow is characterised by a nested Stewartson shear layer along the tangent cylinder $\mathcal{C}$ where the meridional circulation is essentially concentrated. This narrow shearing region separates two regions of quasi-geostrophic flow: the volume inside and outside $\mathcal{C}$. Inside $\mathcal{C}$, the no-slip conditions on the inner core impose an almost rigid rotation, while outside $\mathcal{C}$ a columnar differential rotation appears as a consequence of the surface stress.

\subsubsection{Flow outside the core tangent cylinder $\mathcal{C}$}\label{sec:BL}

We now seek for an analytical solution for both the primary and secondary flows outside $\mathcal{C}$ ($s > \eta$). In the limit of small Ekman numbers, the divergenceless velocity statisfies the Taylor-Proudman theorem, we thus seek a geostrophic solution for both primary and secondary flows outside the tangent cylinder $\mathcal{C}$. Such a solution usually does not satisfy the viscous boundary conditions, however. We thus decompose the dynamical variables as

\begin{equation}
    \buu = \overline{\buu} + \tilde{\buu} \ , \andet     p = \overline{p}+ \tilde{p} \ ,
\end{equation}
where overlined variable correspond to the interior geostrophic fields, and tilded variables are their boundary layer corrections. Introducing the $O(1)$ stretched boundary layer coordinate $\zeta = (1-r)/\sqrt{E}$, and using the geostrophic equilibrium relation $\be_z \times \overline{\buu} = - \bnabla \overline{p}$, the equation of motions, or Ekman boundary layer equation, reads

\begin{equation}\label{eq:ekBLeq}
    \be_z \times \tilde{\buu} = - \frac{\partial \tilde{p}}{\partial \zeta} \boldsymbol{n} + \frac{\partial^2 \tilde{\buu}}{\partial \zeta ^2} \ ,
\end{equation}
where $\boldsymbol{n} \equiv \be_r$ is the outwardly directed normal vector at the outer shell. Integrating the azimuthal component of (\ref{eq:ekBLeq}) from $0$ to $\infty$ yields, at the lowest order,

%\corr{
%\begin{equation}\label{eq:ekBLeqphi}
% \cos \theta \tilde{u}_\theta = \frac{\partial^2 \tilde{u}_\phi}{\partial \zeta^2}    \ ,
%\end{equation}
%and we write the outer Ekman layer $\theta$-directed volume flux}

%\corr{
%\begin{equation}
%    \tilde{Q}= \int_{-\infty}^1 \tilde{u}_\theta dr = \sqrt{E} \int_0^{\infty} \tilde{u}_\theta d \zeta %\ .
%\end{equation}}

%\corr{Integrating (\ref{eq:ekBLeqphi}) from $0$ to $\infty$, yields}

\begin{equation}\label{eq:tildeQ1}
  \cos \theta \tilde{Q} = - \frac{\partial \tilde{u}_\phi}{\partial \zeta}\Bigg|_{\zeta=0} \  \with \tilde{Q}=\int_0^{\infty} \tilde{u}_\theta d \zeta \; .
\end{equation}
$\tilde{Q}$ is the outer Ekman layer $\theta$-directed volume flux.
%\corr{
%\begin{equation}
%    \tilde{Q}= \int_{-\infty}^1 \tilde{u}_\theta dr = %\sqrt{E} \int_0^{\infty} \tilde{u}_\theta d \zeta \ . 
%\end{equation}}
Using the boundary condition for the prescribed tangential stress on the outer shell (\ref{eq:BC}), (\ref{eq:tildeQ1}) can be rewritten, at the lowest order

\begin{equation}\label{eq:tildeQ}
\cos \theta \tilde{Q} = \sqrt{E} \left[\tau(\theta) - \frac{\partial}{\partial r} \left( \frac{F(s)}{r} \right)\Bigg|_{r=1} \right] \ ,
\end{equation}
where $F(s)= \overline{u}_\phi$ is the geostrophic solution for the azimuthal velocity in the interior. 
To ensure mass conservation, the divergence of $\tilde{\buu}$ in the boundary layer generates further interior motion by establishing a weak amplitude secondary normal flow $\tilde{u}_r$ known as Ekman pumping. This radial velocity can be determined by integrating the mass conservation equation in the outer Ekman boundary layer

\begin{equation}\label{eq:pumping}
\begin{aligned}
\tilde{u}_r&= -\frac{\sqrt{E}}{\sin \theta} \frac{\partial \sin \theta \tilde{Q}}{\partial \theta}    \\
 &= -\frac{E}{\sin \theta} \frac{\partial}{\partial \theta} \frac{\sin \theta}{\cos \theta} \left[ \tau(\theta) - \frac{\partial}{\partial r}\left(\frac{F(s)}{r}\right)\Bigg|_{r=1} \right] \ .
\end{aligned}
\end{equation}

This equation shows that, near the outer shell, and in the Ekman layer, for the  boundary condition on the tangential stress to be ensured, there must exist a $O(\sqrt{E})$ tangential flow that, in turn, generates a $O(E)$ circulation in the interior.  We now express the geostrophic azimuthal velocity $F(s)$ as a function of the prescribed tangential surface-stress $\tau(\theta)$ driving differential rotation. To do so, we ensure that the no-penetration boundary condition $u_r=0$ at $r=1$ is enforced, namely

\begin{equation}\label{eq:ur0}
    s u_s + z u_z + \tilde{u}_r = 0 \qquad \rm at \; r=1\ ,
\end{equation}
where $u_s$ and $u_z$ are obtained from the azimuthal component of the momentum equation and from the mass conservation equation, respectively. Eq.~(\ref{eq:ur0}) finally yields a third order ordinary differential equation for $F(s)$

\begin{equation}
%\begin{aligned}
    -\frac{1-s^2}{s^2} \ds{} s\left( \nabla^2 - \frac{1}{s^2}\right)F + s\left( \nabla^2 - \frac{1}{s^2}\right)F + \ds{}\left(s^2 \ds{}\frac{F}{s}\right)  + sq\ds{} \frac{F}{s} = -K(1+q(s)) \ ,\label{eqF}
  %  \end{aligned}
\end{equation}
where $K=A \sqrt{3/4\pi}$. The solutions of~(\ref{eqF}) can be expressed as an integral functional of the prescribed stress \citep{friedlander76}, that is

\begin{equation}\label{eq:sol}
 F(s)=-s \int \frac{ds}{s^3 \sqrt{1-s^2}} \int s ds \int \frac{\partial}{\partial s} \left( \frac{-K s^2 }{\sqrt{1-s^2}}\right) ds    \ .
\end{equation}

The solution for the geostrophic azimuthal velocity outside the tangent cylinder $\mathcal{C}$,  avoiding singularities of the vorticity at $s = 1$, and accounting for the no-slip boundary condition at the inner shell finally reads

\begin{equation}
F(s)=-\frac{K}{3} \left( s \ln s - \frac{1}{s} + \alpha(\eta) s \right)   \ ,
\label{the_sol_F}
\end{equation}
with

\begin{equation}
    \alpha(\eta)= \frac{1}{\eta^2} - \ln \eta \ ,
\end{equation}
and for which

\begin{equation}\label{eq:usuz}
    u_s = E\left(\Delta - \frac{1}{s^2} \right)F(s)=- \frac{2KE}{3s} \ , \andet u_z=- \int_0^{z} \frac{1}{s}\frac{\partial s u_s}{\partial s} dz = 0 \ .
\end{equation}

%\corr{The differential rotation in the rotating frame of reference and outside the tangent cylinder $\mathcal{C}$ therefore reads}

%\corr{
%\begin{equation}\label{eq:omega_an}
%\delta \Omega=-\frac{K}{3} \left(  \ln (s/\eta) +\frac{1}{\eta^2}- \frac{1}{s^2}  \right)   \ .
%\end{equation}}

\begin{comment}

There, the geostrophic azimuthal velocity induced by the prescribed tangential stress on the outer sphere, is non-zero in the rotating frame of reference attached to the inner shell, and the secondary flow results from mass conservation in the Ekman layer, whose thickness is $O(\sqrt{E})$, at $r=1$. The detailed derivation of the geostrophic solution for the 3D flow outside the tangent cylinder $\mathcal{C}$  is presented in Appendix \ref{sec:A1} and yields the geostrophic solution for the differential rotation  outside $\mathcal{C}$

\begin{equation}\label{eq:omega_an1}
\delta \Omega=-\frac{K}{3} \left(  \ln (s/\eta) +\frac{1}{\eta^2}- \frac{1}{s^2}  \right)   \ .
\end{equation}
The radial cylindrical component of the velocity is
\begin{equation}\label{eq:us1}
    u_s= E(F'' + F'/s - F/s^2) \ ,
\end{equation}
where $F=s \delta \Omega$ is the geostrophic solution for the azimuthal velocity in the rotating frame of reference. The  $z$-component

\begin{equation}\label{eq:uz1}
    u_z=- \int_0^{z} \frac{1}{s}\frac{\partial s u_s}{\partial s} dz = -\frac{E}{s} \left[(sF')'' - (F/s)'\right]z \ . 
\end{equation}

\end{comment}

\begin{figure}
     \includegraphics[width=.5\textwidth]{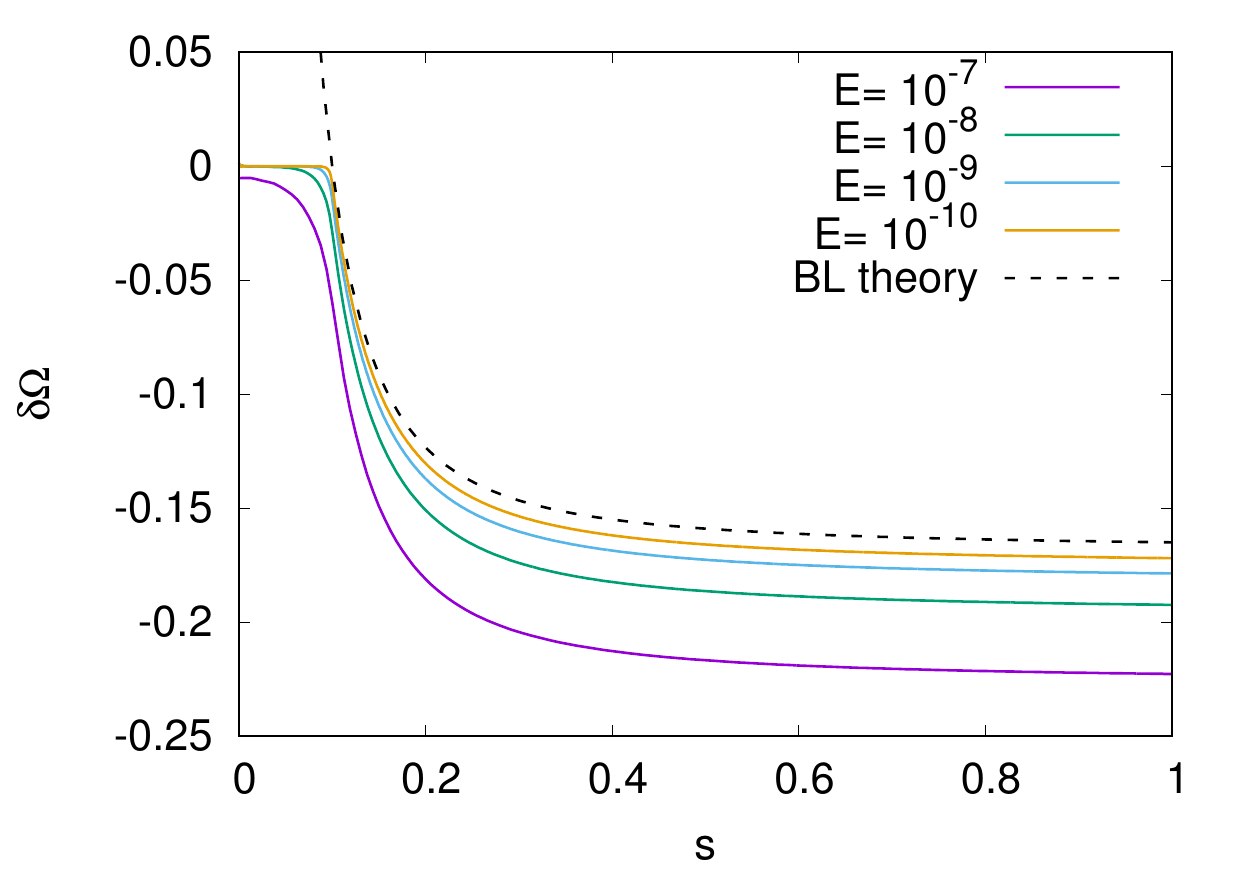}\hfill
     \includegraphics[width=.5\textwidth]{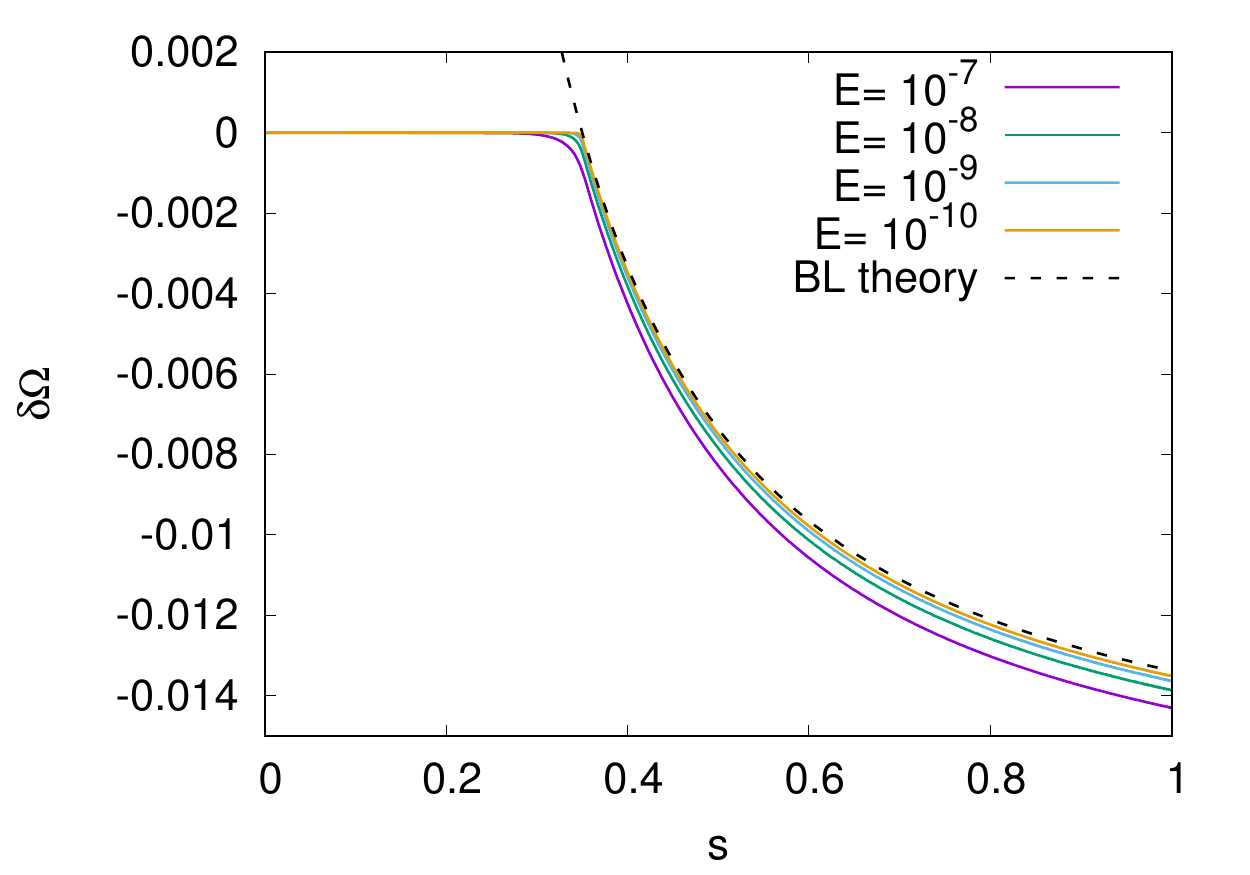}
   \caption{Angular velocity in the rotating frame of reference as a function of the cylindrical radial coordinate $s$ for various Ekman numbers, $r=0.7$, $\eta=0.1$ (left) and $\eta=0.35$ (right), and $A=0.01$. The  black dashed line corresponds to the analytical solution % from Ekman boundary layer analysis (see Eq.~\ref{eq:omega_an}).
   $\delta \Omega = F(s)/s$ (see Eq.~\ref{the_sol_F}).}
\label{fig:compar}
\end{figure}

\begin{figure}
\centering
     \includegraphics[width=0.6\textwidth]{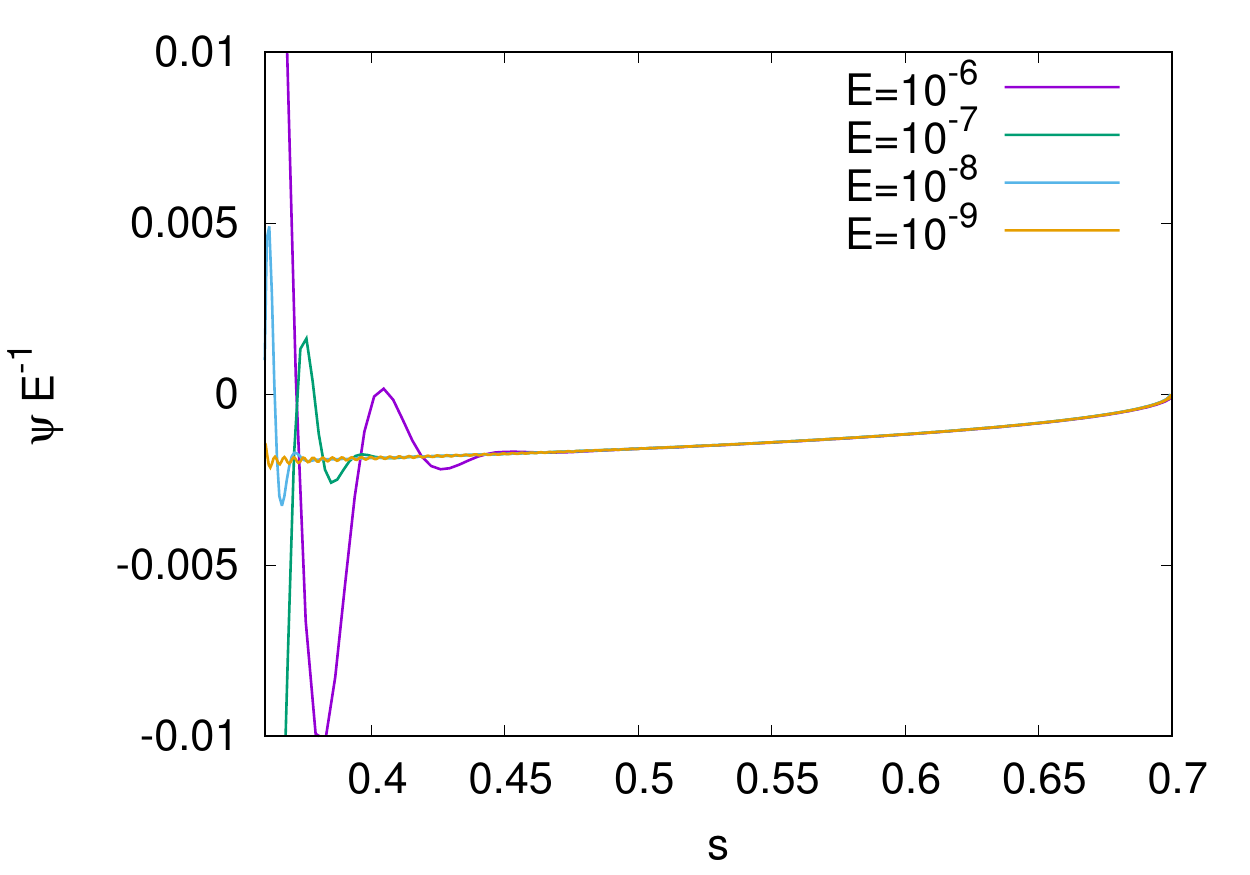}\hfill
   \caption{$\psi E^{-1}$ along a meridian at $r=0.7$ as a function of the cylindrical radial coordinate $s$ for various Ekman numbers, $\eta=0.35$, and $A=0.01$. The predicted $O(E)$ scaling of the secondary flow amplitude is verified outside the Stewartson layer, which is located at $s=0.35$.}
\label{fig:upol_stress}
\end{figure}

We compare the angular velocity profiles from full numerical solutions to the analytical expression $\delta \Omega = F(s)/s$, for various Ekman numbers and two inner core radii ($\eta=0.1$ and $0.35$) in Fig.~\ref{fig:compar}. The analytical asymptotic solution reproduces rather well our numerical results when the Ekman number is  below $10^{-9}$. Our boundary layer analysis also predicts the secondary meridional flow amplitude to be $O(E)$ outside the tangent cylinder $\mathcal{C}$. In Fig.~\ref{fig:upol_stress} we show the stream function, defined as

\begin{equation}\label{eq:streamf1}
    \frac{\partial \psi}{\partial r} = r \sin \theta u_\theta, \quad \frac{\partial \psi}{\partial \theta} = -r^2 \sin \theta u_r  \ ,
    \end{equation}
as a function of the cylindrical radial coordinate $s$ for various Ekman numbers. We note that $\psi$ indeed scales as $E$ outside the Stewartson layer, which appears as oscillations of $\psi$ near $s=\eta$. Furthermore, using (\ref{eq:usuz}), the stream function associated with the geostrophic solution for the velocity field reads

\begin{figure}
%\centering
%\includegraphics[width=0.5\textwidth]{G_mask_E1e-8_stress_r07_eta_01.pdf}\hfill
%\includegraphics[width=0.5\textwidth]{G_an_mask_E1e-8_stress_r07_eta_01.pdf}
\includegraphics[width=0.5\textwidth]{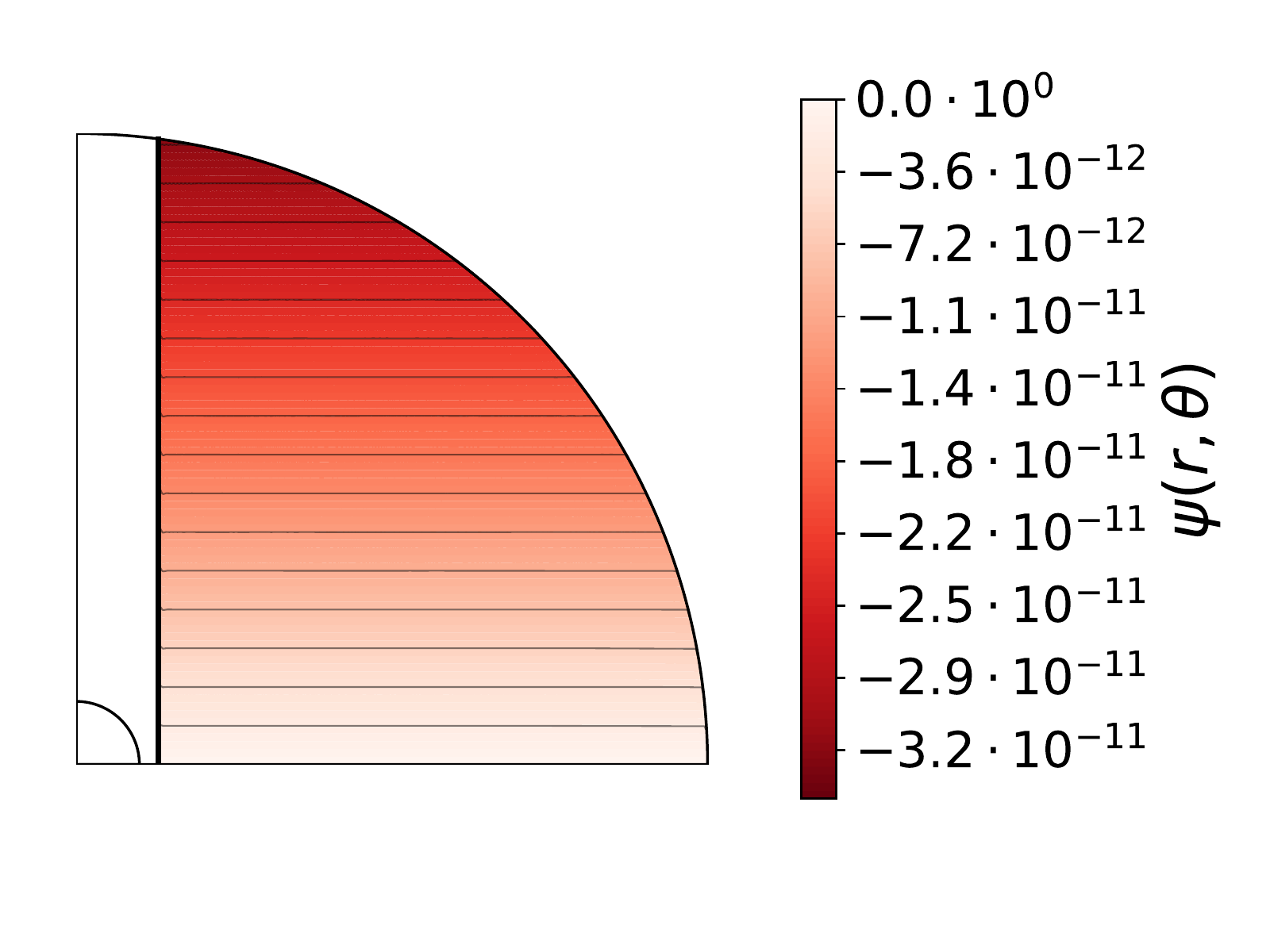}\hfill
\includegraphics[width=0.5\textwidth]{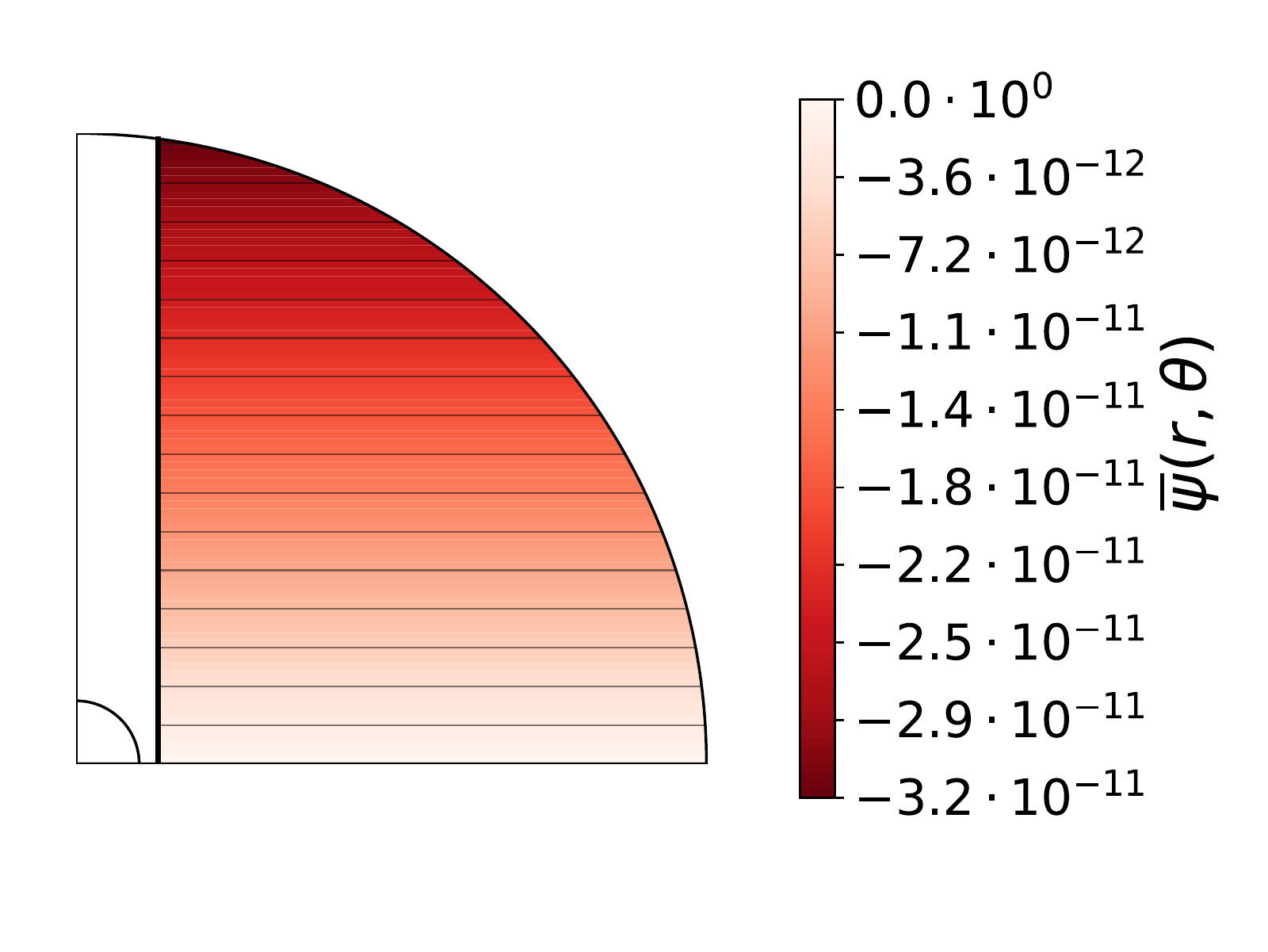}
   \caption{Meridional view of the stream function and streamlines from full numerical simulations (left), and from the boundary layer analysis (right), for $E=10^{-8}$, $\eta=0.1$, and $A=0.01$. The tangent cylinder $\mathcal{C}$ as well as the adjacent reconnecting layer are masked.}
\label{fig:G_compar}
\end{figure}

\begin{comment}

Now inserting the expression (\ref{eq:omega_an1}) in (\ref{eq:us1}) and (\ref{eq:uz1}) gives the meridional flow outside the tangent cylinder $\mathcal{C}$, namely

%The fact that this scaling remains valid close to the vertical shear layers, in particular for small Ekman numbers, points out the limited extent of the reconnecting layer where the $O(E)$ flow in the $s$-direction matches the quasi-vertical flow in the outer $O(E^{1/4})$ shear layer. This indicates the (spatially) limited impact of the vertical shear layers on the geostrophic analytical solution. We can actually go further and compare the stream functions from full numerical solutions to the analytical one in the asymptotic regime of small Ekman numbers. Replacing the expression for the geostrophic solution $F(s)$ in (\ref{eq:us}) and (\ref{eq:uz}) yields $u_z=0$ outside the tangent cylinder $\mathcal{C}$, 

\begin{equation}
    u_s= -\frac{2KE}{3s} \andet u_z=0\ ,
    \label{mer_circ_out}
\end{equation}
and the associated stream function reads

\end{comment}

\begin{equation}\label{eq:psi}
    \overline{\psi}(s,z)=-\frac{2KE}{3}z \ ,
\end{equation}
implying that streamlines are $z={\rm Cst}$ lines in a meridional plane. Hence, the outer Ekman boundary layer expels fluid in the $s$-direction, towards the Stewartson shear layer. The stream function $\psi$ as well as the corresponding streamlines, from Eq.~(\ref{eq:psi}) and from full numerical solutions are represented in Fig.~\ref{fig:G_compar}, where we have masked the Stewartson layer. We see that the analytical and numerical solutions nicely match. Interestingly, the flow (\ref{eq:usuz}) reconnects with the Stewartson layer and partly sources the upwelling flow in this region (see below Sect.~\ref{sec:Stew_inc}).

The foregoing solution shows that the stress-driven spin-down flow is rather different from the spherical Taylor-Couette flow outside the tangent cylinder. For such a flow, the amplitude of the meridional circulation in this region is vanishing exponentially away from the Stewartson layer \citep{Proudman1956,Dormy1998}. In our case there is a residual flow, exactly perpendicular to the rotation axis, directed to it, and which scales as the Ekman number.

%While we have recovered the asymptotic dimensionless spin-up timescale to be $E^{-1/2}$ for incompressible models with almost rigid rotation in the previous section, we find the latter to be $E^{-1}$, that is equal to the viscous diffusion timescale, when the flow is driven by an anisotropic surface stress instead.

\subsubsection{Flow inside the tangent cylinder $\mathcal{C}$}\label{sec:332}

Let us now investigate the geostrophic flow inside the tangent cylinder 
$\mathcal{C}$. Writing $u_\phi = \ell/(r \sin \theta)$ where $\ell$ is the specific angular momentum, the azimutal component of the momentum equation and vorticity equation respectively read \citep[e.g.][]{goldstein1938,Proudman1956,stewar66}

\begin{equation}\label{eq:Goldstein}
    \begin{cases}
    \begin{aligned}
    &\frac{\partial \psi}{\partial r}\cos \theta - \frac{1}{r}\frac{\partial \psi}{\partial \theta}\sin \theta = E D^2 \ell \\
   \\ &-\frac{\partial \ell}{\partial r}\cos \theta + \frac{1}{r}\frac{\partial \ell}{\partial \theta}\sin \theta = E D^4 \psi 
        \end{aligned}
    \end{cases}
\end{equation}
where

\begin{equation}
    D^2 = \frac{\partial^2}{\partial r^2}+ \frac{\sin \theta}{r^2} \frac{\partial}{\partial \theta} \left(\frac{1}{\sin \theta}\frac{\partial}{\partial \theta} \right) \ .
\end{equation}

The boundary conditions (\ref{eq:BC}) in terms of streamfunction and specific angular momentum in turn read

\begin{subequations}\label{eq:BC_stream}
\begin{equation}
    \psi = \frac{\partial \psi}{\partial r} = \ell = 0  \qquad  {\rm at \quad } r=\eta 
\end{equation}
\begin{equation}\left\{ \begin{array}{l}
\displaystyle{\psi=\frac{\partial^2 \psi}{\partial r^2} - 2 \frac{\partial \psi}{\partial r}=0} \ ,  \\
\\
\displaystyle{\frac{\partial \ell}{\partial r}-2\ell=-K \sin^2 \theta  }
\end{array}\right. \qquad {\rm at \quad } r=1 \ .
\end{equation}
\end{subequations}

In the Ekman boundary layers, (\ref{eq:Goldstein}) may be rewritten

\begin{equation}\label{eq:Goldstein_CL}
    \begin{cases}
    \begin{aligned}
    &\frac{\partial \psi}{\partial r}\cos \theta = E \frac{\partial^2 \ell}{\partial r^2}\\ \\
    &-\frac{\partial \ell}{\partial r}\cos \theta  = E \frac{\partial^4 \psi}{\partial r^4} \ .
            \end{aligned}
    \end{cases}
\end{equation}
We integrate the first equation over $r$ to obtain the fourth-order differential equation on $\psi$

\begin{equation}\label{eq:ED}
    \frac{\partial^4 \psi}{\partial r^4}= - \frac{\cos^2 \theta}{E^2} (\psi-\overline{\psi}) \ ,
\end{equation}
where $\overline{\psi}$ is a function of $\theta$ only that can be determined by the boundary conditions \citep{Proudman1956}. Let us first focus on the outer Ekman layer localised near $r=1$. The solution of (\ref{eq:ED}) satisfying the boundary conditions (\ref{eq:BC_stream}) as well as the condition $\psi (\xi \to \infty) = \overline{\psi}$, where $\xi=(1-r)\sqrt{\cos \theta/ 2E}$, reads

\begin{equation}\label{eq:sol1}
    \psi=\overline{\psi} \left( 1- e^{-\xi} \cos \xi \right) \ .
\end{equation}

From the first equation of (\ref{eq:Goldstein_CL}), we may further write

\begin{equation}
    \ell - \overline{\ell}= \overline{\psi}\alpha e^{-\xi}\sin(\xi-\pi/4) \ ,
\end{equation}
where $\alpha=\sqrt{\cos \theta /(2E)} \gg 1$ in the asymptotic limit of small Ekman numbers. The boundary condition (\ref{eq:BC_stream}b) then implies

\begin{equation}
   \frac{\partial \ell}{\partial r}\Bigg|_{r=1}- 2\ell(1)= -2\alpha^2\overline{\psi}- 2\overline{\ell} +2\alpha\overline{\psi} = -K\sin^2\theta \ ,
\end{equation}
where $\overline{\ell}$ is also a function of $\theta$ only that can be determined by the boundary conditions. Hence, the azimuthal and meridional components of the quasi-geostrophic velocity outside the Ekman and Stewartson shear layers, are not independent and satisfy the outer Ekman jump condition at the lowest order

\begin{equation}\label{eq:cond1}
     \overline{\psi}= \frac{E}{\cos\theta}(K\sin^2\theta-2\overline{\ell})   \ .
\end{equation}

In a similar way, the study of the inner Ekman layer with boundary conditions (\ref{eq:BC_stream}a) yields

\begin{equation}\label{eq:soll}
    \psi= \overline{\psi}\left(1-e^{-\xi'}(\cos \xi' + \sin \xi') \right) \ ,
\end{equation}
where $\xi'=(r-\eta)\alpha$, and the Ekman jump condition at the inner shell reads

\begin{equation}\label{eq:cond2}
    \overline{\ell}=-2 \overline{\psi} \alpha \ .
\end{equation}

Hence (\ref{eq:cond2}) implies $\overline{\psi}= O(\overline{\ell}\sqrt{E})$ and thus (\ref{eq:cond1}) implies $\overline{\psi}= O(E)$ and $\overline{\ell}= O(\sqrt{E})$. We may therefore rewrite (\ref{eq:cond1}) as

\begin{equation}
     \overline{\psi}= \frac{E}{\cos\theta}K\sin^2\theta + O(E^{3/2}) \ .
\end{equation}

Finally, the geostrophic solution for the velocity inside the tangent cylinder $\mathcal{C}$ ($s < \eta $) reads, at the lowest order,

\begin{equation}\label{eq:geostroph}
    \overline{\psi}(s) = \frac{EK s^2}{\sqrt{1-s^2}} \andet  \delta\Omega = \frac{\overline{\ell}(s)}{s^2} = - \frac{ \sqrt{2E}K}{\sqrt{1-s^2}} \left(1-\frac{s^2}{\eta^2}\right)^{1/4} \ ,
\end{equation}
and the meridional components of the geostrophic velocity at the lowest order

\begin{equation}\label{eq:geostroph_mer}
u_s =  \frac{1}{s} \frac{\partial \overline{\psi}}{\partial z} = 0 \andet    u_z =  -\frac{1}{s} \frac{\partial \overline{\psi}}{\partial s} = -\frac{E K}{\sqrt{1-s^2}} \left( 1 + \frac{1}{1-s^2}\right) \ .
\end{equation}

\begin{comment}
We may set the azimuthal velocity to zero at the outset. Indeed, within $\mathcal{C}$ the velocity meets no-slip boundary conditions on the inner shell, and this zero-velocity condition propagates in the whole tangent cylinder thanks to the Taylor-Proudman theorem, but up to viscous correction;  in other words

\begin{equation}
    (\be_z \cdot \bnabla ) \buu \simeq O(E) \ .
\end{equation}
Hence, in the considered region, the ageostrophic component of the velocity field is of order $E$ in the rotating frame. From the $\phi$-component of the momentum equation in cylindrical coordinates

\begin{equation}
    u_s= E (\Delta - 1/s^2) u_\phi \ ,
\end{equation}
we find the amplitude of the velocity in the cylindrical radial direction $u_s$ to be of order $E^2$. As in the study of the meridional circulation outside the tangent cylinder $\mathcal{C}$ presented in Sect.~\ref{sec:BL}%Appendix \ref{sec:A1}
, we enforce the no-penetration boundary condition $u_r=0$ on the outer shell. In this region, this condition reads (see Eq.~\ref{eq:ur0})

\begin{equation}
u_z \simeq - \frac{\tilde{u}_r}{\cos \theta } + O(E^2) \ ,
\end{equation}
since $u_s$ is \od{E^2}. Using the expression of Ekman pumping (\ref{eq:pumping}), we finally get

\begin{equation}\label{eq:uz_inner}
u_z = -\frac{E K}{\sqrt{1-s^2}} \left( 1 + \frac{1}{1-s^2}\right) + O(E^2)\quad {\rm for} \quad s< \eta \ .
\end{equation}

\end{comment}

\begin{figure}
\centering
          \includegraphics[width=0.49\textwidth]{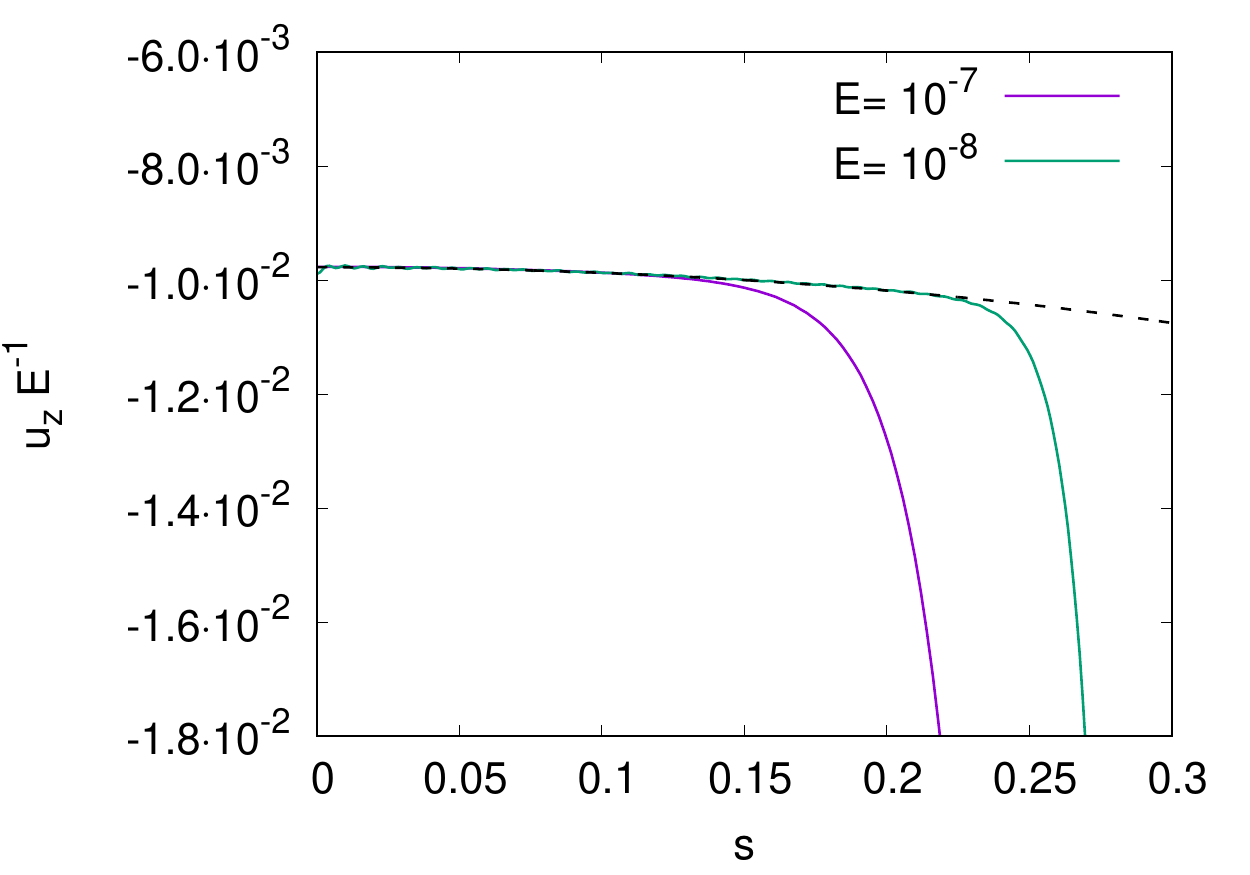}
           \includegraphics[width=0.49\textwidth]{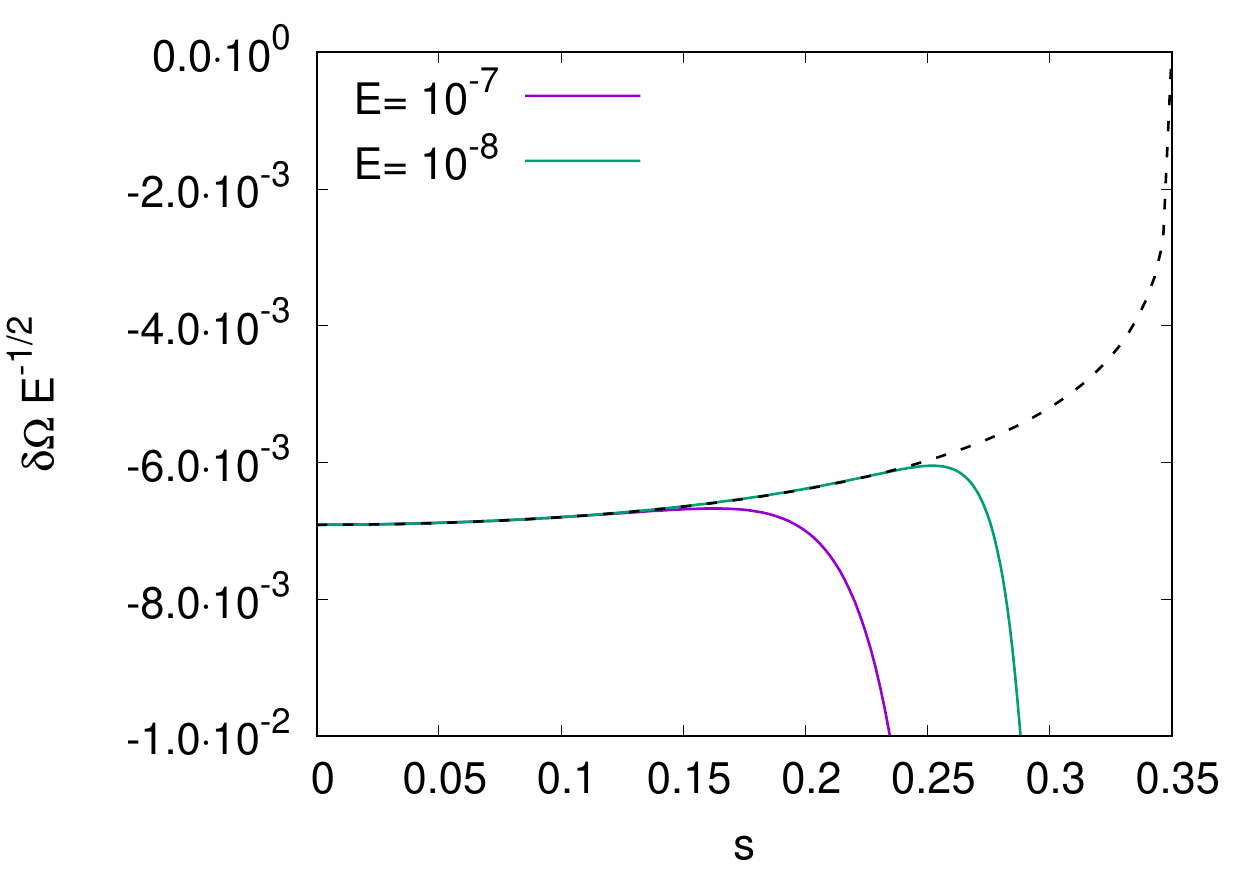} \hfill
    \caption{$z$-directed velocity $u_z E^{-1}$ (left) and angular velocity in the rotating frame of reference (right) inside the tangent cylinder $\mathcal{C}$ as a function of the cylindrical radial coordinate $s$, for $E=10^{-7}$ and $E=10^{-8}$, $r=0.7$, and $\eta=0.35$. The black dashed line corresponds to the analytical geostrophic solutions (\ref{eq:geostroph}) and (\ref{eq:geostroph_mer}).}
    \label{fig:uz_inside}
\end{figure}

\begin{figure}
\centering
     \includegraphics[width=0.6\textwidth]{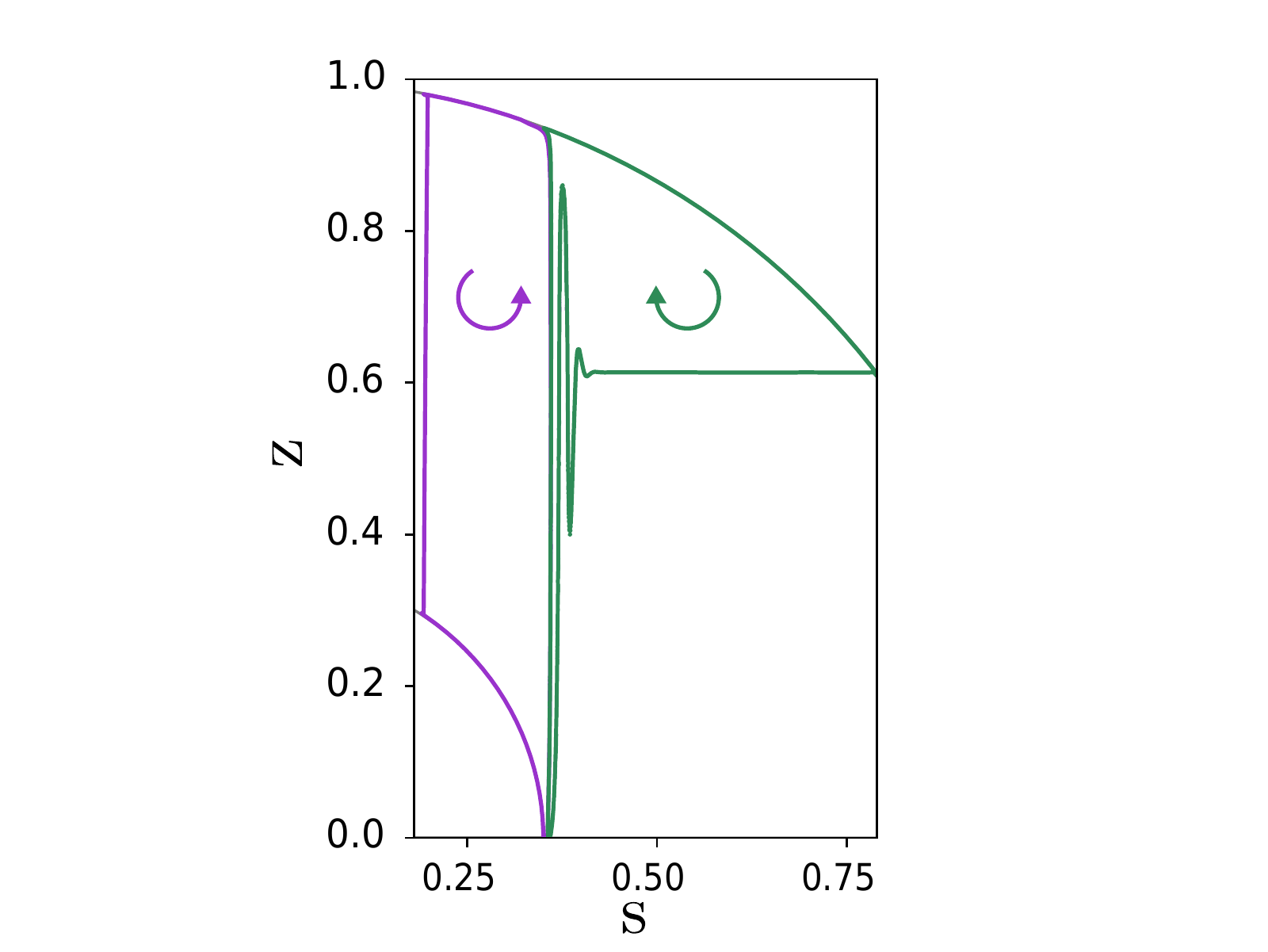}\hfill
   \caption{Meridional view of two arbitrarily selected streamlines, inside and outside the tangent cylinder $\mathcal{C}$. The arrows indicate the direction of propagation of the fluid along the streamlines.}
\label{fig:psi_single}
\end{figure}

 We compare these analytical $z$-directed velocity and angular velocity profiles with full numerical solutions  in Fig.~\ref{fig:uz_inside}, for various $E$. We find our analytical expression to be in good agreement with the numerical solutions. Remarkably, the meridional flow inside the tangent cylinder is parallel to the rotation axis and directed towards the inner core, which is quite different from the outer-$\calC$ meridional flow. 
 Once inside the inner Ekman layer, the flow heads towards the equator where the Ekman layer thickens and eventually changes scale at the equatorial singularity \citep[e.g.,][]{RS63,stewar66,Holl94a, Dormy1998,Marcotte2016}. The fluid  then returns to the outer Ekman layer following the Stewartson layer. Fig.~\ref{fig:psi_single} shows the meridional view of two arbitrarily selected streamlines, one inside and one outside the tangent cylinder $\mathcal{C}$. It illustrates the different shape of the secondary flow in the two regions, as well as the reconnecting shear layer redirecting the $s$-direction flow outside $\mathcal{C}$ to the Stewartson layer where it flows parallel to the rotation axis towards the outer boundary.

\subsubsection{Flow in the nested Stewartson shear layer}\label{sec:Stew_inc}

\begin{figure}
    \includegraphics[width=0.5\textwidth]{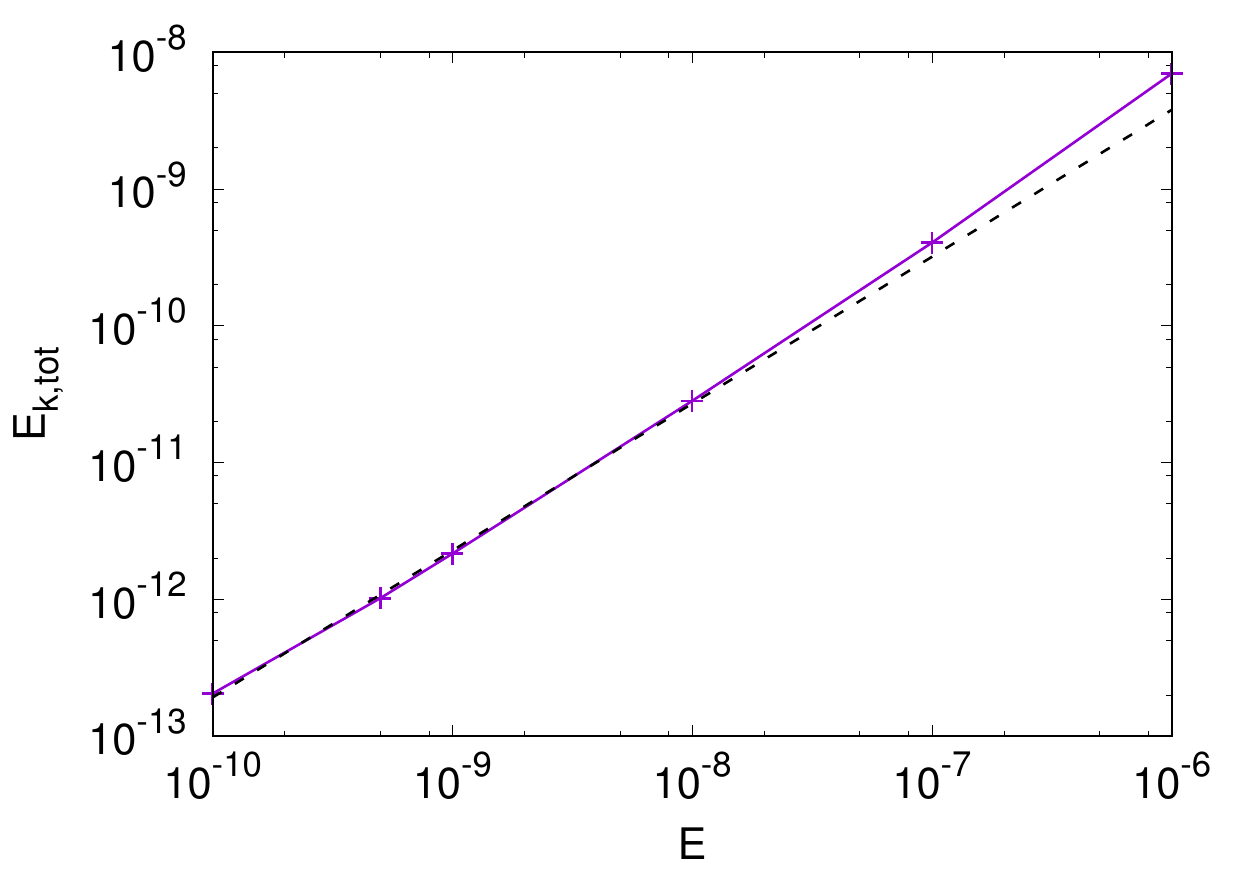}\hfill
     \includegraphics[width=0.5\textwidth]{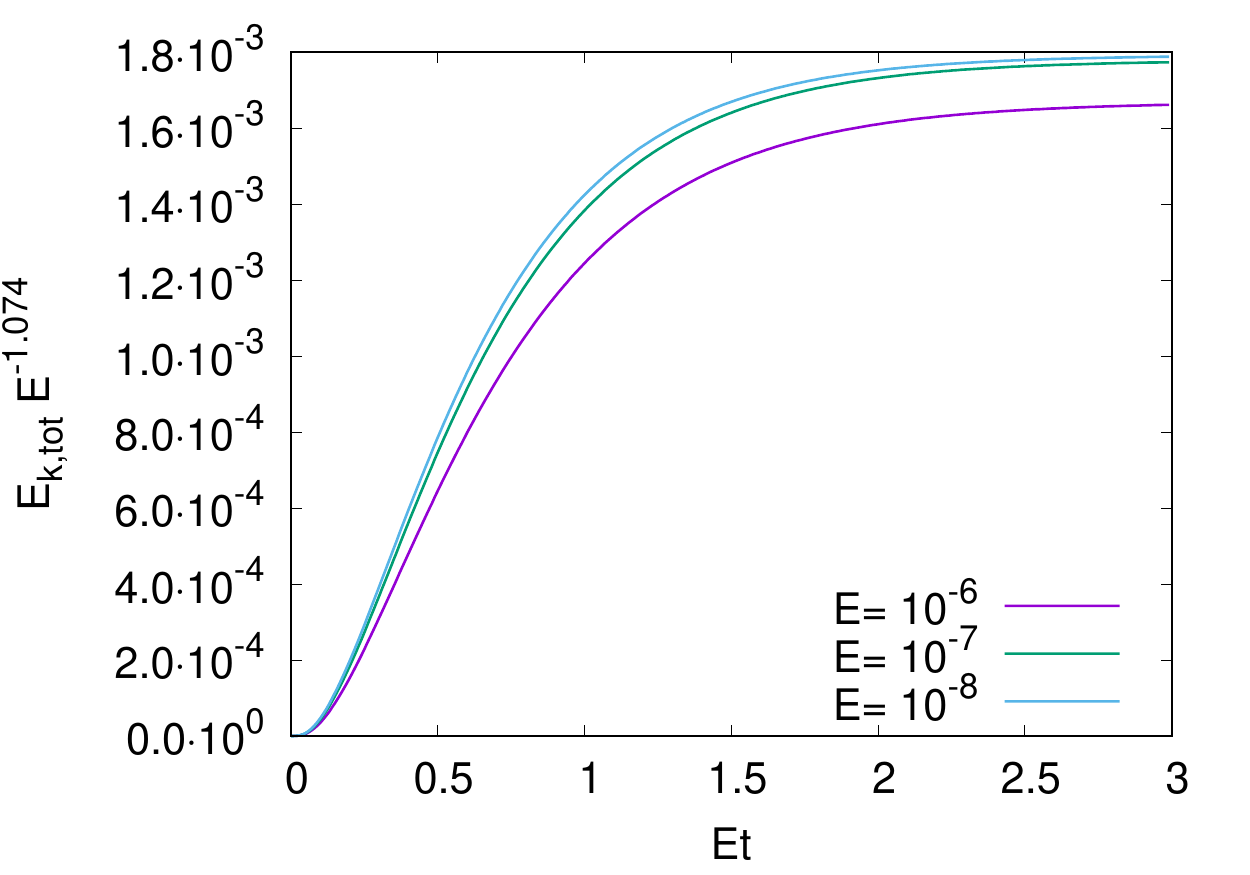}

    \caption{Left: Total meridional kinetic energy of the stationary flow as a function of the Ekman number, for $A=0.01$ and $\eta=0.1$. The black dashed line corresponds to the fit in the asymptotic regime ($E \lesssim 10^{-9}$) and yields $E_{k,\rm tot} \propto E^{1.074}$. Right: $E_{k,\rm tot} E^{-1.074}$ as a function of the reduced time $Et$ for various Ekman numbers, $A=0.01$, and $\eta=0.35$. }
\label{fig:ekmer}
\end{figure}

 In Fig.~\ref{fig:upol_stress} and Fig.~\ref{fig:uz_inside} we see that the meridional flow strongly deviates from the analytical solution as one nears the Stewartson layer. From the numerical solution, the amplitude of the meridional circulation turns out to be much stronger in this layer than outside it. This is obvious if we compute the total meridional kinetic energy of the secondary flow

\begin{equation}
E_{k,\rm tot}= \int V_m^2 dV     \ , 
\end{equation}
where $V_m^2= u_r^2 + u_\theta^2$. In Fig.~\ref{fig:ekmer}a we show $E_{k,\rm tot}$ as a function of the Ekman number. We find $E_{k,\rm tot}$ to scale as $E$ in the asymptotic regime $E\to 0$, whereas the contribution of the flows outside the Stewartson layer remains  $O(E^2)$. Hence, the Stewartson layer deserves some investigation.

We first recall that in the Taylor-Couette flow the Stewartson layer is a nested shear layer that can be split into three layers of width $E^{2/7}$, $E^{1/3}$ and $E^{1/4}$ \citep{stewar66}. The $E^{1/3}$-layer is the central layer, while the $E^{2/7}$-layer is on the inner side and the $E^{1/4}$ on the outer side. We now wish to find out if these three nested layers are present in our models. %Let  us start with the central ageostrophic layer of width $O(E^{1/3})$. 
%and the equilibrium between the two terms requires $\gamma=1/3$, indicating a vertical boundary layer of thickness $O(E^{1/3})$. 
Let us start by noticing that, according to (\ref{eq:pumping}), the Ekman pumping  may be written

\begin{equation}\label{eq:pumping_scal}
    \tilde{u}_r = - \sqrt{E}\left[\bnabla \times (\tilde{Q} \be_\phi)\right]\cdot\be_r \ . 
\end{equation}
Hence, introducing the adimensional cartesian stretched coordinate of order $O(1)$ in a shear layer parallel to the rotation axis, outside the tangent cylinder, and of thickness $O(E^{\gamma})$, $\xi=E^{-\gamma}(s-\eta)$, we get

\begin{equation}
\tilde{u}_r =O\left[ \left(E \bnabla \times \left(E^{-\gamma}\frac{\partial F}{\partial \xi} \be_\phi\right) \right)\right] \cdot \be_r = O(E^{1-2\gamma}) \ ,
\end{equation}
where $F=\overline{u}_\phi$. We conclude that, in such layers, the radial velocity resulting from the Ekman pumping/suction is of order $O(E^{1-2\gamma})$.
The mass conservation equation, in turn, reads

\begin{equation}\label{eq:masscons}
E^{-\gamma} \frac{\partial u_s}{\partial \xi} + \frac{\partial u_z}{\partial z}= 0 \ ,
\end{equation}
implying that in such a shear layer $u_z = O(E^{-\gamma}u_s)$, that is $u_z \gg u_s$ in the asymptotic regime of small Ekman numbers. 

Simple manipulations of the momentum and vorticity equations allow us to eliminate the velocity $\boldsymbol{u}$ and lead to the following sixth order partial differential equation for the pressure \citep{Green69}

\begin{equation}
\frac{\partial^2 p}{\partial z^2} + E^{2-6\gamma} \frac{\partial^6 p}{\partial \xi^6} = 0 \ ,
\end{equation}
representing a balance between Coriolis and viscous forces. The equality of these two terms thus requires $\gamma=1/3$, and indicates a  shear layer of thickness $O(E^{1/3})$ as expected. Hence, in this central ageostrophic layer, we expect 

\begin{equation}\label{eq:scaling13}
    u_s = O(E^{2/3}) \ ,  \andet u_z = O(E^{1/3})  \ .
\end{equation}

Let us now recall that the quasi-geostrophic $E^{1/4}$ layer is associated with the equilibrium between the Ekman pumping flow, and the internal friction of the $O(1)$ azimuthal flow \citep[e.g.,][]{stewar66,Barcilon1968}. This equilibrium is, in fact, exactly that which is in place in the entire region outside the tangent cylinder $\mathcal{C}$, and the $E
^{1/4}$-layer is therefore not needed. Indeed,  the azimuthal component of the momentum equation in the boundary layer of thickness $O(E^{\gamma})$ reads

\begin{equation}\label{eq:us_stew}
    u_s = E^{1-2\gamma} \frac{\partial^2 u_\phi}{\partial \xi^2} \ ,
\end{equation}
that is $u_s=O(E^{1-2\gamma})$ if we assume $u_\phi=O(1)$. In addition, (\ref{eq:masscons}) further implies $u_z=O(E^{1-3\gamma})$. However, the Ekman pumping still demands $u_z=O(E^{1-2\gamma})$ (see Eq.~\ref{eq:pumping_scal}). Hence, $\gamma=0$ and the equilibrium between the Ekman pumping flow and the internal friction of the $O(1)$ azimuthal flow does indeed occur in a $O(1)$ ``layer'' that is the entire $s>\eta$ region. Note that if we had considered no-slip boundary conditions at the outer shell, that is the case of a spherical Taylor-Couette flow, the $z$-directed velocity resulting from the Ekman pumping would be of order $O(\sqrt{E} \bnabla \times \overline{u}_\phi \be_\phi )= O(E^{1/2 - \gamma})$. The matching of the $u_z$-flux and the Ekman pumping implies that $1/2 - \gamma=1-3\gamma$, and thus $\gamma=1/4$ as expected.

Let us finally consider the case of a quasi-geostrophic $z$-directed free shear layer of thickness $O(E^{\gamma'})$ located inside the tangent cylinder $\mathcal{C}$ and near $s= \eta$. It follows that in such a region, $\ell$ is a function of $s$ only. Note that in the case of a spherical Taylor-Couette flow, $\gamma'=2/7$ and $u_\phi(s=\eta) = O(E^{1/28})$ \citep[e.g.][]{stewar66,Marcotte2016}.
In this region, the azimuthal component of the  moment equation (\ref{eq:Goldstein}a) integrated over $z$ reads

\begin{equation}\label{eq:cyl1}
    \psi = E z \frac{\partial^2 \ell }{\partial s^2} + f(s) \ ,
\end{equation}
where $f(s)$ is determined by applying the inner Ekman layer jump condition  (\ref{eq:cond2}) at $z=0$ and $s - \eta \ll 1$ \citep{stewar66}, yielding

\begin{equation}
    f(s) \simeq -\sqrt{\frac{E}{2}} \left[2 \left(1-\frac{s}{\eta}\right)  \right]^{-1/4} \ell \ .
\end{equation}

In addition, at $z=\sqrt{1-\eta^2}$, that is at $r=1$ and $s=\eta$, (\ref{eq:cond1}) reads

\begin{equation}
    \psi = \frac{E K \eta^2}{\sqrt{1-\eta^2}} \ ,
\end{equation}
and (\ref{eq:cyl1}) finally becomes

\begin{equation}\label{eq:diff27}
 E^{1-2\gamma'} \frac{\sqrt{1-\eta^2}}{\eta^4} \frac{d^2 \ell}{dx^2}-  E^{1/2 - \gamma'/4} (2x)^{-1/4} \ell =\frac{EK\eta^2}{\sqrt{1-\eta}}  \ ,
\end{equation}
where we have introduced the stretched $O(1)$ shear layer coordinate $x=(1-s/\eta)/E^{\gamma'}$. At this point, it is interesting to determine the correct balancing in (\ref{eq:diff27}). We note that the only balance implying the existence of an asymptotically narrow shear layer is that of the first two terms, the third one being of higher order. This balance enforces $\gamma' = 2/7$, that is a standard inner Stewartson shear layer of  thickness $O(E^{2/7})$. Hence, the second order ordinary differential equation for the specific angular momentum at the lowest order and in the $E^{2/7}$ shear layer reads

\begin{equation}
\frac{d^2 \ell}{dx^2}- \beta x^{-1/4}\ell=0\ ,
\end{equation}
where

\begin{equation}\label{eq:diff2}
    \beta = \frac{\eta^2}{2^{3/4}\sqrt{1-\eta^2}} \ .
\end{equation}

This equation can be transformed into a Bessel equation, and the solution asymptotically matching the solution (\ref{eq:geostroph}), that is decaying to zero as $x \to + \infty$, reads \citep[see also][]{stewar66,MS68,Marcotte2016}

\begin{equation}\label{eq:Ksol}
    \ell(x) =    \ell(0)C \sqrt{x}K_{4/7}\left(\frac{8}{7}\sqrt{\beta}x^{7/8}\right) \ ,
\end{equation}
where $K_{\nu}(z)$ is the modified Bessel function of second kind and $C$ is a constant to be determined. We  note that 

$$\lim_{x\to 0} \sqrt{x}K_{4/7}\left(\frac{8}{7}\sqrt{\beta}x^{7/8}\right)= \frac{ \Gamma\left( 4/7\right)}{2\beta^{2/7}} \left(\frac{7}{4}\right)^{4/7} + O(x)$$  \\
and thus, at the lowest order

\begin{equation}
    C= \frac{2\beta^{2/7} }{\Gamma\left( 4/7\right)} \left(\frac{4}{7}\right)^{4/7} \ .
\end{equation}

Finally, $d \ell / ds $ is rendered continuous across $s= \eta$ in the two quasi-geostrophic regions, that is 

\begin{equation}\label{eq:continuity}
   \frac{ d \ell}{ds}\Bigg|_{s=\eta^-} =  \frac{ d \ell}{ds}\Bigg|_{s=\eta^+} \ ,
\end{equation}
which yields the specific angular momentum at $s=\eta$

\begin{equation}
    \ell(0) =    -\frac{K\eta^2 E^{2/7}}{3\beta^{4/7}}  \left( \frac{2}{\eta^2} +1 \right) \left(\frac{7}{4}\right)^{1/7}  \frac{\Gamma(4/7)}{\Gamma(3/7)} = O(E^{2/7})\ .
\end{equation}

\begin{figure}
\centering
    \includegraphics[width=0.5\textwidth]{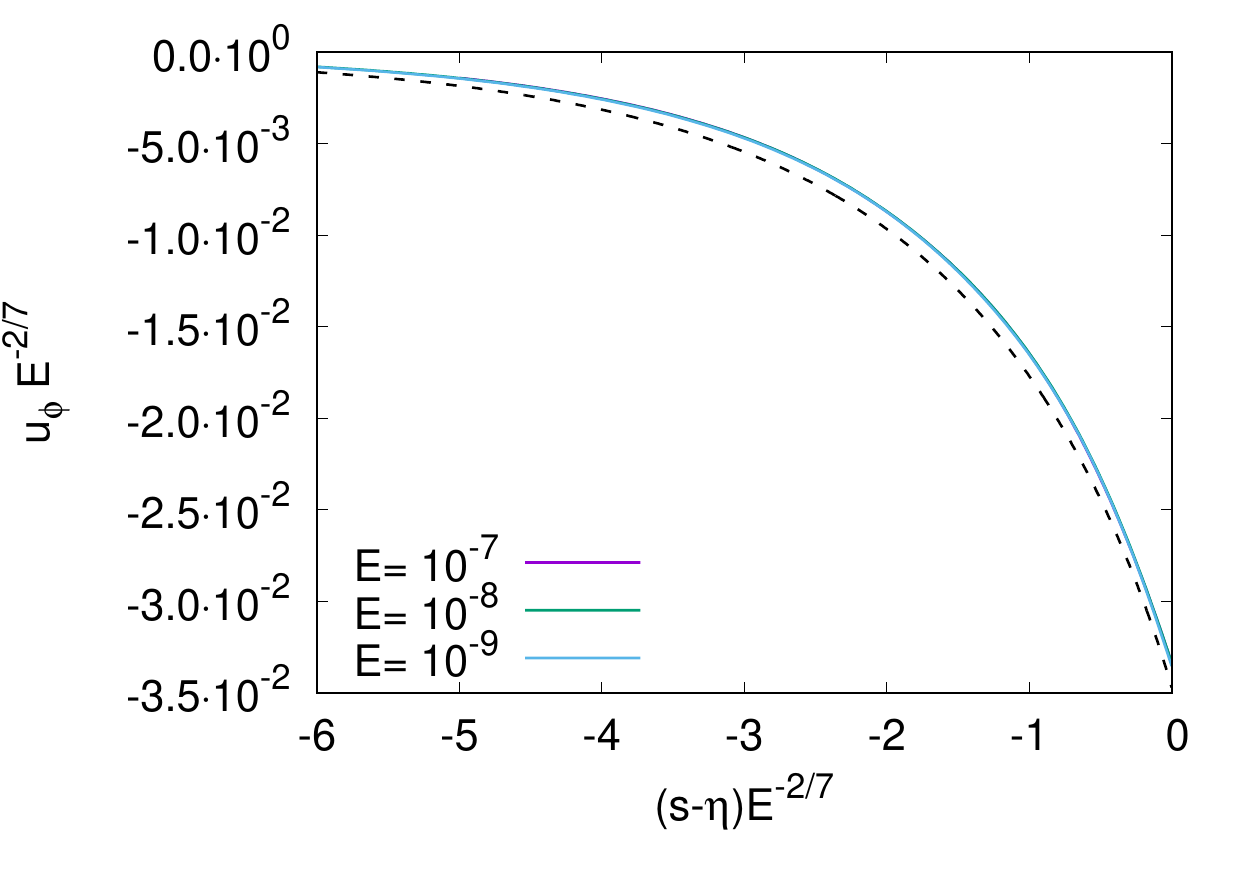}\hfill
        \includegraphics[width=0.5\textwidth]{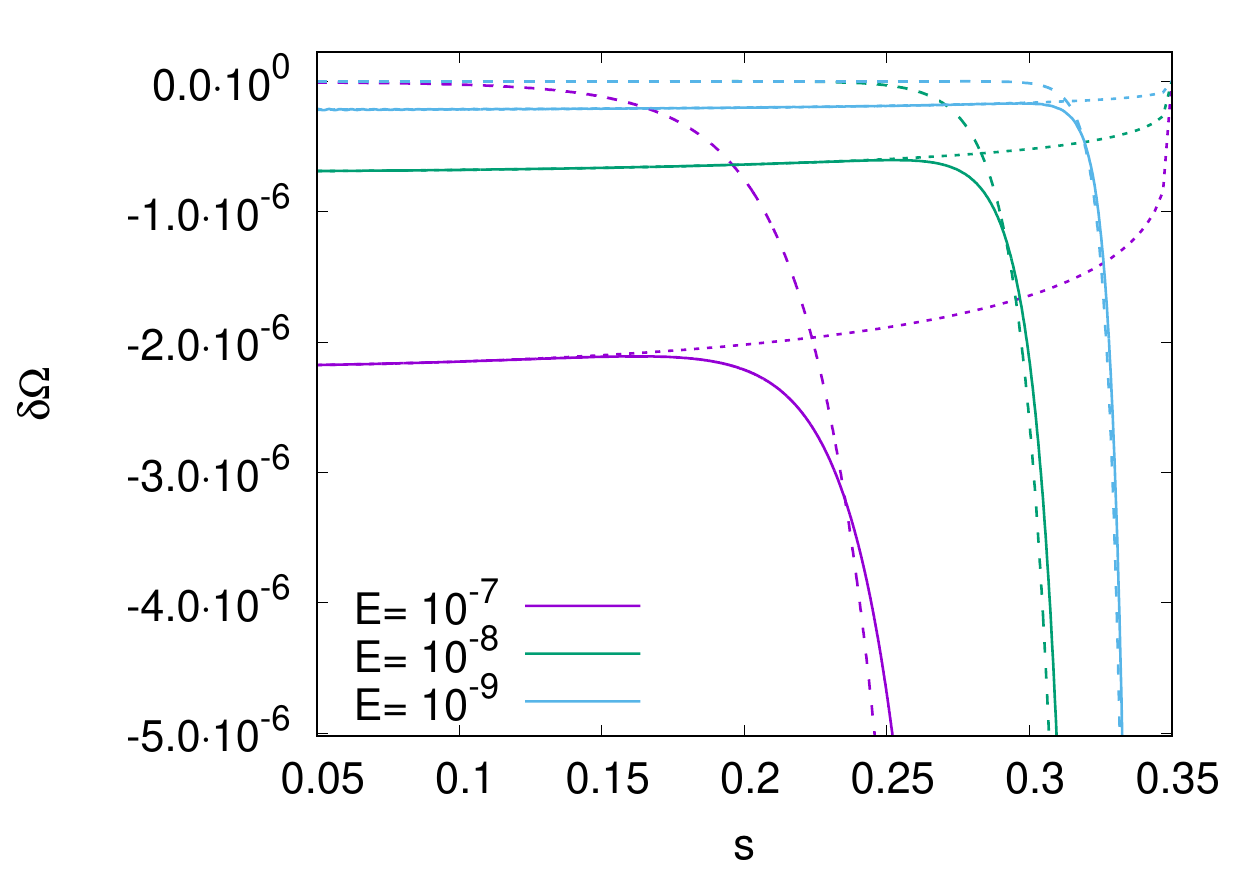}\hfill
   \caption{Left: $u_{\phi} E^{-2/7}$ as a function of the stretched cylindrical radial coordinate $(s-\eta)E^{-2/7}$ for various Ekman numbers, $\eta=0.35$, $z=0.7$, and $A=0.01$. The black dashed line corresponds to the analytical geostrophic solution (\ref{eq:Ksol}). Right: Angular velocity in the rotating frame of reference as a function of the cylindrical radial coordinate $s$. The dashed lines correspond to the analytical solutions (\ref{eq:geostroph}) and (\ref{eq:Ksol}).}
\label{fig:uphi2}
\end{figure}

We verify the shear layer thickness as well as the analytical azimuthal velocity profile (\ref{eq:Ksol}) comparing it with full numerical solutions in Fig.~\ref{fig:uphi2}a. The lowest order solution for the azimuthal velocity in the entire  $s \leq \eta$ domain is shown in Fig.~\ref{fig:uphi2}b. Finally, using (\ref{eq:cyl1}), the corresponding stream function reads

\begin{equation}\label{eq:psi27}
    \psi(x)= -\frac{2K\eta^2 E^{5/7} }{3\beta^{2/7}\Gamma(3/7)} \left( \frac{2}{\eta^2} +1 \right)  \left(\frac{4}{7}\right)^{3/7} x^{1/4} K_{4/7} \left( \frac{8}{7}\sqrt{\beta}x^{7/8}\right) \left[ \frac{\beta z}{\eta^2} - \frac{1}{2^{3/4}} \right] \ ,
\end{equation}
and thus 

\begin{equation}\label{eq:scaling27}
\begin{aligned}
    u_s &=-\frac{B_s E^{5/7}}{1-xE^{2/7}} x^{1/4} K_{4/7} \left( \frac{8}{7}\sqrt{\beta}x^{7/8}\right) \\
    u_z &= \frac{B_z E^{3/7}}{x^{3/4}(1-xE^{2/7})}\left(K_{4/7}\left( \frac{8}{7}\sqrt{\beta}x^{7/8}\right) + 4 \sqrt{\beta}x^{7/8}K_{3/7}\left( \frac{8}{7}\sqrt{\beta}x^{7/8}\right)\right)  \ ,
    \end{aligned}
\end{equation}
in the $E^{2/7}$-layer, where 
\begin{equation}
    B_s =  \frac{2K \beta^{5/7} }{3 \eta \Gamma(3/7)} \left( \frac{2}{\eta^2} +1 \right)  \left(\frac{4}{7}\right)^{3/7} 
\end{equation}
and 

\begin{equation}
    B_z =   \frac{K\left( 2 + \eta^2 \right)  }{6 \eta^2 \beta^{2/7}\Gamma(3/7)} \left(\frac{4}{7}\right)^{3/7} \left( \frac{\beta z}{\eta^2} - \frac{1}{2^{3/4}} \right) \ .
\end{equation}
%\begin{equation}\label{eq:scaling27}
%    u_s = O(E^{5/7}) \ ,  \andet u_z = O(E^{3/7})  \ ,
%\end{equation}

\begin{figure}
\centering
     \includegraphics[width=0.7\textwidth]{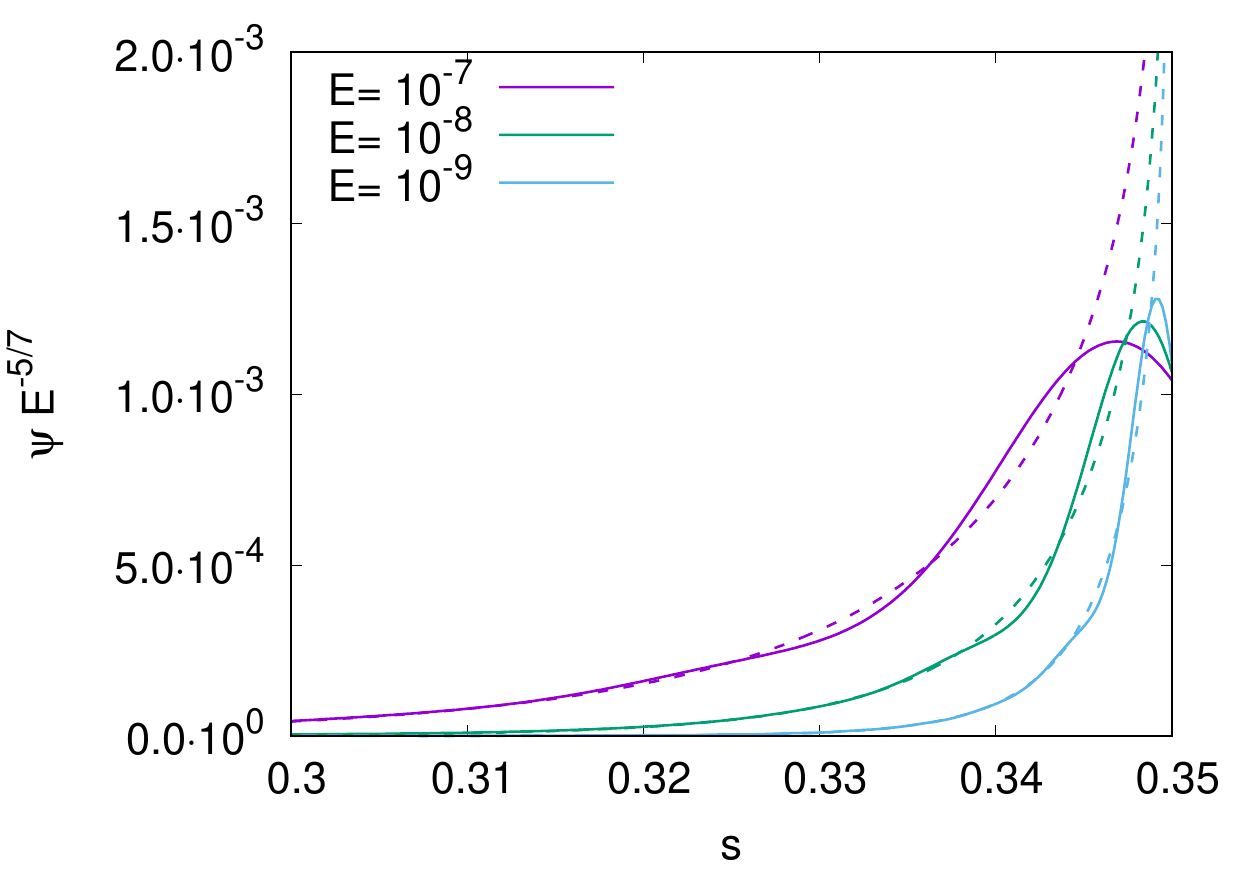}
   \caption{$\psi E^{-5/7}$ as a function of the cylindrical radial coordinate $s$ for various Ekman numbers, $\eta=0.35$, $z=0.7$, and $A=0.01$. The dashed lines correspond to the analytical geostrophic solutions (\ref{eq:psi27}).}
\label{fig:3}
\end{figure}

 We verify the validity of the solution (\ref{eq:psi27}) in the $E^{2/7}$-layer away from the $E^{1/3}$-layer in Fig.~\ref{fig:3}. We note that $\psi$ and its derivatives remain discontinuous across $s=\eta$, hence, contrary to the differential rotation (\ref{eq:Ksol}), the solution (\ref{eq:psi27}) is not valid in the close vicinity of $s=\eta$, and it is the ageostrophic $E^{1/3}$-layer that smoothes them out. Furthermore, since $3/7>1/3$, the $z$-component of the velocity is expected to be maximum in the $E^{1/3}$ layer.

%The balance between the first and third term assuming the second one to be of higher order implies $\gamma'=0$ which violates our assumption that the second term is of higher order.

%The azimuthal component of the momentum equation (\ref{eq:us_stew}) finally implies 

%\begin{figure}
%\centering
%    \includegraphics[width=0.7\textwidth]{uphi_left.pdf}\hfill
%   \caption{$u_{\phi} E^{-2/7}$ as a function of the stretched cylindrical radial coordinate $(s-\eta)E^{-2/7}$ for various Ekman numbers, $\eta=0.35$, $z=0.7$, and $A=0.01$.}
%\label{fig:uphi}
%\end{figure}

\begin{figure}
    \includegraphics[width=0.5\textwidth]{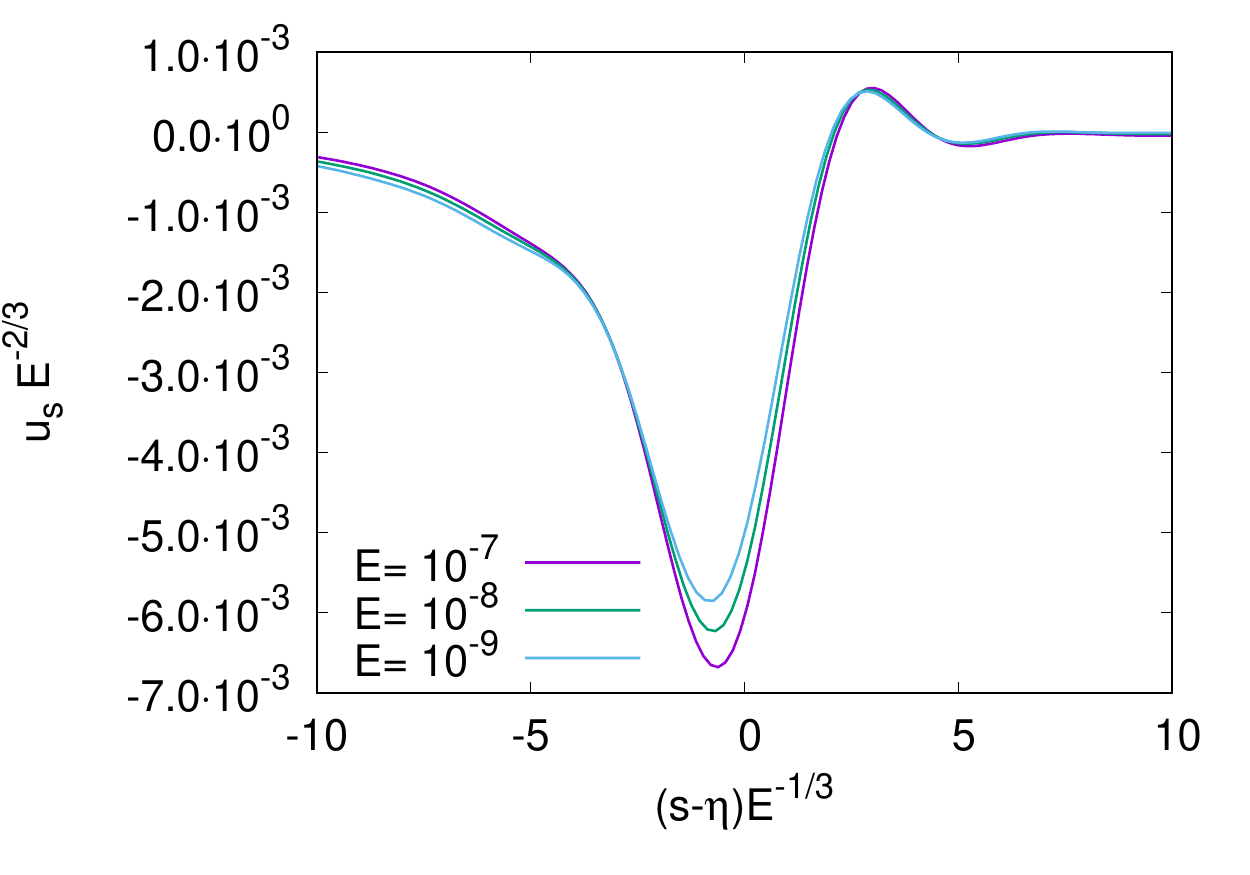}\hfill
     \includegraphics[width=0.5\textwidth]{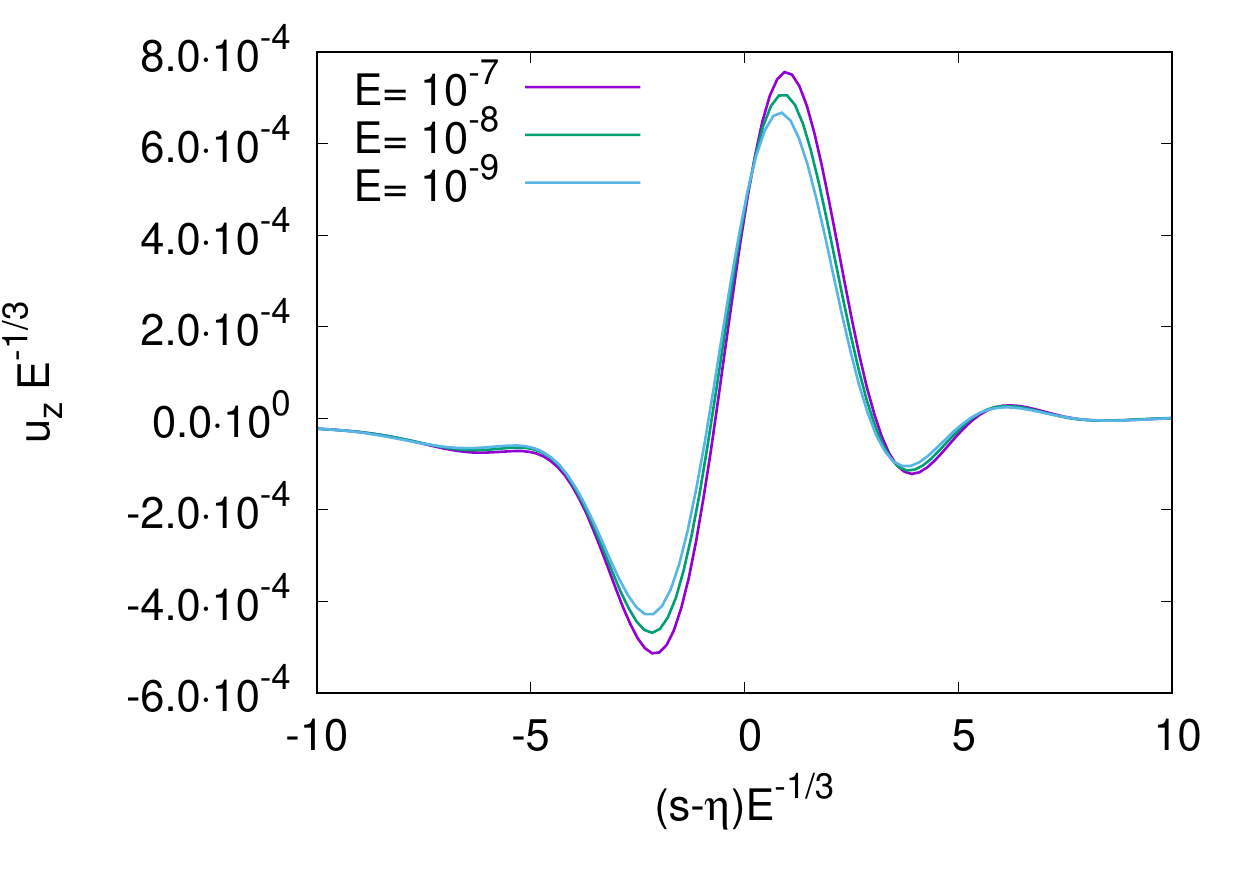}
   \caption{$u_s E^{-2/3}$ (left) and $u_z E^{-1/3}$ (right) as a function of the stretched cylindrical radial coordinate $(s-\eta)E^{-1/3}$ for various Ekman numbers, $\eta=0.35$, $z=0.7$, and $A=0.01$.}
\label{fig:uz_us_centre}
\end{figure}

\begin{figure}
    \includegraphics[width=0.5\textwidth]{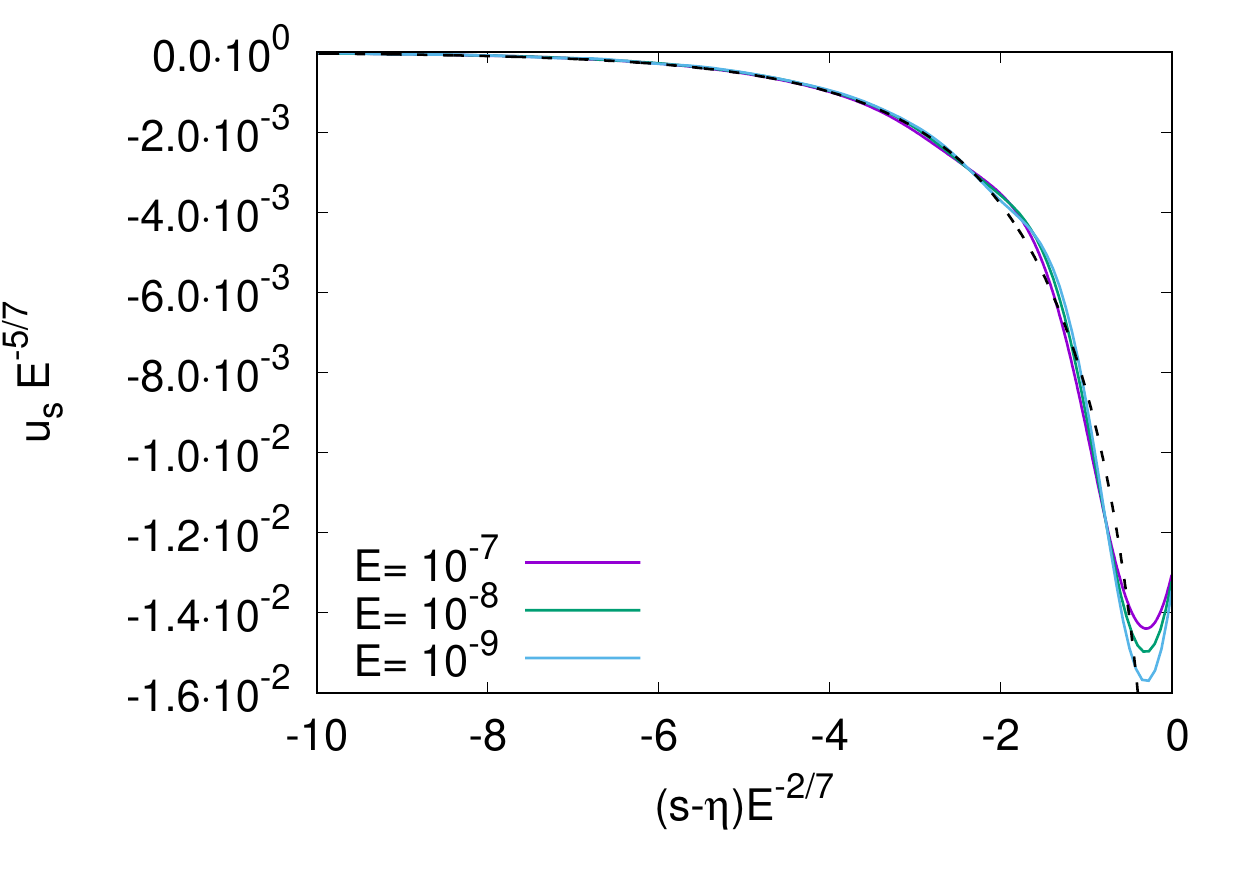}\hfill
     \includegraphics[width=0.5\textwidth]{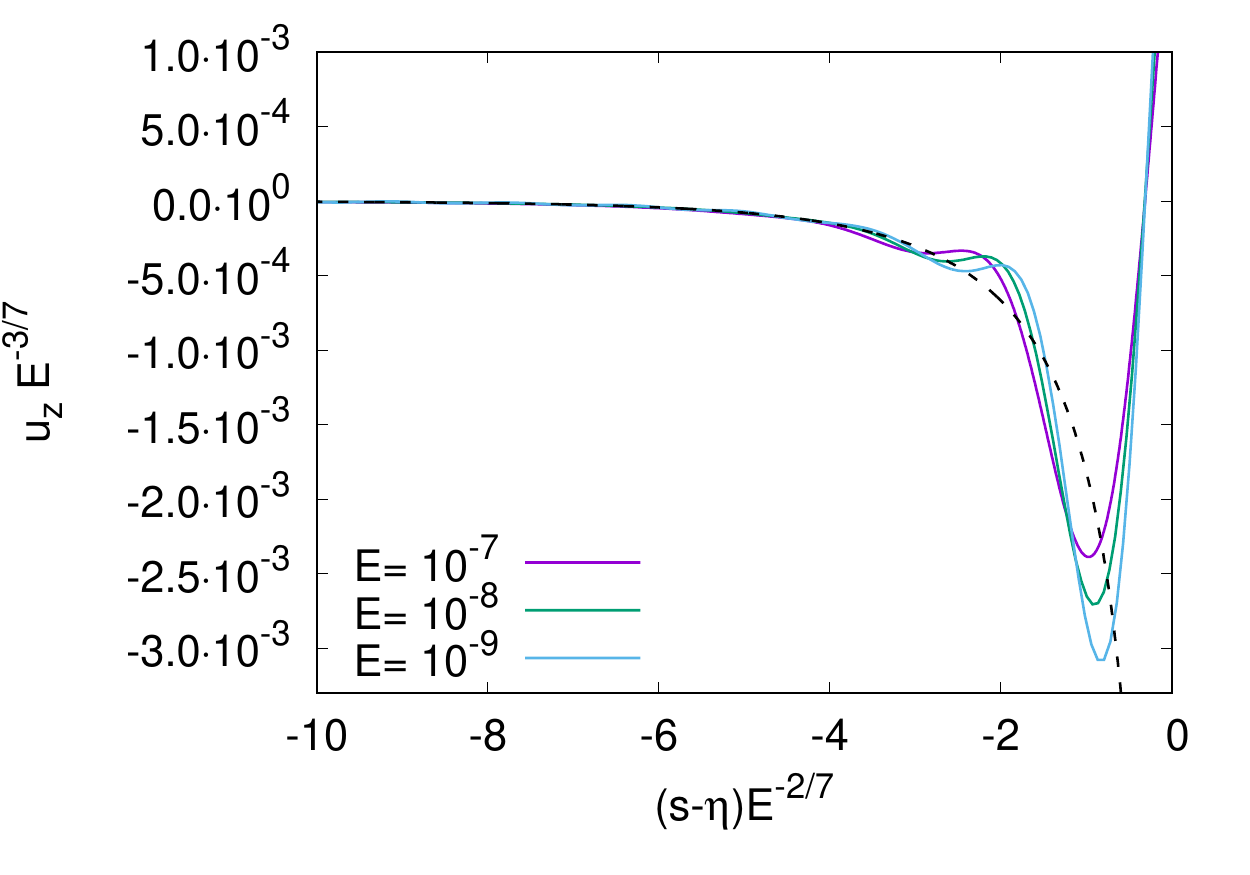}
   \caption{$u_s E^{-5/7}$ (left) and $u_z E^{-3/7}$ (right) as a function of the stretched cylindrical radial coordinate $(s-\eta)E^{-2/7}$ for various Ekman numbers, $\eta=0.35$, $z=0.7$, and $A=0.01$. The black dashed lines correspond to the solution (\ref{eq:scaling27}).}
\label{fig:uz_us_left}
\end{figure}

We further verify the scalings for the thickness of the two Stewartson shear layers, as well as that of the amplitude of their associated $s$- and $z$-directed velocity numerically, in Fig.~\ref{fig:uz_us_centre} and in Fig.~\ref{fig:uz_us_left}.
Though the amplitude of the $z$-directed velocity is, as expected, maximum in the central Stewartson layer, we find a slight discrepancy with the expected power laws for the velocity near $s=\eta$. This discrepancy is, in fact, not surprising as these two Stewartson shear layers are nested. The velocity in the region where the two coexist, that is the $E^{1/3}$-layer, is thus expected to follow both  (\ref{eq:scaling13}) and (\ref{eq:scaling27}). In practice, we measure

\begin{equation}\label{eq:scaling_c}
    u_s \simeq O(E^{0.7}) \ , \andet u_z \simeq O(E^{0.37})  \ ,
\end{equation}
which lies in-between the velocity scalings in the two layers. 
For simplicity's sake, we may consider the amplitude scaling of meridional velocity in the $E^{1/3}$-layer to be that of Eq.~\ref{eq:scaling13}, that is $u_s=O(E^{2/3})$ and $u_z=O(E^{1/3})$.

Ultimately, we find our results to be consistent with the scaling of the kinetic energy $E_{k,\rm tot}\propto E$, which suggest that the  main contribution comes from a $E^{1/3}$-velocity field spread over a free shear layer of width $E^{1/3}$. This $E^{1/3}$ amplitude scaling of the velocity field in the Stewartson layer is actually verified throughout the transient, as can be seen in Fig.~\ref{fig:ekmer}b. This implies that despite a steady-state seemingly not being reached in the lifetime of massive stars, the time relevant for the advective transport of angular momentum and of chemicals within their radiative envelope scales as $E^{-2/3}$, since the the whole meridional velocity grows on a viscous time scale (e.g. Fig.~\ref{fig:psicentr}).

\color{green}
\color{black}
%---------------------------------------------
%---------------------------------------------
%---------------------------------------------
%---------------------------------------------

\subsection{Conclusions}
The picture which results from the  foregoing discussion is rather neat. The spherical shell is split into three domains:

\begin{enumerate}
\item The domain outside the tangent cylinder $\calC$, $s>\eta$ where at first order, and outside any layer, 

\[u_s = -\frac{2KE}{3s},  \quad u_\phi = -\frac{K}{3} \left( s \ln s/\eta - \frac{1}{s} + \frac{s}{\eta^2} \right), \quad u_z=0\]
Thus for a braking torque, $u_\phi<0$ and $u_s<0$ so that there is a weak radial flow towards the Stewartson layer.

\item The domain inside the tangent cylinder, $s<\eta$ where we find that

\[u_s=0,  \quad u_\phi=- \frac{ \sqrt{2E}Ks}{\sqrt{1-s^2}} \left(1-\frac{s^2}{\eta^2}\right)^{1/4}, \quad u_z = -\frac{E K}{\sqrt{1-s^2}} \left( 1 + \frac{1}{1-s^2}\right)\] 
\item In between, the Stewarston layer is composed of two nested shear layers of thickness $O(E^{2/7})$ and $O(E^{1/3})$, and is dominated by a flow parallel to the rotation axis and directed towards the outer shell. In the $E^{2/7}$-layer, 
%\item In between, the Stewarston layer is dominated by a flow parallel to the rotation axis and directed towards the outer shell. Its components scale like
\[u_s\sim E^{5/7}, \qquad u_\phi \sim E^{2/7}, \qquad u_z \sim E^{3/7}\]
and in the $E^{1/3}$-layer,

\[u_s\sim E^{2/3}, \qquad u_\phi \sim E^{1/3}, \qquad u_z \sim E^{1/3}\]
Hence the meridional circulation is completely dominated by the shear flow of the $E^{1/3}$ Stewartson layer.
\end{enumerate}

%The study of the incompressible flow driven by a prescribed tangential surface-stress thus underlines some important points regarding the secondary flow affecting the rotation-induced mixing in an astrophysical (and geophysical) context:

%\begin{enumerate}
%\item The properties of the secondary flow strongly vary in the three important regions: inside and outside a tangent cylinder $\mathcal{C}$ of radius $\eta$, and in vertical shear layers located around $\eta$.
%\item The secondary flow is (almost) purely in the $s$-direction outside $\mathcal{C}$ and (almost) purely parallel to the rotation axis inside $\mathcal{C}$, both their amplitude scaling as $O(E)$.
%\item The foregoing meridional circulation is actually completely dominated by that inside the Stewartson nested shear layer separating the two aforementioned regions, in particular in the central layer of width $O(E^{1/3})$. Inside the latter, the circulation consists in a $O(E^{1/3})$ vertical returning flow and a $O(E^{2/3})$ flow in the $s$-direction.
%\item Both the secondary flows inside and outside the tangent cylinder $\mathcal{C}$ contribute to the sourcing of the upwelling flow in the central vertical shear layer, respectively through the equatorial Ekman singularity at the inner container wall and through a reconnecting shear layer.
%\end{enumerate}

\section{The role of thermal stratification}\label{sec:thermal}
\subsection{Description}

Stable stratification introduces a restoring buoyancy force that acts to inhibit vertical motions and may partially or entirely eliminate the control of the flow exercised by the Ekman layers. Indeed, the tendency for viscous layers to generate secondary circulation is counteracted by the stable stratification. Thus, we may wonder to what extent the flow inside the stably stratified radiative envelope of a massive star losing angular momentum can remain properly modelled with a geostrophic solution. To investigate this question we introduce a radially stable stratification and use the Boussinesq approximation. We  still neglect the centrifugal acceleration. The linearised and dimensionless vorticity, mass conservation, and heat equations read
%We focus here on the limit of both weakly and strongly stably stratified stress-driven spin-down flows and determine to what limit the stellar application belongs.

\begin{equation}\label{eq:eq_strat}
\begin{cases}
\begin{aligned}
&\frac{\partial}{\partial t} \left(\bnabla \times \buu \right)+\bnabla \times (\be_z \times \buu - \delta T \br) = E \Delta \bnabla \times \buu \ , \\
\\
&    \bnabla \cdot \buu = 0 \ , \\
\\
& \frac{\partial \delta T}{\partial t} + \Pran S r u_r  = E \Delta \delta T \ ,
\end{aligned}
\end{cases}
\end{equation}
where we have used $\Delta T (2\Omega_c)^2/ \calN^2$ as the temperature scale. $\Delta T$ is the temperature difference between outer and inner shells, and $\mathcal{N}$ is the the Brunt-V\"ais\"al\"a frequency defined as

\begin{equation}
\calN^2= \frac{\alpha g \Delta T}{R}    \ ,
\end{equation}
where $\alpha$ is the coefficient of thermal expansion and $\boldsymbol{g}= -g \br$ is the gravitational field. We have assumed the equilibrium temperature gradient  to be produced by a uniform distribution of heat source \citep{chandra61}, which implies that $\bnabla T_{\rm eq, *} = (\Delta T / R^2) r \be_r$. $\delta T$ is the temperature perturbation from such thermal equilibrium, and $\Pran= \nu/\kappa_T$ is the Prandtl number, which only enters the equations of motion combined with $S= \calN^2/(2 \Omega_c)^2$, as actually noted by \cite{Barcilon1967}. $\kappa_T$ is the heat diffusivity that we have assumed to  be constant. In what follows, we use the dimensionless parameter introduced by \cite{garaud02}, namely

\begin{equation}
\lambda= \Pran\frac{\calN^2}{(2\Omega_c)^2}  \ ,
\end{equation}
as the parameter characterising the effective strength of thermal stratification.

We complete these equations with the boundary conditions (\ref{eq:BC}) for the velocity field, and we impose $\delta T=0$ at the boundaries. The initial conditions are $\buu=\boldsymbol{0}$, and $\delta T=0$. Projecting (\ref{eq:eq_strat}) on spherical harmonics yields the following system of equations for radial parts

\begin{equation}\label{eq:syst_strat}
\begin{cases}
\begin{aligned}
&\frac{\partial w^l}{\partial t} - \left[ \Alm r^{l-1}\frac{\partial}{\partial r}\left(\frac{r \ulmm}{r^{l-1}}\right) + \Alp r^{-l-2}\frac{\partial}{\partial r}(r^{l+3} \ulp) \right]=   E  \Delta_l \wl \ , \\
&\frac{\partial}{\partial t}\left( \Delta_l r u^l \right) +\Blm r^{l-1}\frac{\partial}{\partial r}\left(\frac{w^{l-1}}{r^{l-1}}\right) + \frac{\Blp}{r^{l+2}} \frac{\partial}{\partial r}(r^{l+2} \wlp) + \Lambda T^l = E\Delta_l \Delta_l r u^l \ , \\
\\
& \frac{\partial T^l}{\partial t}+  \lambda r u^l = E \Delta_l T^l \ ,
\end{aligned}
\end{cases}
\end{equation}
where

\begin{equation}
\delta T(r,\theta)=  \sum_{l=0}^{+\infty} T^l(r) Y_l (\theta) \ .
\end{equation}

\begin{figure}
\centering
     \includegraphics[width=1\textwidth]{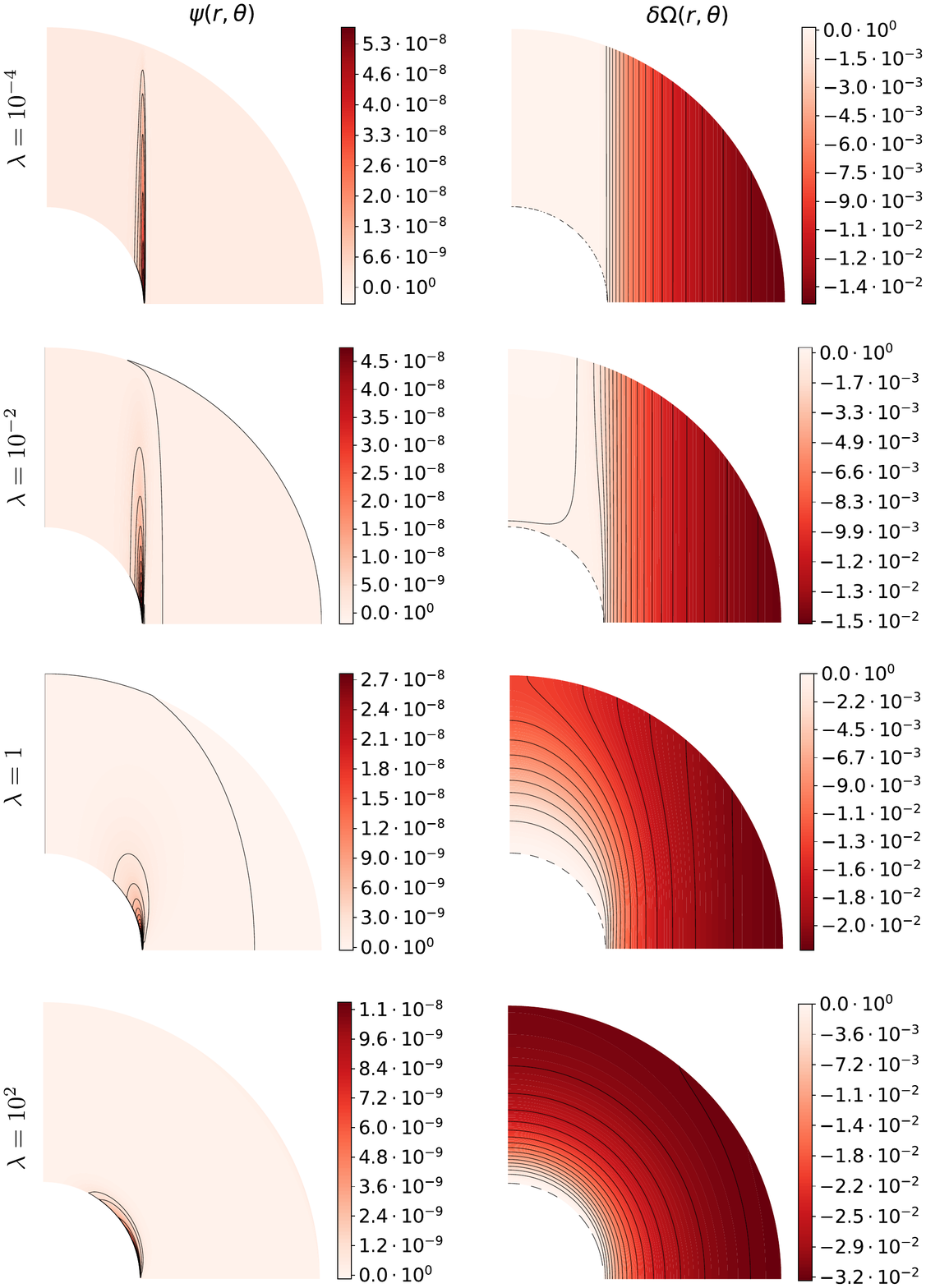}\hfill
     \caption{Meridional view of the stream function $\psi$ (left) and of the differential rotation in the rotating frame of reference $\delta \Omega = (\Omega-\Omega_c)/ (2\Omega_c$)  (right) for $E=10^{-7}$, $\eta=0.35$, $A=0.01$, and various $\lambda$.}
     \label{fig:diff_lambda}
\end{figure}

Fig.~\ref{fig:diff_lambda}  shows the differential rotation in the rotating frame of reference and the meridional circulation for $E=10^{-7}$ and various $\lambda$ once the steady-state is settled. We see that the flow departs from the quasi-geostrophic equilibrium as $\lambda$ increases. At high $\lambda$ the angular velocity profile thus becomes shellular, namely it only depends on the radial distance. %We now focus on the two asymptotic regimes of strong ($\lambda \gg 1$) and weak ($\lambda \ll 1$) stratification. 

\subsection{Horizontal boundary layers}\label{sec:HBL}

In the case of a thermally homogeneous fluid, we have seen that horizontal boundary layers, and specifically Ekman layers play a crucial role on the interior dynamics of the flow. Hence, one may wonder how such layers impact the interior flow in a thermally stratified fluid. We first consider the condition of existence of such horizontal boundary layers and determine their thickness scaling with the relevant adimensional parameters.

Inside boundary layers, horizontal length scales are much larger than the transverse one. Radial derivatives are therefore expected to prevail.  We thus write the simplified equations, assuming $\partial_r \gg \partial_\theta$ in horizontal boundary layers

\begin{equation}
\begin{cases}
\begin{aligned}
- u_\phi \sin \theta & \simeq - \frac{\partial p}{\partial r} + \delta T r + E \frac{\partial^2 u_r}{\partial r^2} \\ 
- u_\phi \cos \theta & \simeq - \frac{1}{r}\frac{\partial p}{\partial \theta} + E \frac{\partial^2 u_\theta}{\partial r^2} \\ 
u_\theta \cos \theta + u_r \sin \theta &\simeq E \frac{\partial^2 u_\phi}{\partial r^2} \\
\frac{\partial u_r}{\partial r} + \frac{1}{r}\frac{\partial u_\theta}{\partial \theta} &\simeq 0 \\
r u_r &\simeq \frac{E}{\lambda} \frac{\partial^2 \delta T}{\partial r^2} \ .
\end{aligned}
\end{cases}
\end{equation}

Combining these equations, only keeping the highest order derivatives for each coefficient ($E^2$, $\lambda$ and 1) yields the general horizontal boundary layer equation

\begin{equation}\label{eq:BLeq}
    E^2 r\frac{\partial^6 u_\theta}{\partial r^6} + \lambda \frac{\partial^2 u_\theta}{\partial \theta^2} - r \cos^2 \theta \frac{\partial^2 u_\theta}{\partial r^2} \simeq 0 \ .
\end{equation}

Introducing the $O(1)$ stretched radial coordinate $\zeta= (1-r)/\sqrt{E}$, (\ref{eq:BLeq}) can be re-written, in the outer boundary layer and in the asymptotic regime of small Ekman numbers,

\begin{equation}\label{eq:BLeq1}
    \frac{1}{E} \frac{\partial^6 u_\theta}{\partial \zeta^6} + \lambda \frac{\partial^2 u_\theta}{\partial \theta^2} - \frac{1}{E}\cos^2 \theta \frac{\partial^2 u_\theta}{\partial \zeta^2} \simeq 0 \ .
\end{equation}

We see that if $\lambda \ll 1/E$, we get the actual Ekman layer where the Coriolis force is balanced by the viscous shear. As remarked by \cite{Barcilon1967}, we note that, whenever Ekman layers are present, their structure is independent of the stratification strength because the forces in balance in such layers  essentially involve horizontal motions. Similarly, if $\lambda\gg1$, introducing the stretched radial coordinate $\gamma= (1-r)\sqrt{\lambda}$, (\ref{eq:BLeq}) can be re-written, in the outer boundary layer,

\begin{equation}\label{eq:BLeq2}
    E^2 \lambda^2 \frac{\partial^6 u_\theta}{\partial \gamma^6} + \frac{\partial^2 u_\theta}{\partial \theta^2} - \cos^2\theta \frac{\partial^2 u_\theta}{\partial \gamma^2} \simeq 0 \ .
\end{equation}

Hence, if $1 \ll \lambda^2 \ll 1/E^2$, the viscous term drops out and buoyancy balances the Coriolis force. This is typical of a thermal boundary layer of width $\delta_\lambda= O(1/\sqrt{\lambda})$. We note that in the parameter regime $\lambda \ll 1$, only Ekman layers are present at the boundaries, while in the regime $ 1 \ll \lambda \ll 1/E$ both thermal and Ekman boundary layers coexist. In this latter case thermal boundary layers are always much thicker than Ekman layers.
%in the considered parameter regime, the total horizontal boundary layers radial width is expected to  be that of the thermal layer, that is of order $1/\sqrt{\lambda}$.

\subsection{Massive stars interior flows}

%In stellar fluids, the Prandtl number is usually small and if one considers rapidly rotating stars, we expect $\lambda \ll 1$. 
From the foregoing discussion, it turns out that the dynamics of the flows may be quite different whether $\lambda\gg1$ or $\lambda\ll1$. Let us now place the case of rapidly rotating massive stars.

We note that, in the envelope of massive stars, the radiative viscosity and radiative heat diffusion largely dominate diffusion of collisional origin \citep{ELR13}. These two quantities read

\begin{equation}
    \nu_{\rm rad} = \frac{4aT^4}{15 c \kappa_R} \ , \quad \ \kappa_{\rm rad} = \frac{4 a c T^3  }{3 \kappa_R c_p} \ ,
\end{equation}
where  $T$ is the temperature, $\kappa_R$ is the Rosseland mean opacity, $a = 4 \sigma /c$ is the radiation density constant with $\sigma$ the Stefan-Boltzmann constant, $c$ the speed of light and $c_p$ is the specific heat capacity at constant pressure. In such a radiation-dominated system, the radiative Prandtl number reads
\begin{equation}
    \Pran_{\rm rad} = \frac{3}{10}\frac{ c_s^2}{c^2} \ ,
\end{equation}
where $c_s$ is the adiabatic sound speed, showing that, naturally, $\Pran_{\rm rad}\ll1$.

\begin{figure}
\centering
     \includegraphics[width=0.5\textwidth]{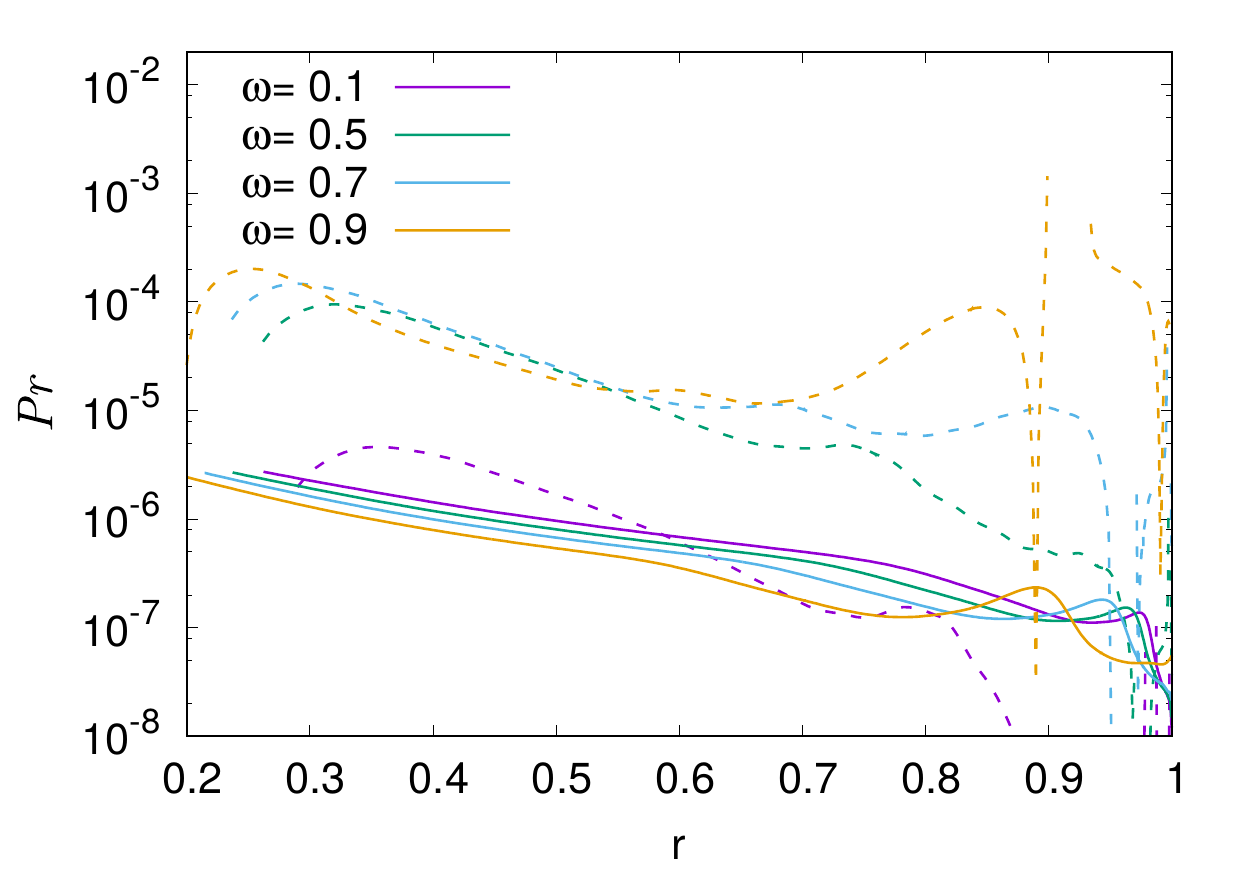}\hfill
     \includegraphics[width=0.5\textwidth]{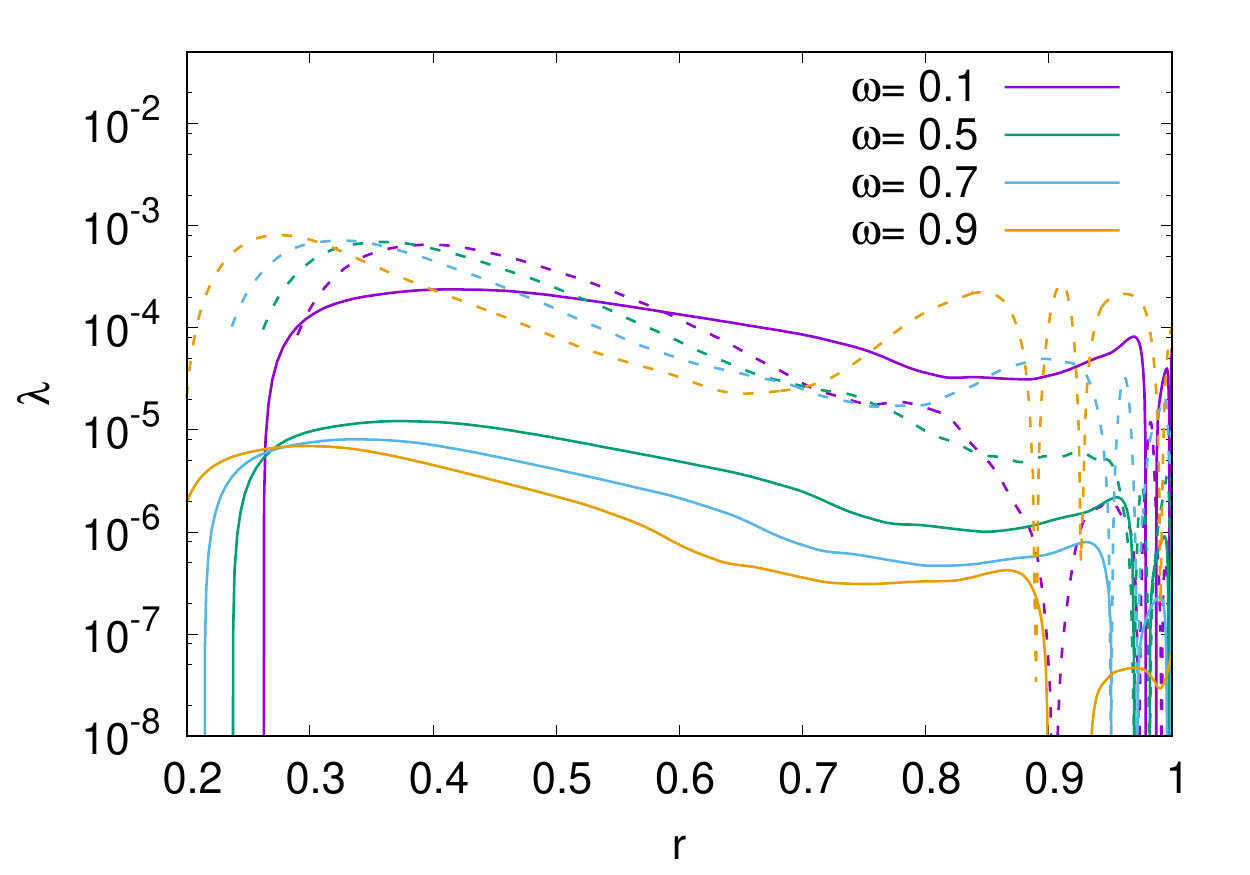}
          \caption{$\Pran$ and $\lambda$ radial profiles at the equator measured from ESTER 2D stellar models of a 15~$M_\odot$ star, for various angular velocity ratios $\omega$. The radiative Prandtl number $\Pran_{\rm rad}$ and the corresponding $\lambda$-parameter are represented in full lines, and their turbulent counterparts are represented in dashed lines.}
    \label{fig:lambda_ester}
\end{figure}

However, the differential rotation of the radiative envelope is expected to drive some small-scale turbulence. \cite{zahn92} proposes that the  turbulent vertical kinematic viscosity associated with marginal shear stability reads as 

\begin{equation}
    \nu_t = \frac{\Ri_c \kappa_T}{3} \left( \frac{s}{\calN} \frac{d \Omega}{ds}\right)^2 \ ,
\end{equation}
where $Ri_c \simeq 1/4$ is the critical Richardson number. In Fig.~\ref{fig:lambda_ester}, we plot the radiative and turbulent Prandtl numbers radial profiles at the equator of a 15 solar mass star as given by a 2D-ESTER model \citep{Rieutordal2016} for various rotation rates defined as the ratio $\omega$ between the equatorial angular velocity and the equatorial Keplerian angular velocity. The associated radial profiles of $\lambda$ are also shown. Although the turbulent Prandtl number is a few orders of magnitude larger than the radiative one in fast rotating stars, turbulent $\lambda$ is just one order of magnitude larger.
%it is of the same order as its radiative counterpart in the case of slow rotation where the differential rotation is weak.  We can further argue that in such case, differential rotation may be stable regarding shear instability.

We find that the turbulent $\lambda$-parameter is roughly independant of $\omega$, and never exceeds $10^{-3}$ in the considered models. The radiative $\lambda$-parameter, on the other hand, increases for decreasing $\omega$ to the point it may locally exceed $10^{-4}$ when $\omega \lesssim 0.1$. As massive stars are often considered to be fast rotators, we conclude that the thermal stratification regime of their radiative envelope corresponds to $\lambda \ll 1$.

%We conclude that the radiative envelope of rotating massive stars

%\corr{We conclude that the thermal stratification regime in the radiative envelope of rotating massive stars has a weak/negligible influence on the primary and secondary quasi-geostrophic flows.}

%{\bf Finalement jusqu'ou peut-on descendre $\omega$ pour que $\lambda$ reste petit et la stratif négligeable?}

\subsection{Asymptotically weak thermal stratification regime}

We now study the case of weak temperature stratification, relevant to the radiative envelope of rotating massive stars. In other context the $\lambda\ll1$ regime may also be reached if the Brunt-V\"ais\"al\"a frequency is just small.
%, that is either due to a small difference of equilibrium temperature between inner and outer shell, to rapid rotation, or to a small Prandtl number.
We have seen that in this regime, only Ekman layers characterised by an $O(E)$ radial velocity at their edge, are present at the boundaries. We thus assume $\lambda\ll1$ for the weak stratification, but also suppose that $E\ll\lambda$ as expected in stars.

\subsubsection{The steady flow}

 We first focus on the steady flow. The radial component of the steady  momentum equation reads

\begin{equation}
    - u_\phi \sin \theta = -\frac{\partial p}{\partial r} + \delta T r + E \left[ \Delta u_r -\frac{2}{r^2}\left( u_r + \frac{1}{\sin \theta} \frac{\partial \sin \theta u_\theta}{\partial \theta}\right) \right] \ ,
\end{equation}
where differential rotation is driven by the $O(1)$ surface stress, that is $u_\phi=O(1)$. In the asymptotic regime of small Ekman number, the temperature deviation from equilibrium  is therefore, at most, $O(1)$. If that is the case, namely if $\delta T = O(1)$, the heat equation (\ref{eq:eq_strat}) implies that the interior radial velocity $u_r$ is $O(E/\lambda)$, namely larger than the $O(E)$ Ekman pumping (\ref{eq:pumping}), in the considered stratification regime. Because of the boundary conditions imposed by Ekman layers, an $O(E/\lambda)$ radial velocity must vanish at $r=\eta$ and $r=1$. Furthermore, the radial component of the vorticity equation reads

%\begin{equation}\label{eq:ur_ekpump}
%\tilde{u}_r = \mp  \frac{E}{\sin \theta} \frac{\partial }{\partial \theta} \frac{1}{\cos \theta}\bigg( \tau(\theta) - r \frac{\partial}{\partial r} \frac{u_\phi }{r} \bigg) \ ,
%\end{equation}
%at the outer and inner boundary respectively, in the considered stratification regime. Because of the boundary conditions imposed by Ekman layers, an $O(E/\lambda)$ radial velocity must vanish at $r=\eta$ and $r=1$. Furthermore, the radial component of the vorticity equation reads

\begin{equation}
\frac{\partial u_r}{\partial z} = O(E)  \ ,
\end{equation}
implying that $u_r$ is $z$-independent up to $O(E)$ corrections \citep[see][]{Barcilon1967}. Hence, a consistent solution for the interior flow is $u_r = O(E)$ and consequently $\delta T = O(\lambda)$, which is completed by the classical Ekman boundary layer with $\tilde{u}_r=O(E)$ and $\tilde{u}_\theta=O(\sqrt{E})$.

Let us now write the variables of the interior flow, as a power expansion of the small parameter $\lambda$, in the asymptotic regime of small Ekman numbers. They read

\begin{equation}\label{eq:powerexp_lamb}
\begin{aligned}
u_r &= E(u_{r,0} + \lambda u_{r,1} + ...)\\
u_\phi &= u_{\phi,0} + \lambda u_{\phi,1} + ... \\
\delta T &= \lambda \delta T_1 + \lambda^2 \delta T_2 + ... \\
p &= p_0 + \lambda p_1 + ... \ .
\end{aligned}
\end{equation}

Injecting (\ref{eq:powerexp_lamb}) into the momentum equation then yields, at zeroth and first order
%the $O(1)$ and $O(\lambda)$ interior equations, that is

\begin{equation}\label{eq:mom0}
u_{\phi,0} = \frac{\partial p_0}{\partial s}, \quad {\rm and} \quad 0=\frac{\partial p_0}{\partial z} \ , 
\end{equation}
and

\begin{equation}\label{eq:momlamb}
u_{\phi,1} = \frac{\partial p_1}{\partial s} - s \delta T_1, \quad {\rm and} \quad z \delta T_1=\frac{\partial p_1}{\partial z} \ .
\end{equation}

Hence, at the lowest order, and identically to the homogeneous case, the Taylor-Proudman theorem is satisfied and the $O(1)$ geostrophic flow dynamics is consequently entirely controlled by the Ekman pumping. This is indeed verified in Fig.~\ref{fig:diff_lambda}. On the other hand, the $O(\lambda)$ flow is affected by temperature deviation from equilibrium produced by the $O(E)$ radial velocity. We conclude that the thermal stratification regime in the radiative envelope of rotating massive stars has a negligible influence on the primary and secondary quasi-geostrophic and stationary flows.

\subsubsection{The transient flow}

 As in the unstratified case, we now wish to determine the  scaling of the transient time with the relevant non-dimensional parameters, that is, the Ekman number and the $\lambda$-parameter, in the $E\ll\lambda \ll 1$ limit. We write the radial component of the vorticity equation
 
\begin{equation}\label{eq:eqstrat_time}
  \frac{\partial}{\partial t}\left( \frac{1}{r \sin \theta}\frac{\partial}{\partial \theta} \sin \theta u_\phi\right)  -\frac{\partial u_r}{\partial z} - \frac{\sin \theta}{r} u_\theta = \frac{E}{r \sin \theta}\frac{\partial}{\partial \theta}\sin \theta \left( \nabla^2 -\frac{1}{r^2 \sin^2 \theta}\right) u_\phi \ ,
\end{equation}
as well as the $z$-derivative of the heat equation

\begin{equation}\label{eq:dzheat_time}
 \frac{\partial}{\partial t} \left( \frac{\cos \theta}{\lambda} \frac{\partial }{\partial z} \frac{\delta T}{ z}\right)+   \frac{\partial u_r}{\partial z} = \cos \theta \frac{E}{\lambda} \frac{\partial}{\partial z}\frac{1}{z}\Delta \delta T + \frac{\sin^2 \theta}{r \cos \theta} u_r \ .
\end{equation}

Adding (\ref{eq:eqstrat_time}) and (\ref{eq:dzheat_time}), and using the $\phi$-component of the momentum equation, we get

%\begin{equation}
%   \frac{\partial u_\phi}{\partial t} + \cos \theta u_\theta + \sin \theta u_r = E \left( \nabla^2 -\frac{1}{r^2 \sin^2 \theta}\right) u_\phi \ ,
%\end{equation}
%finally reads
% 
 
\beqan
\frac{\partial}{\partial t} \left( \frac{z}{\lambda} \frac{\partial }{\partial z} \frac{ \delta T}{z} + \cotan\theta \frac{\partial}{\partial \theta}\tan \theta u_\phi \right) = \nonumber \\
E \cotan \theta \frac{\partial }{\partial \theta} \tan \theta \left( \nabla^2 -\frac{1}{r^2 \sin^2 \theta}\right) u_\phi + \frac{E}{\lambda}z \frac{\partial}{\partial z}\frac{1}{z}\Delta \delta T  \ .
\eeqan{eq:syst_fin_time}

Hence, if $\delta T$ remains $O(\lambda)$ during the transient, we expect the quasi-geostrophic steady state to be reached on the $O(E^{-1})$ viscous time scale in the limit $\lambda \ll 1$.

\begin{figure}
\centering
     \includegraphics[width=0.6\textwidth]{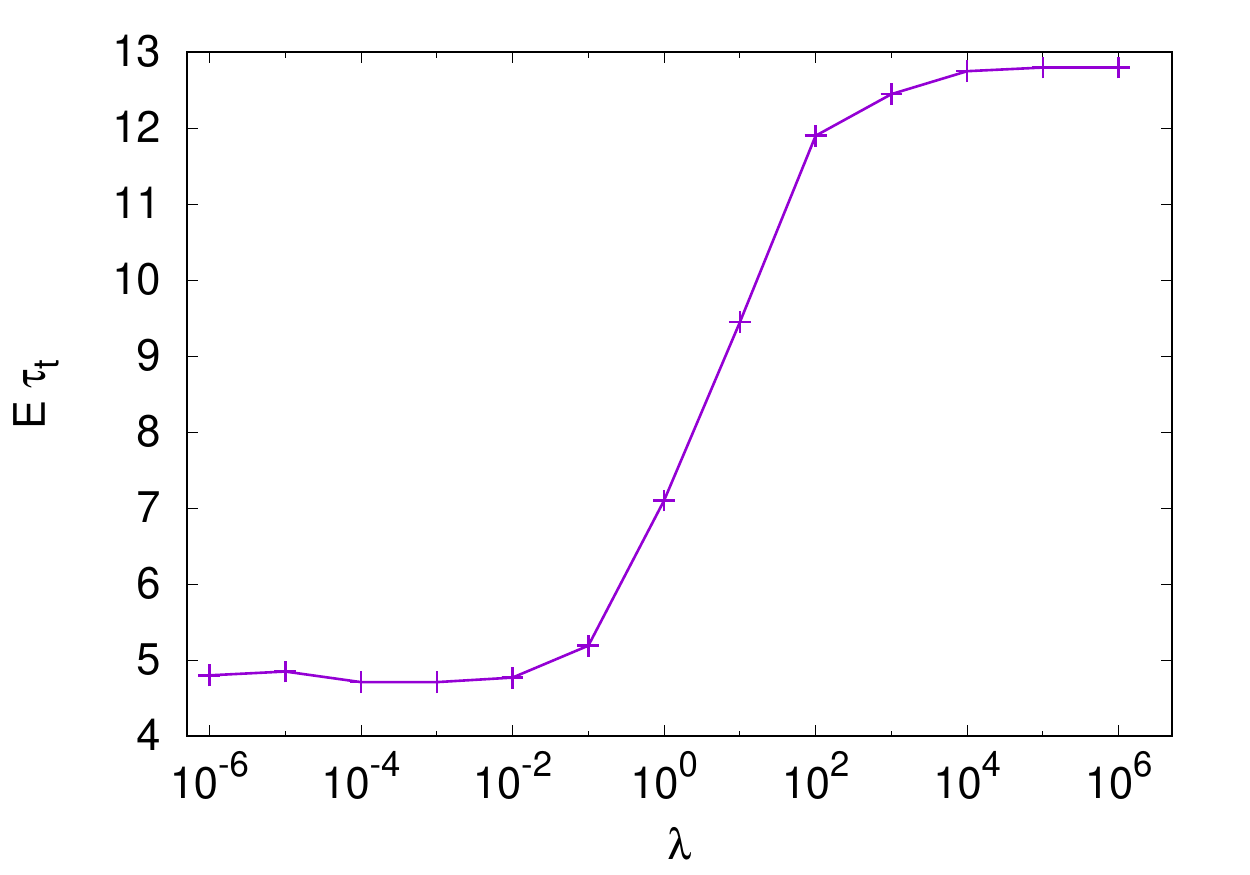}\hfill
          \caption{The scaled transient time scale  $E\tau_t$ given by the numerical solution as a function of $\lambda$ for $E=10^{-6}$. The steady state is reached on a $O(E^{-1})$ time scale, both in the strongly and weakly stratified limits.}
    \label{fig:tauss_lamb}
\end{figure}

We verify this conclusion with our numerical solution and measure the transient time scale $\tau_t$  as the time required for the relative difference between the torques at the boundaries to be less than $0.01\%$. We show the scaled values $E\tau_t$ as a function of $\lambda$ for $E=10^{-6}$ in Fig.~\ref{fig:tauss_lamb}. We find that indeed, as for the thermally homogeneous case, the transient time scale is $O(E^{-1})$ in the weakly stratified limit.

The thermal stratification regime in the interior of slowly rotating stars may, on the other hand, lie in the $\lambda \gg 1$ limit. In this regime, it is well known that the Eddington-Sweet time scale associated with the angular momentum redistribution by meridional circulation  is so long that it is unlikely that the system would ever relax to a steady-state. Hence, the study of the time-dependent interior dynamics of slowly rotating stars, which may largely depend on initial conditions and on the damping rate of the baroclinic modes, is outside the scope of this work.
%which aims at constraining wind-driven flows in rotating massive stars.
However, for the sake of completeness and to appreciate some effects of a strong thermal stratification in the stress-driven barotropic spin-down flow, we give a short account of the $\lambda\gg1$ regime in Appendix~\ref{sec:A3}.

\section{The flow in a polytropic envelope}\label{sec:anelastic}

\subsection{The background}

The next step towards a realistic model of the radiative envelope of a massive star is to include the strong density variations of the fluid between the convective core and the stellar surface. Typically density varies by a factor of $10^9$ in the envelope of a star of 15 solar masses. To take this density distribution into account in a simple way we assume that the gas can be described by a polytropic equation of state as usually done in stellar physics \cite[e.g.][]{maeder09}. Hence the hydrostatic background state verifies:

\begin{equation}\label{eq:polytrope}
    -\frac{d p_{*}}{dr_{*}}-\frac{GM}{r_*^2}\rho_{*}=0 \quad {\rm and}  \quad \p_{*}=\kappa\rho_{*}^\Gamma \ ,
\end{equation}
where $p_{*}$, $\rho_*$, and $r_*$ are the dimensional pressure, density, and radial coordinate, respectively. As previously, we also use the Roche approximation and assume that the core gathers all the mass $M$ of the star.
%Here it is also assumed that the core gathers all the mass of the star $M$ (usually referred to as the Roche approximation).
In (\ref{eq:polytrope}) $G$ is the gravitational constant,
$\kappa$ is a constant related to the thermal conditions at the core boundary, and $\Gamma=1+1/n$, where $n$ is the so-called polytropic index. In the following we shall set $n=3$, which is typical for radiative envelopes. We note that for such an index the fluid is stably stratified \cite[e.g.][for instance]{DR01,RD02}, but as concluded from the previous section the small value of the Prandtl number allows us to neglect buoyancy effects. The following results therefore assume no buoyancy force.

Eq.~(\ref{eq:polytrope}) can be easily solved and gives the density profile

\begin{equation}\label{eq:rho}
    \rho_{*}(r_{*})=\rho_c(1+B j(r_{*}))^n \ ,
\end{equation}
where

\begin{equation}
  B= \frac{\rho_s^{1/n}-1}{1/\eta - 1}, \qquad j(r_{*})=\frac{1}{\eta} - \frac{R}{r_{*}} \ .
\end{equation}

Here $\rho_c = \rho_{*}(\eta R)$ is the density at the core boundary and $\rho_s=\rho_{*}(R)/\rho_c$ is the adimensional surface density. Note that such a background is also used in numerical simulations of convection in stellar envelope \citep{raynaud_etal18} or planetary atmospheres \citep{gastine_wicht12}. 
Hence, in this section, we solve equations (\ref{eq:nucst}) with the background polytropic density profile (\ref{eq:rho}), the boundary conditions (\ref{eq:BC}), and with the initial condition $\buu= \boldsymbol{0}$.

\subsection{The transient phase}

\begin{figure}
\centering
     \includegraphics[width=0.6\textwidth]{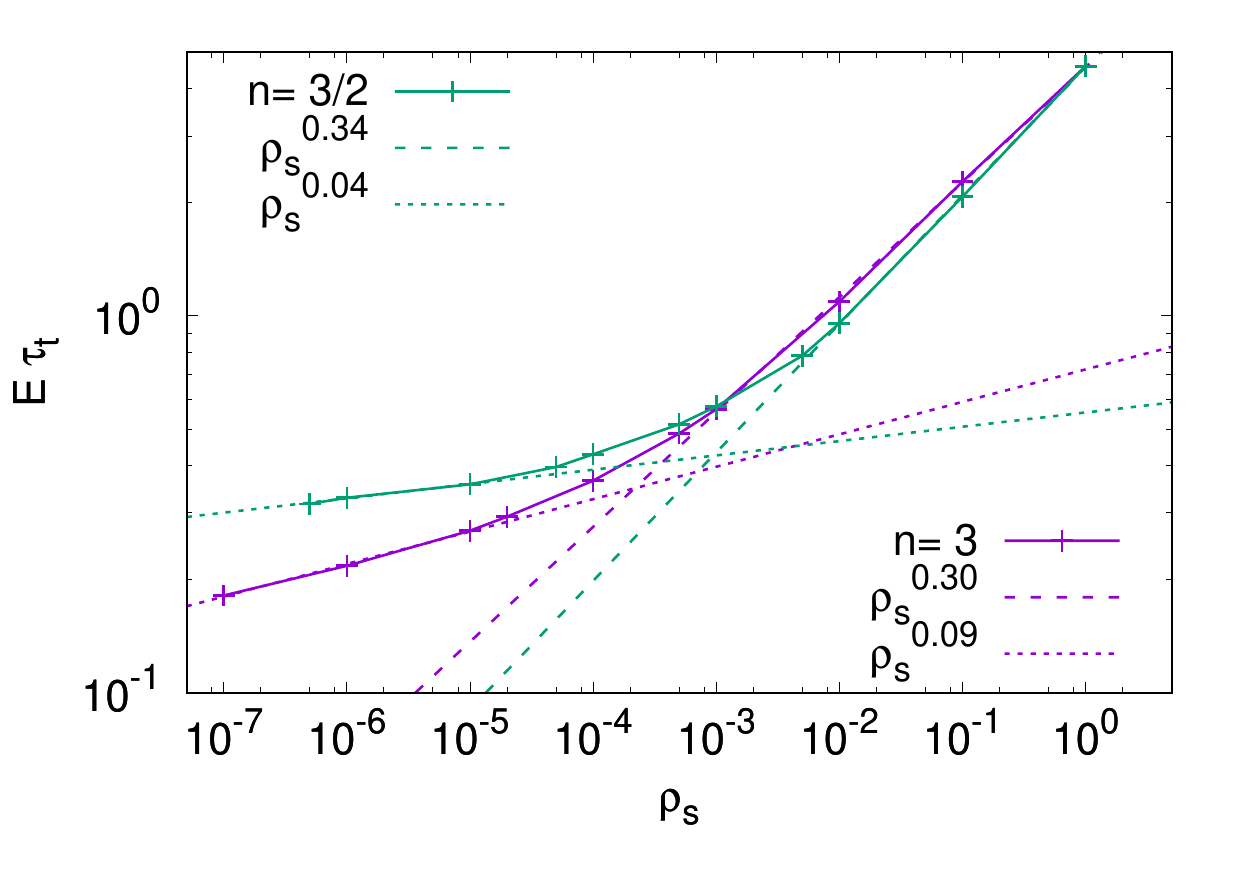}\hfill
      \caption{Scaled steady-state time scale  $E\tau_t$ as a function of the surface density $\rho_s$, for $E=10^{-6}$ models with various values of $\rho_s$, $n=3/2$ and $n=3$. The dashed and dotted lines correspond to the steady-state time scalings with $\rho_s$.}
    \label{fig:tauss_rhos}
\end{figure}

Again, we first focus on the transient phase preceding the settling of a steady-state. In particular, we aim at determining whether the scaling of the governing time scales are modified by density variations of the background. We measure the transient time as the time for which relative difference between the torque exerted by the fluid  on the stationary inner sphere and the torque applied on the outer sphere, is less than $0.01 \%$. We monitor the evolution of the relative difference between the torques $\Delta \Gamma / \Gamma(1)$ given by (\ref{eq:torque_diff}), for various surface density ratios. We show the resulting steady-state times in Fig.~\ref{fig:tauss_rhos} for the polytropic indexes $n=3$ and $n=3/2$. The latter index corresponds to an isentropic monatomic gas, hence to a neutral thermal stratification. We find the steady state time to depend on the density ratio between the core and the surface. For the polytropic index $n=3$, relevant to the envelope of massive stars, we find $E \tau_t \propto \rho_s^{0.3}$ for $ 10^{-3} \leq \rho_s$, and $E \tau_t \propto \rho_s^{0.09}$ for $\rho_s  < 10^{-4}$. Of course, these power laws quantitatively depend on the chosen density profile, as can be seen with the $n=3/2$ case. Unlike thermal stratification, density stratification thus (mildly) influences the time scale required to reach a steady-state in a stellar envelope.

\subsection{The steady flow}

We now focus on the steady flow. We may observe that if viscosity, non-linearity and buoyancy are neglected then the steady flow obeys a 
Taylor-Proudman theorem applied to the momentum $\rho\buu$ instead of the velocity, namely

\begin{equation}\label{eq:TP}
(\be_z \cdot \bnabla ) \rho \buu = \mathbf{0} \ .
\end{equation}
Unfortunately we cannot reiterate the boundary layer analysis of the incompressible case since the viscous force (Eq.~\ref{eq:visc}) now includes terms that depend on $r=\sqrt{s^2+z^2}$ and make the partial differential equation not separable. We may however observe that since the density variations do not add any new length scale, the viscous balance in the shear layers remains similar and we can still expect the presence of a Stewartson layer along the tangent cylinder.

\begin{figure}
\centering
\includegraphics[width=0.5\textwidth]{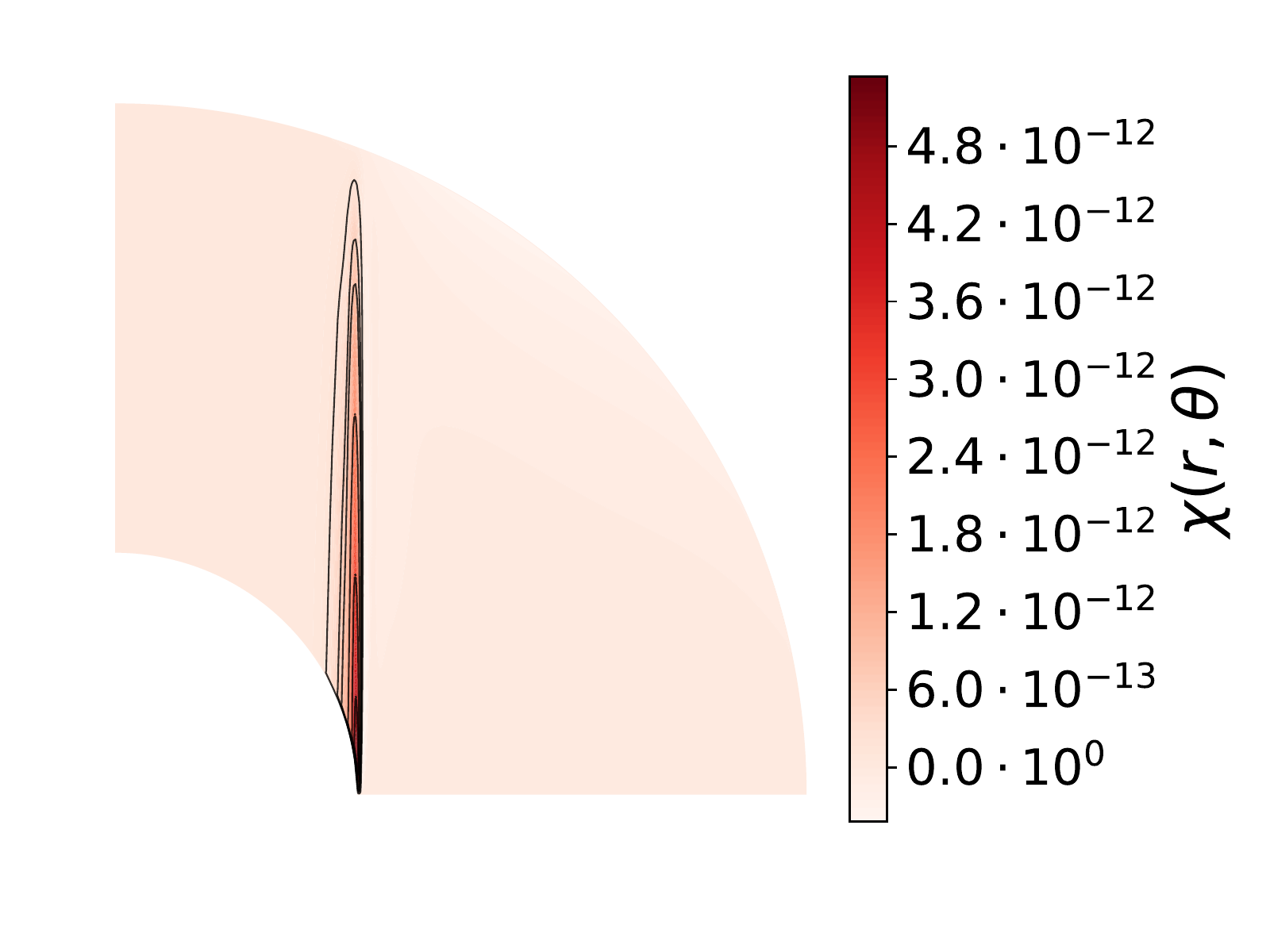}\hfill
 \includegraphics[width=0.5\textwidth]{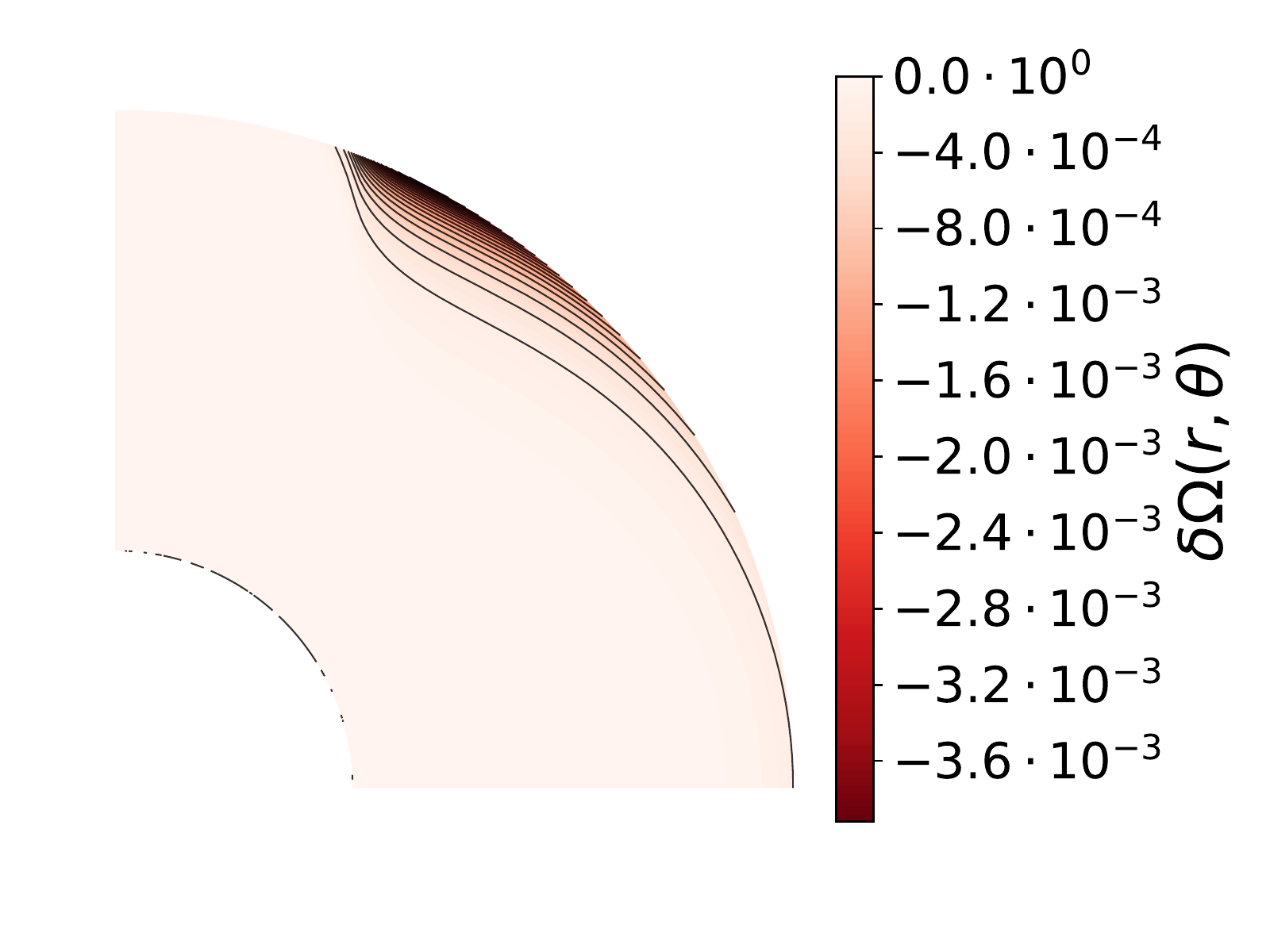}\hfill
   \caption{Meridional view of the stream function $\chi=\rho \psi$ (left) and of the differential rotation in the rotating frame of reference $\delta \Omega = (\Omega-\Omega_c)/2\Omega_c$  (right) for $E=10^{-7}$, $\eta=0.35$, $\rho_s=10^{-4}$, and $A=0.01$.}
\label{fig:compr}
\end{figure}

We now revert to numerical solutions to make progress. We thus solve \eqref{eq:nucst} with boundary condition \eqref{eq:BC}. As already mentioned, we neglect the effects of buoyancy. %, although the n=3-polytrope is a stably stratified configuration. This is justified a posteriori because the meridional flow are slow and impose a time scale that is much longer than the heat diffusion time scale. In other words the driven flows do not disturb the thermal equilibrium of the star.
 We first focus on the differential rotation and the meridional circulation. It is convenient to introduce the stream function $\chi$ associated with the meridional momentum, namely 

\begin{equation}
    \frac{\partial \chi}{\partial r} = r \sin \theta \rho u_\theta, \quad \quad \frac{\partial \chi}{\partial \theta} = -r^2 \sin \theta \rho u_r \ .
\end{equation}
Fig.~\ref{fig:compr} shows the differential rotation and this new stream function in the rotating frame of reference for $E=10^{-7}$.

\begin{figure}
%\centering
\includegraphics[width=0.5\textwidth]{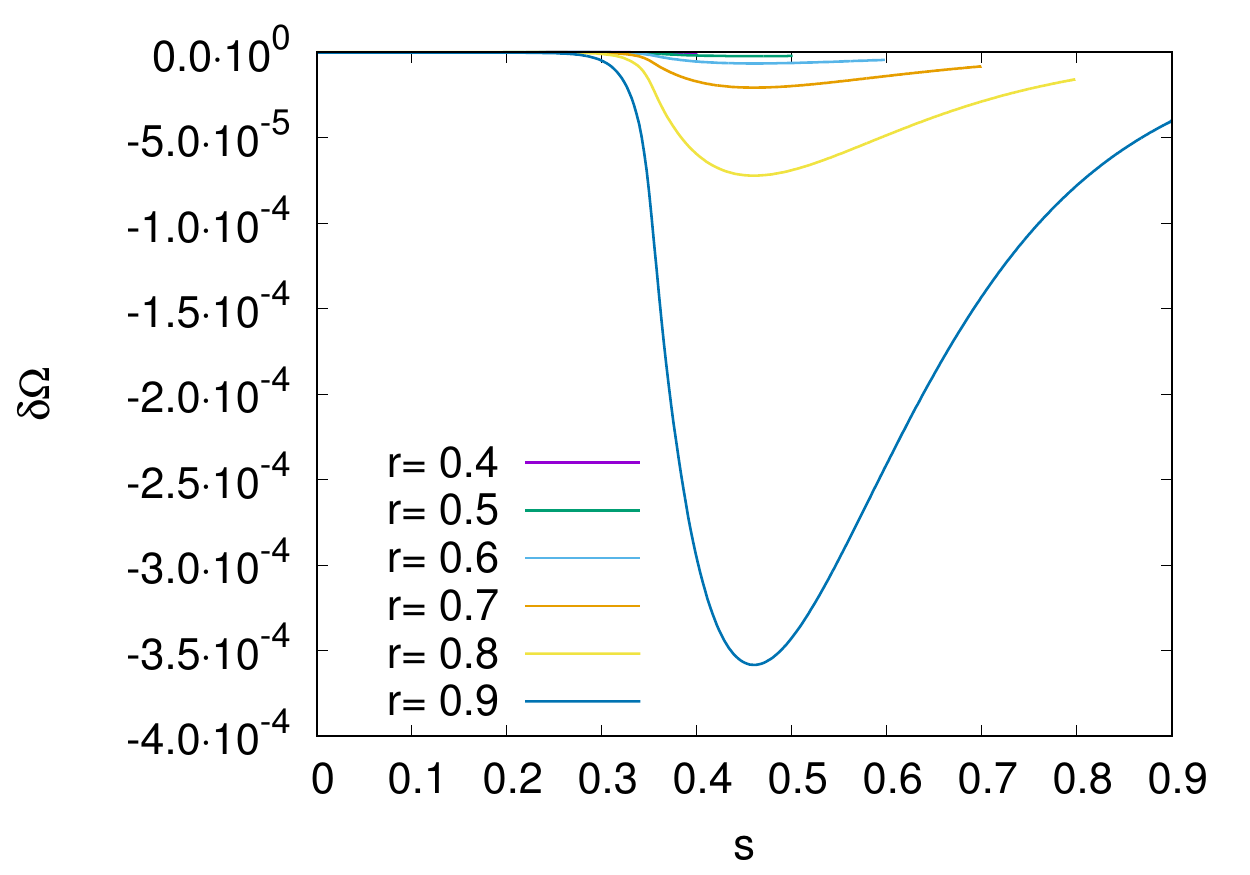}\hfill
 \includegraphics[width=0.5\textwidth]{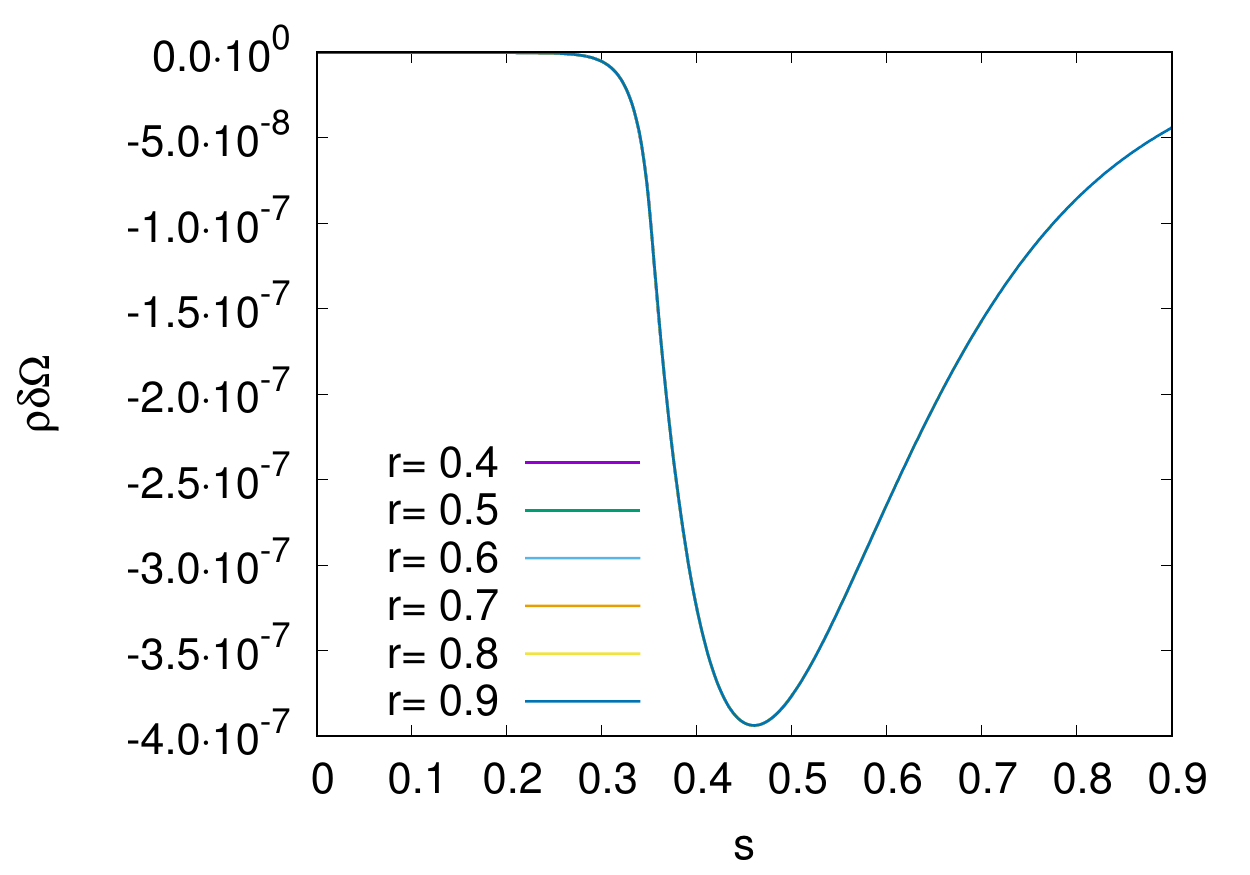}\hfill
   \caption{Normalised angular velocity (left) and normalised angular velocity multiplied by the density $\rho$ (right), and as a function of the cylindrical radial coordinate $s$ for $E=10^{-7}$, $\eta=0.35$, $\rho_s=10^{-4}$, $A=0.01$, and various radius.}
\label{fig:uphi_stress_compr}
\end{figure}

We note that as for the incompressible flow, the amplitude of the secondary flow is dominated by the Stewartson layer located at the edge of the tangent cylinder $\mathcal{C}$. The differential rotation on the other hand, and as expected, is no longer cylindrical and very $z$-dependent, with a maximum value that is close to the outer shell and the tangent cylinder. Fig.~\ref{fig:uphi_stress_compr}a shows the differential rotation as a function of the cylindrical radial coordinate for $E=10^{-7}$ at different radii $r$, illustrating the $z$-dependence of $u_\varphi$, while Fig.~\ref{fig:uphi_stress_compr}b shows that $q_\varphi=\rho(r)u_\varphi$ verifies the Taylor-Proudman constraint.

\begin{figure}
\centering
     \includegraphics[width=0.5\textwidth]{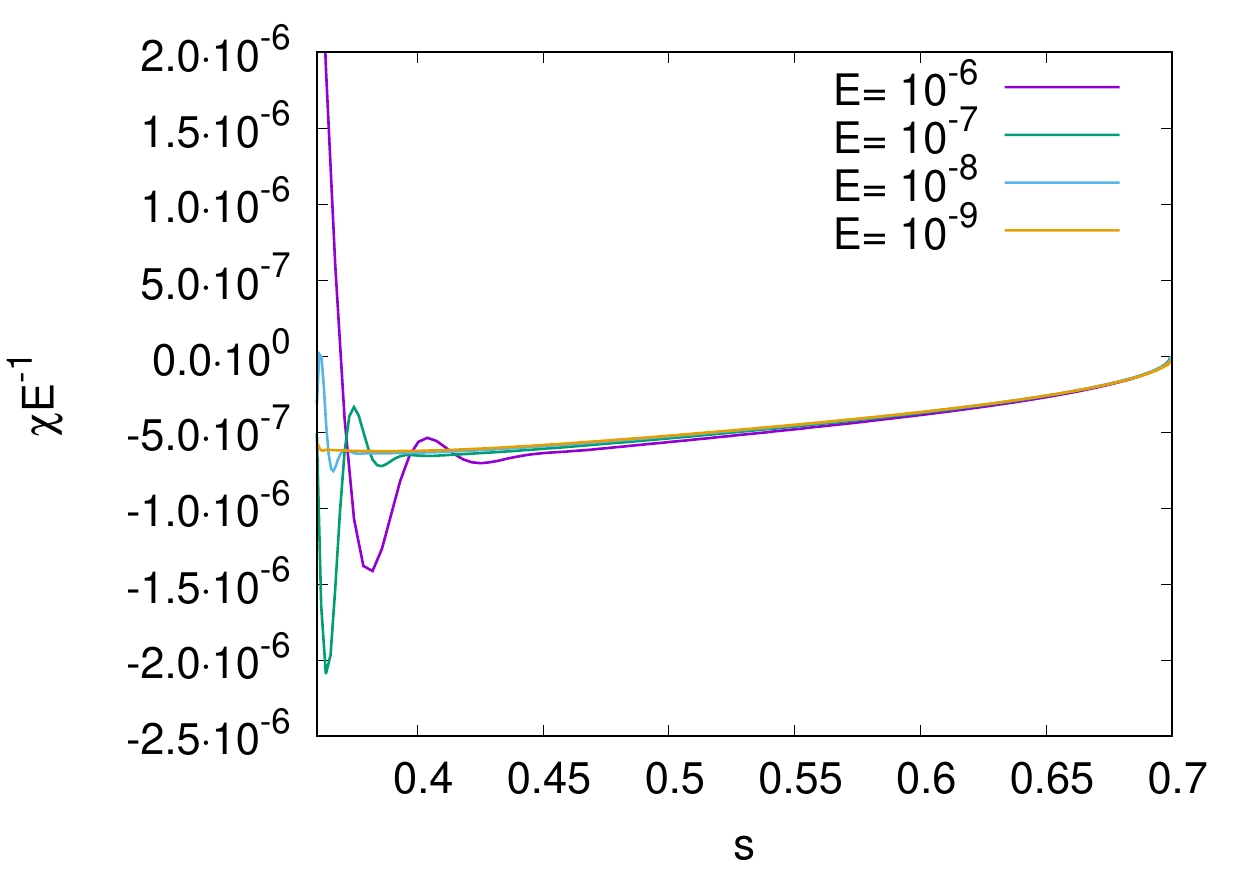}\hfill
     \includegraphics[width=0.5\textwidth]{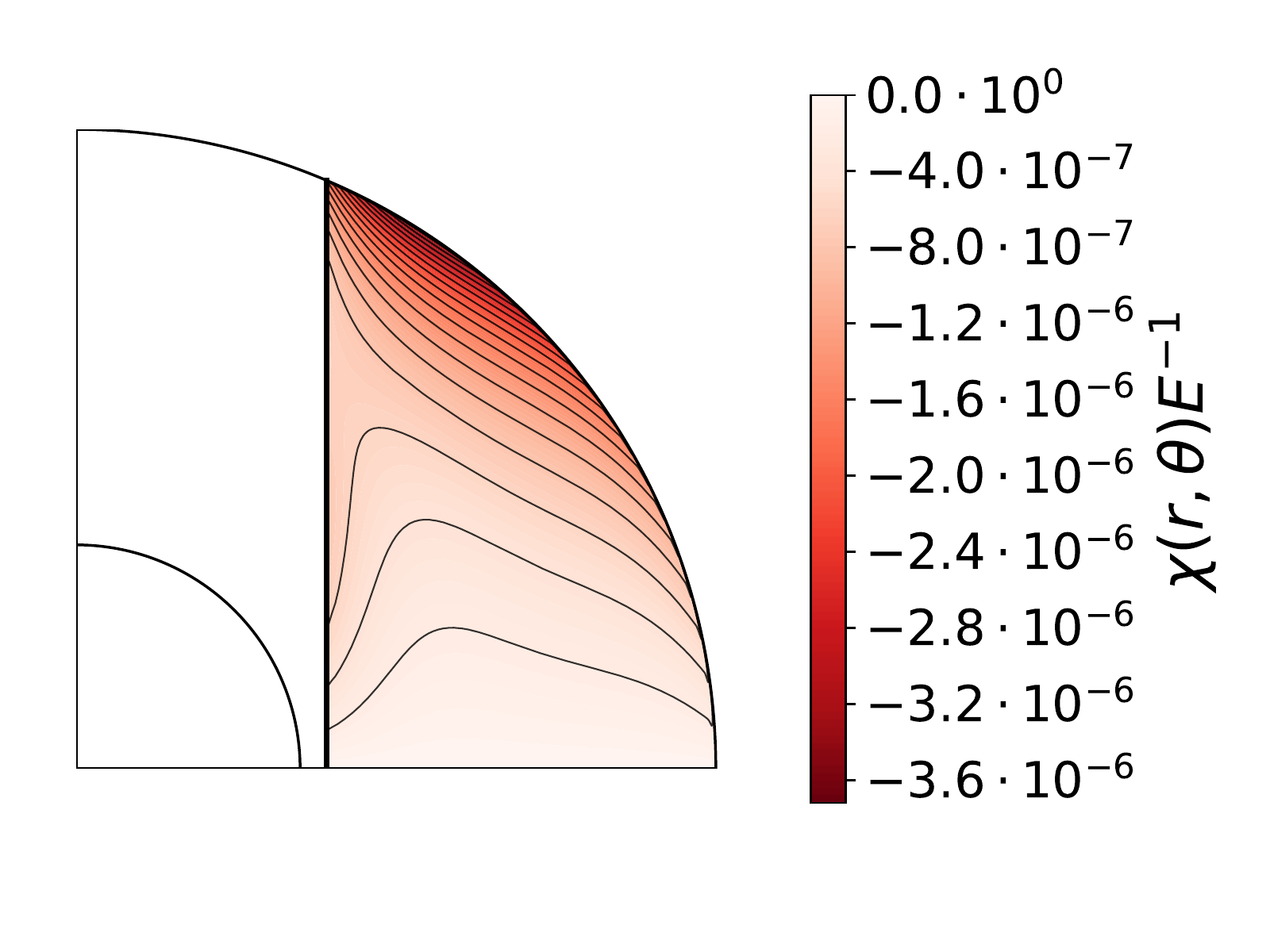}\hfill    
   \caption{Left: $\chi E^{-1}$ as a function of the cylindrical radial coordinate $s$ for various Ekman numbers, $\eta=0.35$, $r=0.7$, $\rho_s=10^{-4}$, and $A=0.01$. The predicted $O(E)$ scaling of the secondary flow is verified away from the Stewartson nested layers. Right: Meridional view of the stream function $\chi E^{-1}$ outside $\mathcal{C}$, for the $E=10^{-9}$ model.}
\label{fig:chi_compr}
\end{figure}

\begin{figure}
\centering
     \includegraphics[width=0.5\textwidth]{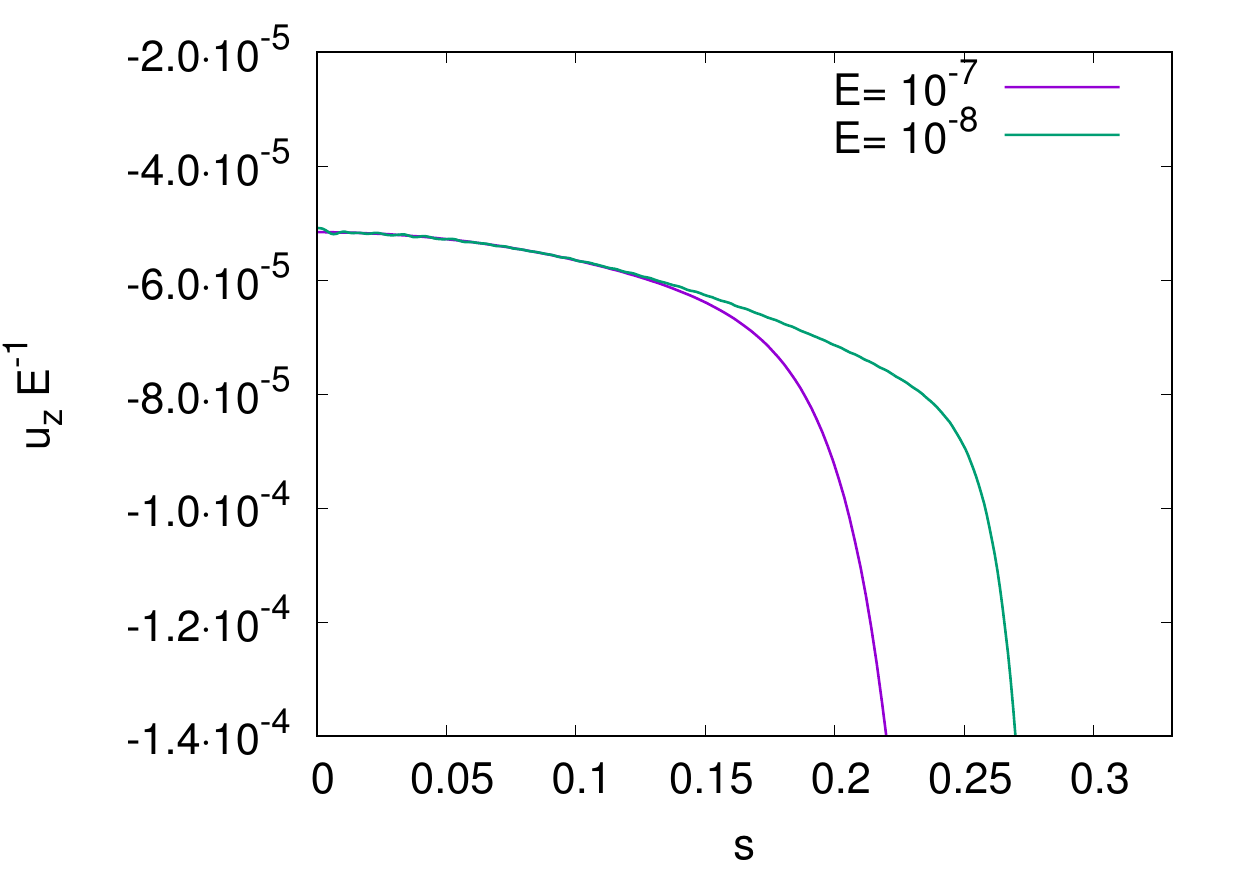}\hfill
     \includegraphics[width=0.5\textwidth]{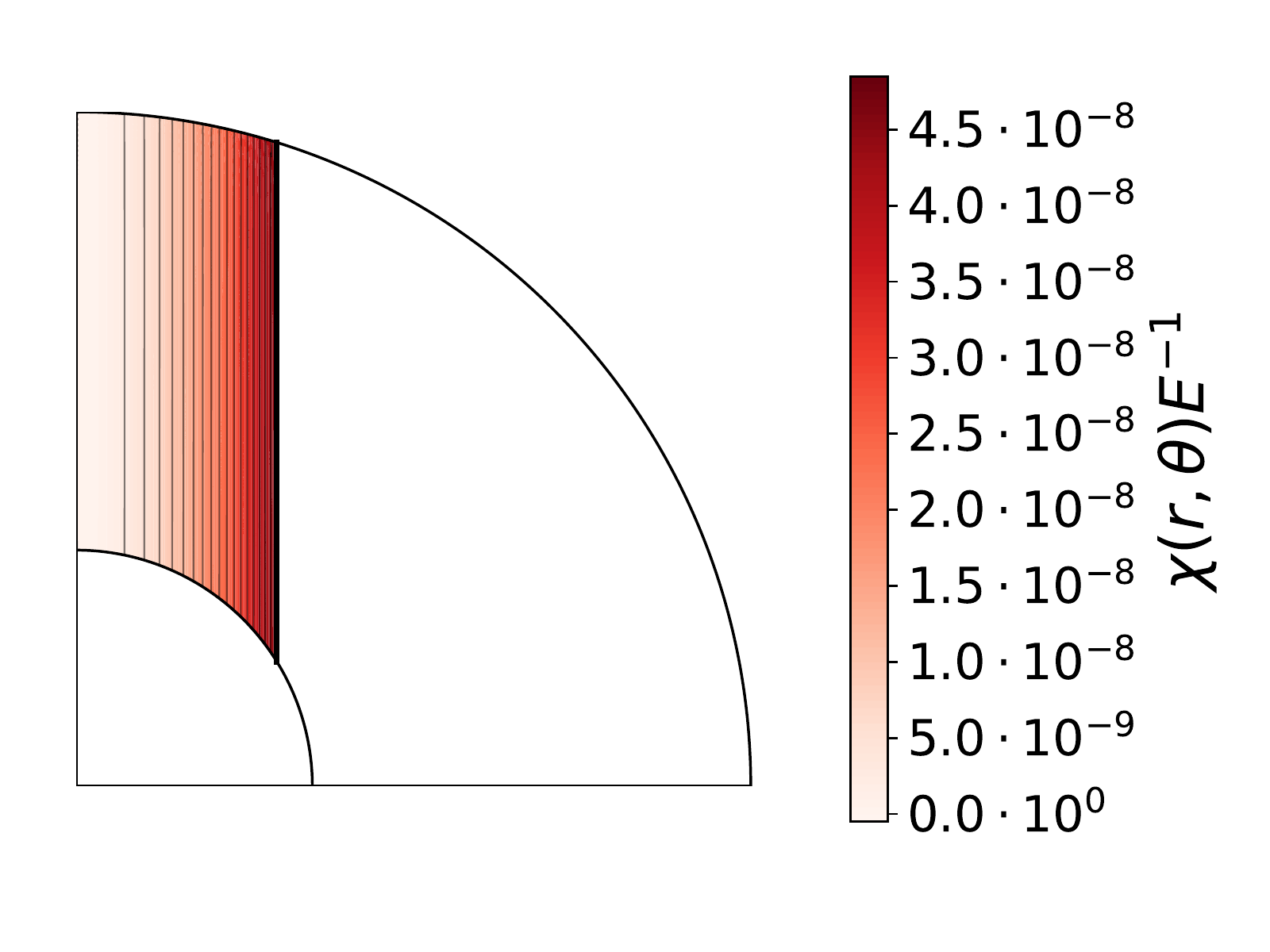}\hfill    
  \caption{Left: $u_z E^{-1}$ as a function of the cylindrical radial coordinate $s$ for various Ekman numbers, $\eta=0.35$, $r=0.7$, $\rho_s=10^{-4}$, and $A=0.01$. Right: Meridional view of the stream function $\chi$ inside $\mathcal{C}$ and away from the $O(E^{2/7})$ Stewartson layer, for the $E=10^{-9}$ model.}
\label{fig:chi_compr_ins}
\end{figure}

As far as meridional circulation is concerned, this is still an \od{E} flow outside the Stewartson layer as shown by Fig.~\ref{fig:chi_compr}a and \ref{fig:chi_compr_ins}a. We note that outside the tangent cylinder, streamlines are no longer straight lines, even for the $\bq$ momentum field. In the inner cylinder and near the rotation axis we still get streamlines parallel to the rotation axis for the momentum $\bq$. This latter feature comes from the fact that the Ekman layer has the same structure as in the constant density case and drives an Ekman circulation which has a unique component along the $z$-axis. From, Eq.~\ref{eq:geostroph_mer} we deduce that

\[q_z = -\frac{E K\rho_s}{\sqrt{1-s^2}} \left( 1 + \frac{1}{1-s^2}\right) + O(E^2)\]
in this region.

\begin{figure}
\centering
     \includegraphics[width=0.5\textwidth]{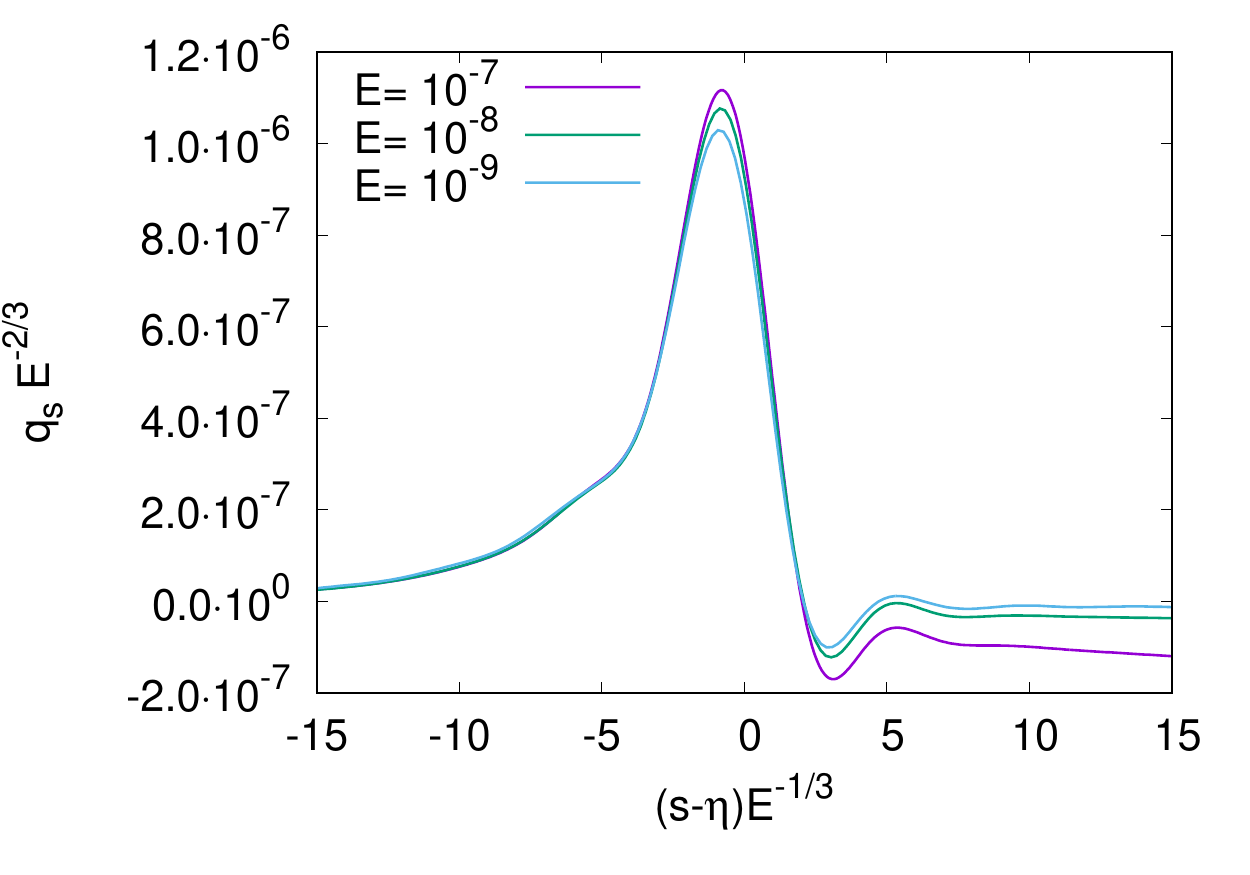}\hfill
     \includegraphics[width=0.5\textwidth]{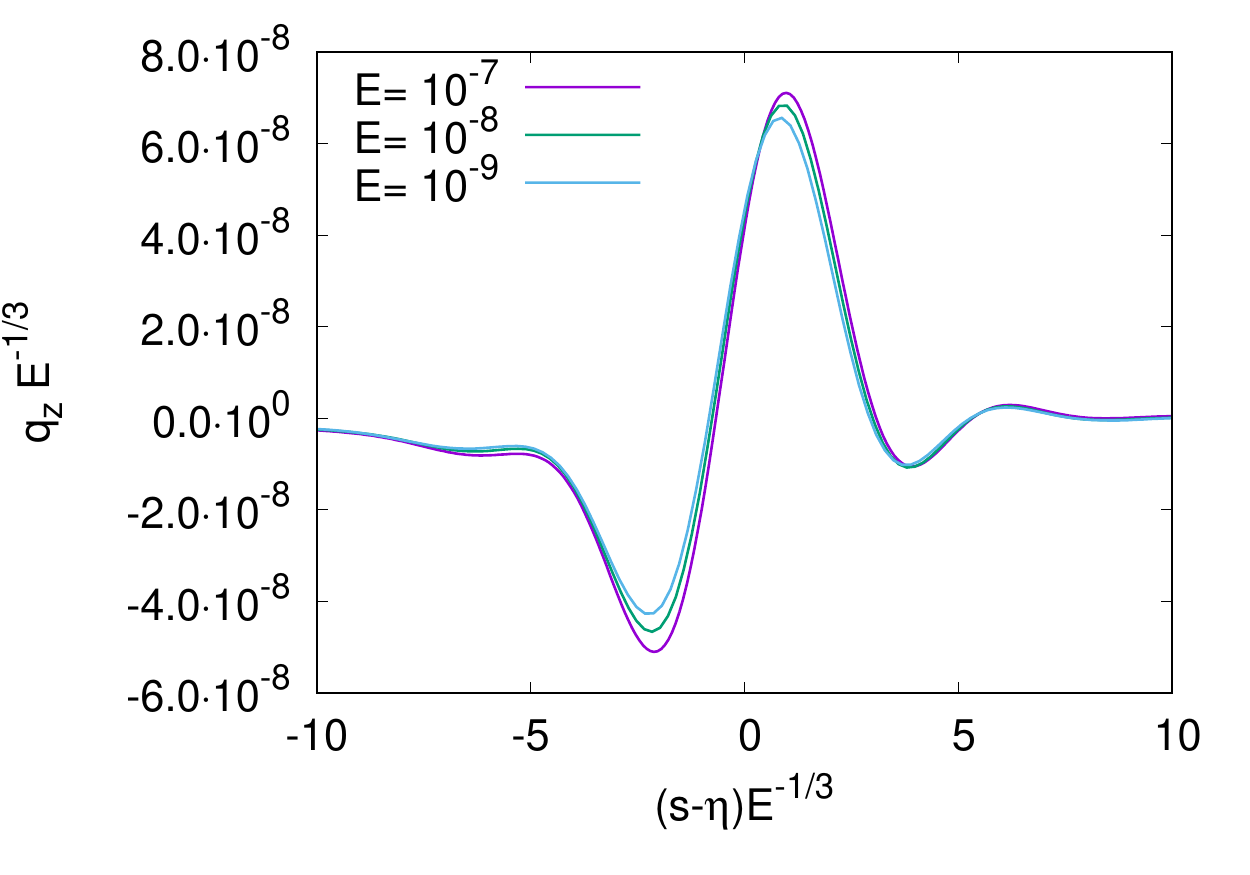}\hfill    
  \caption{$q_s E^{-2/3}$ (left) and $q_z E^{-1/3}$ (right) as a function of the stretched cylindrical radial coordinate $(s-\eta)E^{-1/3}$ for various Ekman numbers, $\eta=0.35$, $z=0.7$, $\rho_s=10^{-4}$,  and $A=0.01$.}
\label{fig:chi_compr_centr}
\end{figure}

\begin{figure}
\centering
     \includegraphics[width=0.5\textwidth]{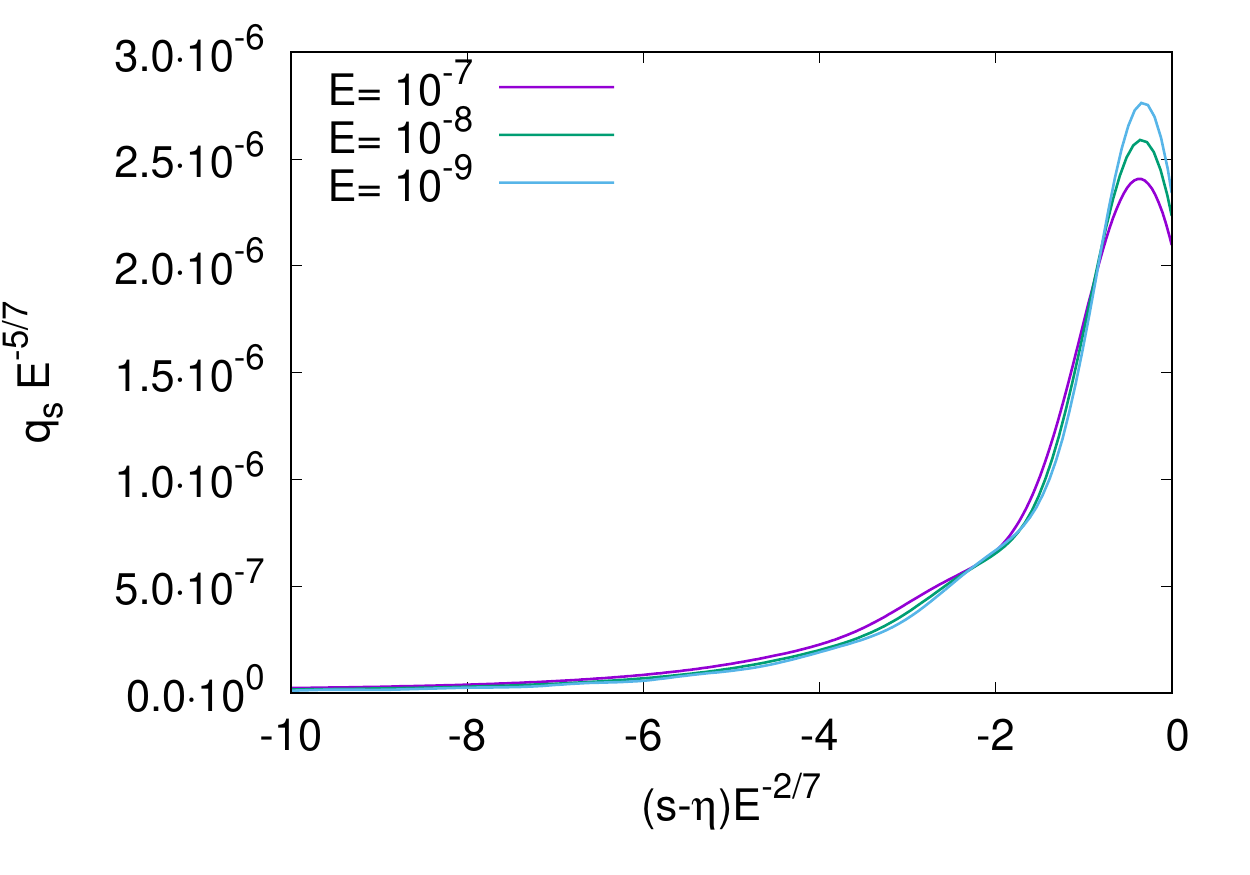}\hfill
     \includegraphics[width=0.5\textwidth]{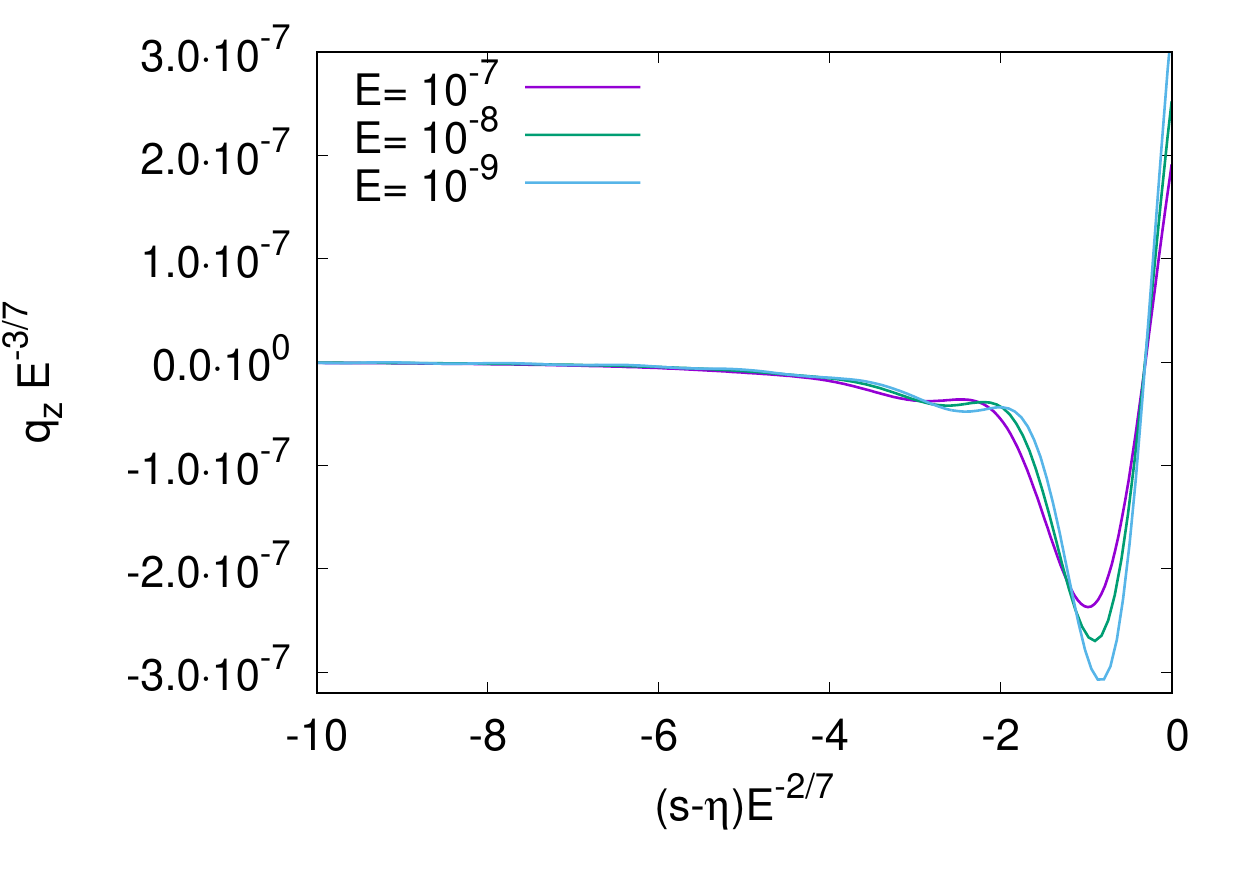}\hfill     
   \caption{$q_s E^{-5/7}$ (left) and $q_z E^{-3/7}$ (right) as a function of the stretched cylindrical radial coordinate $(s-\eta)E^{-2/7}$ for various Ekman numbers, $\eta=0.35$, $z=0.7$, $\rho_s=10^{-4}$, and $A=0.01$.}
\label{fig:uz_us_compr}
\end{figure}

Finally, the Stewartson layer seems to conserve its structure if we focus on the momentum, as shown by Figs~\ref{fig:chi_compr_centr} and \ref{fig:uz_us_compr}. Hence, the $E^{1/3}$ scale is still the dominating scale.

%The study of the anelastic flow driven by a prescribed tangential surface-stress thus brings us to the same conclusions regarding the strength of the circulation as that of the incompressible flow. Indeed, we find most of the meridional kinetic energy to be localised in the central vertical shear layer. In this region, the secondary flow is mostly parallel to the rotation axis and its amplitude scales as $O(E^{1/3})$. The circulation in the adjacent regions scales as $O(E)$.

%However, because the velocity no longer conforms the Taylor-Proudman theorem, the secondary flow outside $\mathcal{C}$ does not consist in $z={\rm Cst}$ streamlines, and a vertical variation of the circulation is dictated by the imposed radial density profile. Because of the imposed no-slip boundary conditions at the inner boundary, the anelastic flow, however, remains parallel to the rotation axis inside $\mathcal{C}$, to $O(E^2)$ terms.
 
\section{Summary and conclusions}\label{sec:conclusions}

Aiming at a better description of the dynamics of the radiative envelopes
of massive stars, which lose both mass and angular momentum through
radiative winds, we have considered the problem of the spin-down flow
of a viscous fluid inside a spherical shell. The spin-down is assumed
to be driven by a tangential stress prescribed  on the outer shell.
This problem is quite close to the classical spherical Couette flow
\citep{zikanov+96,RTZL12}, but includes new features like density and thermal
stratification, or stress driving, that needed new investigations.

To start with, we therefore considered this problem in the case of a mild
driving so as to be able to deal with linear equations. After examining
the case of constant density, we investigated the role of both density
and thermal stable stratification. Since stars are large-size bodies,
the Ekman number is very small. We therefore restricted our analysis to
small Ekman numbers.

Our numerical solutions showed that the meridional kinetic energy is
concentrated in the Stewartson shear layer that is tangent to the inner
shell, both for incompressible and anelastic stationary flows assuming
no thermal stratification. Outside this layer, a boundary layer analysis
of the Ekman layers allowed us to exhibit analytical solutions for
the quasi-geostrophic and incompressible primary and secondary stationary
flows. We found the latter flow to be essentially perpendicular to the
rotation axis outside the tangent cylinder, flowing from the outer Ekman
layer towards the Stewartson layer. Inside the tangent cylinder, we found
this poloidal flow to be parallel to the rotation axis, being pumped into
the Ekman layer attached to the inner core. The mass-flux is then returned
towards the outer Ekman layer through the Stewartson layer. Outside the
Stewartson layer, the amplitude of the meridional flow (Ekman circulation)
scales like $E$, as a consequence of the surface stress driving. In our model, the Stewartson layer is composed of two nested free shear layers of thickness $O(E^{2/7})$ and $O(E^{1/3})$. The analysis of the quasi-geostrophic $E^{2/7}$-layer located inside the tangent cylinder allowed us to derive asymptotic solutions for the primary and secondary flows, and a simple analysis of the ageostrophic $E^{1/3}$-layer indicates that it dominates the entire meridional circulation with a maximum amplitude of the $z$-directed velocity scaling as $E^{1/3}$.%\textbf{In the Stewartson layer, the amplitude of the $z$-directed velocity is maximum in the layer of thickness $O(E^{1/3})$, and scales as $E^{1/3}$.}  

We then accounted for a stable thermal stratification by introducing a
radial temperature gradient using the Boussinesq approximation. Two
limits of the parameter $\lambda=\Pran \calN^2/(2\Omega_c)^2$ show up.
In the limit of strong thermal stratification, $\lambda\gg1$, the
angular velocity profile becomes shellular (only radially dependent),
while the circulation is concentrated in thermal boundary
layers and the radial motion is strongly inhibited outside of them. As
a consequence, the Stewartson layer is suppressed. A more in-depth consideration of this regime, though certainly relevant to slowly rotating stars, has been deliberately left aside. Indeed, such slow rotators may never relax to a steady-state and thus might crucially depend on initial conditions. The study of the $\lambda\gg1$ asymptotic regime, including the time-dependent baroclinic flow, certainly calls for a separate investigation.
%Because the Prandtl number is very small in stars, however, the limit $\lambda\ll1$ is the
%relevant one for the radiative envelope of rotating massive stars.
However, for rapidly rotating stars the relevant limit is $\lambda \ll1$.
In this case both the structure and amplitude of the incompressible and stationary flow are unaffected by the thermal stable stratification. 

Radiative envelopes of massive stars exhibit however strong density
variations between the convective core and the surface, making the
incompressible and Boussinesq approximations too restrictive. But since thermal
stratification has little impact for our stars, we can neglect buoyancy
in the dynamics and concentrate on the effects of density stratification
through the anelastic approximation. We chose to describe the gas of
the stellar envelope by a polytropic equation of state with polytropic
index $n=3$. We found that, assuming no further thermal stratification,
the amplitude scaling of the stationary flow remains $O(E)$ outside
the Stewartson layer and $O(E^{1/3})$ inside. However, outside the tangent cylinder, the streamlines are
no longer straight lines and the flow becomes very dependent on the
coordinate parallel to the rotation axis.

Because of the weakness of the meridional circulation,
the stationary state of the flow is not quite relevant because it
implies very long transients that might actually exceed the lifetime
of the star. Massive stars are indeed short-lived (a few million
years). We therefore checked the time scale associated with the transient
phase and found it to be governed by the viscous diffusion time, that
is $O(E^{-1})$ indeed. The density variations of the hydrostatic
background seem to shorten it slightly. %have little influence on this time scale.
This implies that such a steady flow is
typically reached on a time that is longer than the lifetime of
the star. Fortunately, though the flow structure outside the
tangent cylinder evolves during the transient, the dominating $E^{1/3}$
amplitude scaling of the meridional velocity stands. Hence, for all
the models considered, the time relevant to the advective transport
of chemicals within the radiative envelope of massive stars scales
as $E^{-2/3}$, which is much shorter than the advection of angular momentum acting on $O(E^{-1})$ time scale.

The conclusion of the foregoing study is that the Stewartson layer
is a key feature for the transport of chemical elements between the
core and the surface of a massive star. The advective time scale is
indeed $O(E^{-1/3})$ shorter than the spin-down time scale. In our
case indeed, spin-down is driven by a stress and the steady state is
reached on a viscous diffusion time, which is longer than
the star's life. The much shorter time associated with the
meridional current of the Stewartson layer allows chemical elements
produced in the core to be transported to the surface of the star
and be possibly observable. We have shown that neither the stable
stratification of the envelope, nor its strong density variations
inhibit the rise of the Stewartson layer. The stable stratification
is bypassed thanks to the low Prandtl number, while density
variations have little influence on the mass flux $\rho \buu$
basically because they occur on a large scale.

These conclusions are not the end of the story of course since other
effects might complicate the scenario. The first effect one might
think of is the local anisotropic turbulence of the envelope.
Indeed, the differential rotation induces a local shear that is
unstable. This shear instability is reduced by the stable
stratification (Richarson criterion) but eased by the large heat
diffusion. \cite{zahn92} and \cite{MZ98} have argued that such
instabilities will lead to a strongly anisotropic turbulence, making
horizontal transport much more efficient than the vertical one. It
is therefore an open question whether the Stewartson layer can
resist to an anisotropic turbulent diffusion and in which
circumstances.

If we still wish more realism, the interface between the core and
the envelope will need a more detailed description. In this region,
stable chemical stratification builds up in the course of stellar
evolution. This so-called $\mu$-barrier might isolate the core from
the envelope. However, the thermal gradient is still unstable and
overstable double-diffusive convection is suspected to develop
there \citep{garaud20}. The impact of rotation on the dynamics of
this region is almost unknown and will also motivate future studies. 
Finally, magnetic fields, of fossil or core dynamo origin, may have a dramatic impact on  massive stars interior dynamics. Indeed, depending on the field geometry, it can, for instance, amplify  \citep{Hollerbach1997} or suppress the Stewartson layer and yield a super-rotating jet \citep{Kleeorin1997,Dormy1998}. Their consideration is also a matter for future work.

\begin{acknowledgements}
We are very grateful to the referees for their insightful remarks which allowed us to improve the presentation of our work. The numerical computations have been performed using the HPC resources from CALMIP (Grant 2016-P0107), which is gratefully acknowledged.
\end{acknowledgements}

\appendix

\section{The linear approximation}\label{sec:A2}
\subsection{Validity condition}\label{sec:B1}
This appendix presents an \textit{a posteriori} justification for the use of linear approximation. For simplicity, we focus on the incompressible flow. In spherical coordinates and in the inertial frame, the non-linear dimensional momentum equation reads

\begin{equation}
\frac{\partial \buu_*}{\partial t_*} + (\buu_* \cdot \bnabla)\buu_*  = - \bnabla p_* + \bF_{\rm visc} \ ,
\end{equation}
where

\begin{equation}
    \bF_{\rm visc} = \mu \left(\Delta \buu_* + \frac{1}{3}\bnabla {\rm div} \buu_* \right) + 2(\bnabla \mu \cdot \bnabla)\buu_*  + \bnabla \mu \times (\bnabla \times \buu_*) - \frac{2}{3} \bnabla (\mu {\rm div} \buu_*)
\end{equation}
is the dimensional viscous force, and $\mu=\rho_{*}(r) \nu$ is the dynamical viscosity. Let us  decompose the velocity field as the sum of the bulk component $\Omega_c \be_z \times \br$ and a residual velocity field  $\buu_{*}^r$

\begin{equation}
\buu_* = \Omega_c \be_z \times \br_* + \buu_{*}^r \ .
\end{equation}

For an axisymmetric flow, the non-linear term can be rewritten

\begin{equation}
    (\buu_* \cdot \bnabla)\buu_* = \Omega_c^2 \be_z \times (\be_z \times r_*) + 2 \Omega_c \be_z \times \buu_{*}^r + (\buu_{*}^r \cdot \bnabla)\buu_{*}^r \ .
\end{equation}

We further note that the centrifugal term derives from a potential that can be gathered with the pressure into $\Pi$. Finally, we write the non-linear adimensional momentum equation in the rotating frame

\begin{equation}
\frac{\partial \buu^r}{\partial t} + (\buu^r \cdot \bnabla)\buu^r + \be_z \times \buu^r  = - \bnabla \Pi + E \Delta \buu^r \ .
\end{equation}

 From now on the adimensional velocity field in the rotation frame $\buu^r$ will be written $\buu$ for simplicity. We now explicit the non-linear term as

\begin{align}\label{eq:vgv}
    \buu \cdot \boldsymbol{\nabla} \buu &=  \begin{bmatrix}
           (\buu \cdot \bnabla) u_r - (u_\theta^2 + u_\phi^2)/r \\ \\
           (\buu \cdot \bnabla) u_\theta + (u_\theta u_\phi - u_\phi^2 \cot \theta)/r \\ \\
           (\buu \cdot \bnabla) u_\phi + (u_r u_\phi + u_\theta u_\phi \cot \theta)/r
         \end{bmatrix} \ .
\end{align}

The Ekman boundary layer analysis presented in Sect.~\ref{sec:BL} revealed $u_r$ and $u_\theta$ to be of order $E$ (except in the Stewartson layer). For an axisymmetric flow, and neglecting $O(E^2)$ terms, (\ref{eq:vgv}) can be simplified as

\begin{align}\label{eq:vgv1}
    \buu \cdot \boldsymbol{\nabla} \buu &=  \begin{bmatrix}
           - u_\phi^2/r \\ \\
            - u_\phi^2 \cot \theta/r \\ \\
            (\buu \cdot \bnabla) u_\phi + (u_r u_\phi + u_\theta u_\phi \cot \theta)/r
         \end{bmatrix} \ .
\end{align}

The Coriolis acceleration, in turn, reads

\begin{align}\label{eq:ezv}
    \be_z \times \buu &=  \begin{bmatrix}
           - \sin \theta u_\phi \\ \\
             - \cos \theta u_\phi \\ \\
           \cos \theta u_\theta + \sin \theta u_r
         \end{bmatrix} \ .
\end{align}

Since we work with the vorticity equation, we need
%In this work, so as to get rid of the (modified) pressure gradient, we solve the vorticity equation. Hence, the relevant terms to quantify for this linear approximation justification are 

\begin{align}\label{eq:rotvgv}
    \boldsymbol{\nabla} \times (\buu \cdot \boldsymbol{\nabla}) \buu &=  \begin{bmatrix}
           1/(r \sin \theta) \partial_\theta \left[ \sin \theta\left((\buu \cdot \boldsymbol{\nabla})u_\phi + (u_\phi u_r + u_\phi u_\theta)/r  \right) \right] \\ \\
           -1/r \partial_r \left[ r (\buu \cdot \boldsymbol{\nabla})u_\phi +  u_\phi u_r + u_\phi u_\theta \right]   \\ \\
           A_\phi 
         \end{bmatrix} \ ,
\end{align}
and

\begin{align}\label{eq:rotezu}
   \boldsymbol{\nabla} \times (\be_z \times \buu)&=   \begin{bmatrix}
           1/(r \sin \theta) \partial_\theta \left(\sin\theta \cos\theta u_\theta + \sin^2 \theta u_r \right) \\ \\
           -1/r \partial_r \left(r\cos\theta u_\theta + r \sin \theta u_r \right)  \\ \\
           B_\phi 
         \end{bmatrix} \ ,
\end{align}
where 

\begin{equation}
%\begin{aligned}
A_\phi =\frac{1}{r}\left(\frac{1}{r}\frac{\partial}{\partial\theta} - \cot \theta\frac{\partial}{\partial r} \right)u_\phi^2 
= -\frac{2 u_\phi}{r^2}(s+ z \cot \theta)\frac{\partial u_\phi}{\partial z} \ ,
%\end{aligned}
\end{equation}

and 

\begin{equation}
%\begin{aligned}
B_\phi =\frac{1}{r}\left(\partial_\theta \sin\theta - \partial_r r\cos \theta \right)u_\phi
= - \frac{\partial u_\phi}{\partial z} \ .
%\end{aligned}
\end{equation}

We note that in the steady-state, outside boundary layers $\partial u_\phi / \partial z = O(E)$ according to the Taylor-Proudman theorem, we can thus write the amplitude scaling of both terms  

\begin{equation}
\norm{\boldsymbol{\nabla} \times (\be_z \times \buu)}=\norm{\dz{\buu}} = O(E) \quad {\rm and} \quad  \norm{\boldsymbol{\nabla} \times (\buu \cdot \boldsymbol{\nabla}) \buu} =  O(u_\phi E) \ .
\end{equation}

\begin{figure}
\centering
     \includegraphics[width=0.5\textwidth]{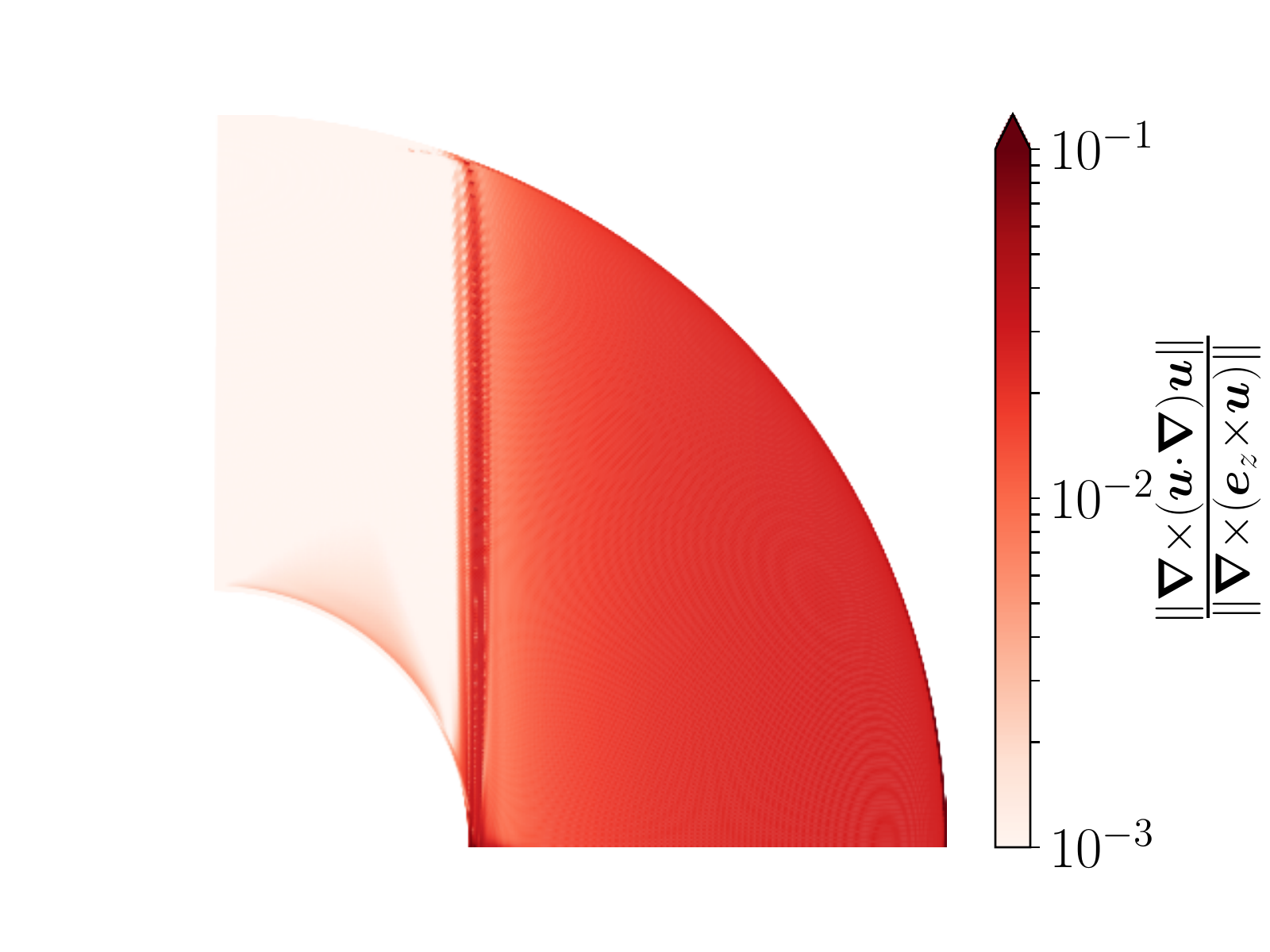}\hfill
     \includegraphics[width=0.5\textwidth]{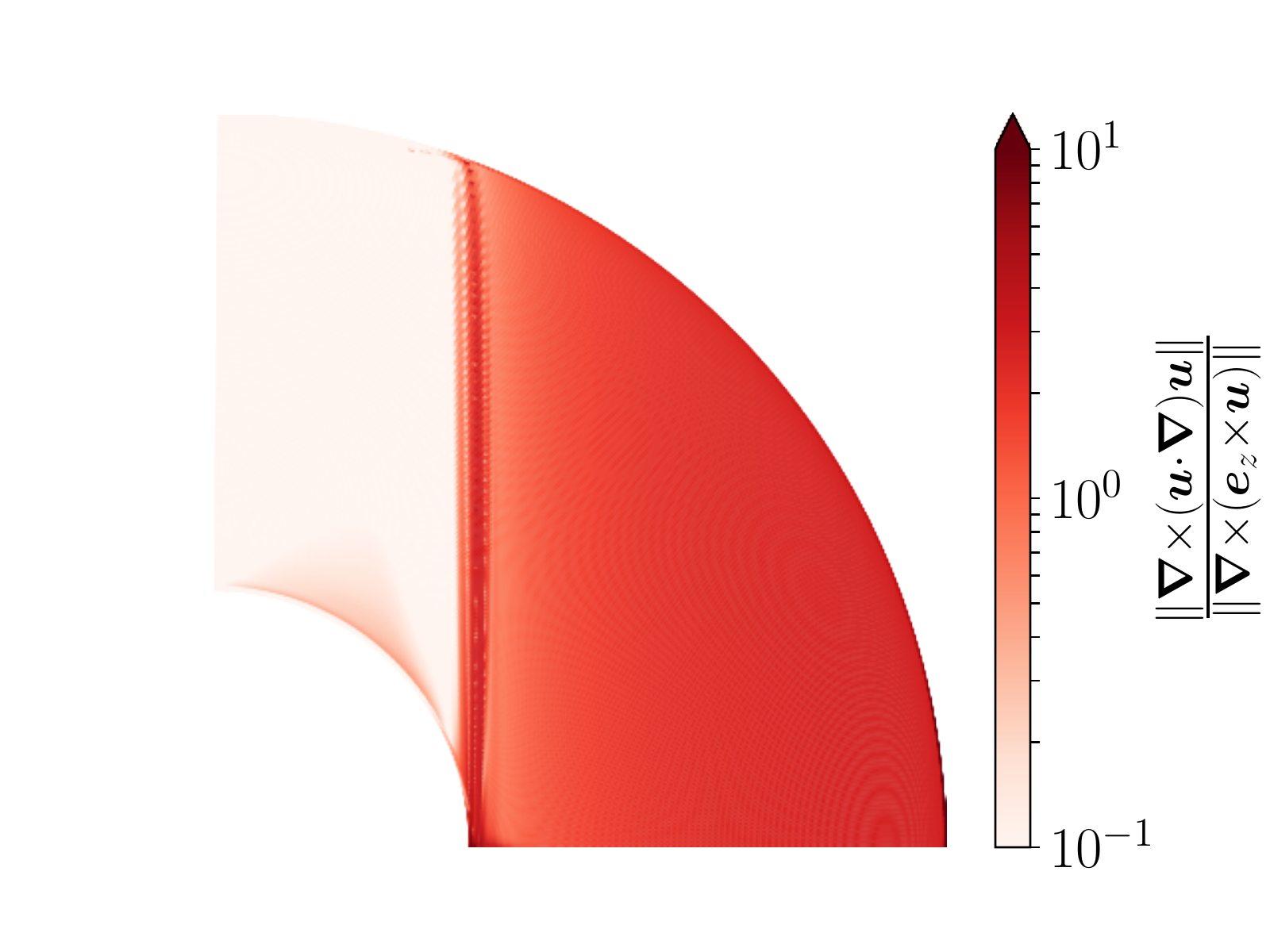}\hfill     
   \caption{Meridional view of the ratio between the Euclidian norm of the vorticity equation non-linear term and that of the curl of the Coriolis acceleration, for $A=0.01$ (left) and $A=1$ (right), and for an incompressible flow.}
\label{fig:ratio}
\end{figure}

Hence, if $u_\phi = O(1)$
%in the rotating frame, thus corresponding to a Rossby number of order unity,
non-linear terms in the vorticity equation are of the same order as the curl of the Coriolis acceleration and therefore cannot be neglected.
However, inside the tangent cylinder, we have found, in Sect.~\ref{sec:332}, that $u_\phi= O(\sqrt{E})$. In that region the non-linear terms can be neglected. Outside the tangent cylinder, the  amplitude of the azimuthal velocity scales as $A$ (see eq.~\ref{the_sol_F}), and the non-linear terms are therefore $O(A E)$. Hence, for sufficiently small $A$, we expect the linear approximation to be relevant to our problem. We show the ratio between the Euclidian norm of (\ref{eq:rotvgv}) and that of (\ref{eq:rotezu}) calculated \textit{a posteriori}, for $A=0.01$ and $A=1$, in Fig.~\ref{fig:ratio}. We find, as expected, this ratio to be of order $A$ outside the tangent cylinder. The linear solution can therefore be considered as a good approximation to the non-linear and incompressible vorticity equation provided $A \ll 1$. 

Similarly, for the anelastic flow, we find

\begin{equation}
    \frac{\norm{\boldsymbol{\nabla} \times \rho(\buu \cdot \boldsymbol{\nabla}) \buu}}{\norm{\boldsymbol{\nabla} \times (\be_z \times \rho \buu)}} = O(u_\phi) \ ,
\end{equation}
where this time $u_\phi = O(A \rho_s)$ outside the tangent cylinder. Since $\rho_s\ll1$, the conditions of linearity are more easily met.
%The individual scaling relations of the two terms are, however, slightly more complicated to determine because they depend on the strength of density stratification as well as on its profile. Hence, the linear solution can be considered a good approximation to the non-linear and  anelastic vorticity equation provided
%$A \ll 1/\rho_s$.

\subsection{Application to massive stars}

\begin{figure}
\centering
     \includegraphics[width=0.7\textwidth]{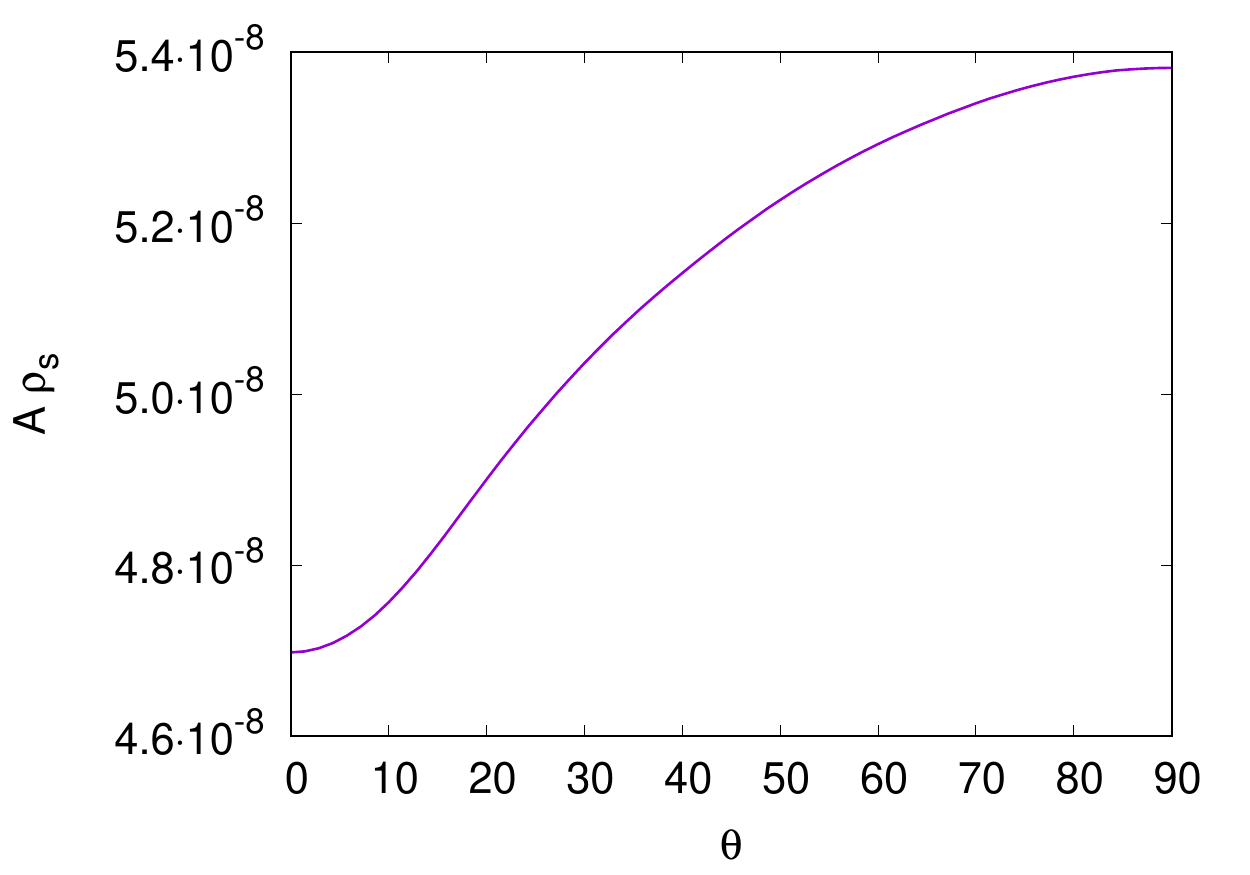}\hfill  
   \caption{Amplitude of the surface stress weighed by the adimensional surface density $A\rho_s$ resulting from the angular momentum outward flux, as a function of co-latitude, for a $15~M_\odot$ 2D ESTER stellar  model rotating with a period of one day.}
\label{fig:A}
\end{figure}

Let us now determine a typical value for $A$ when considering the differential rotation to be driven by a radiative wind at the surface of a massive star. We consider the outwardly accelerated outer layers to exert a viscous stress on the underlying layers of the star, spinning them down. Hence, we assume the local vertical angular momentum flux from the outwardly accelerated flow at the stellar surface to amount for the torque applied to the inner layers of the stars. That is

\begin{equation}\label{eq:astro1}
     \dot{\ell}_z = R \sin \theta \sigma_{r \phi, *} \ , 
 \end{equation}
where $\dot{\ell}_z \equiv \dot{m} \Omega_s (R \sin \theta)^2 $ is the local angular momentum flux,  $\dot{m}$ is the associated local mass-flux that is assumed isotropic, $R$ is the stellar radius, and $[\sigma_*]$ is the dimensional stress tensor. Combining the expression for the imposed azimuthal stress Eq.~(\ref{eq:BC}) with  Eq.~(\ref{eq:astro1}) yields the adimensional amplitude of the surface stress resulting from the outward angular momentum flux

\begin{equation}
    A = \sqrt{\frac{4 \pi }{3}} \frac{\dot{m} \Omega_s(\theta) R(\theta)}{2 \Omega_c \rho_{s,*}(\theta) \nu} \ ,
\end{equation}
where $\rho_{s,*} = \rho_s \rho_c $ is the dimensional surface density. We compute all quantities with the ESTER 2D code \citep{Rieutordal2016,Gagnier2019a} for a $15~M_\odot$ stellar model with a rotation period of one day and we assume a (turbulent)  kinematic viscosity at the surface $\nu = 10^{12} cm^2.s^{-1}$ \citep{zahn92,ELR13}. We plot the amplitude of the surface stress weighed by surface density $A\rho_s $ as a function of co-latitude in Fig.~\ref{fig:A}.  We further note  that rotating stars are not strictly spherically symmetric because of the centrifugal force, in particular when rotation is rapid. Hence, the stellar radius as well as the surface density have latitudinal dependencies.
Besides the resulting slight variation on the stellar surface, we see that $A \rho_s$ is of the order $5 \cdot 10^{-8} \ll 1$ for this model. Hence, according to the results of Sect.~\ref{sec:B1}, the linear approximation can be used to model massive stars losing angular momentum, provided the kinematic viscosity at the surface is no less than $\sim 10^7 cm^2.s^{-1}$, a typical value in radiation-dominated surface layers of massive stars \citep{ELR13}.%, provided $A \rho_s \ll 1$, which is always satisfied

%We see that, as expected, the turbulent viscosity resulting from the acceleration of the outermost layer is zero at the poles and maximum at the equator because its latitudinal dependency is dictated by that of the angular momentum flux. The dimensional value of the turbulent kinetic viscosity at the equator reads $\nu_t(\pi/2) \simeq 1.6 \times 10^{13} cm^2.s^{-1}$, which is consistent with \cite{zahn92} turbulent kinematic viscosity estimates close to the stellar surface \cite[e.g.][]{ELR13}.

\section{Asymptotically strong thermal stratification regime}\label{sec:A3}

In this appendix, we study the case of strong temperature  stratification, that is in the $\lambda \gg 1$ asymptotic regime. We have seen, in Sect.~\ref{sec:HBL}, that in this regime, both Ekman and thermal horizontal boundary layers coexist, with respective thickness of order $\sqrt{E}$ and $1/\sqrt{\lambda}$.% In the following, we consider the resulting steady and transient flows.

\subsection{The steady flow}

In the asymptotic regime of small Ekman numbers, we have seen that because $u_\phi = O(1)$, the temperature deviation from equilibrium is at most $O(1)$. Unlike the weakly stratified case however, the associated $O(E/\lambda)$ interior radial velocity is less than the $O(E)$ stratification independent Ekman pumping, and is therefore a consistent solution for the interior flow.  Let us now write the dynamical variables in the interior, as a power expansion of the small parameter $1/\sqrt{\lambda}$ corresponding to the width of the thermal boundary layer, in the asymptotic regime of small Ekman numbers. They read

\begin{equation}\label{eq:powerexp_lamb-1}
\begin{aligned}
u_r &= E(\frac{1}{\sqrt{\lambda}} u_{r,1} +  \frac{1}{\lambda}u_{r,2} + ...)\\
u_\theta &= E(\frac{1}{\sqrt{\lambda}} u_{\theta,1} +  \frac{1}{\lambda}u_{\theta,2} + ...)\\
u_\phi &= u_{\phi,0} + \frac{1}{\sqrt{\lambda}} u_{\phi,1} + ... \\
\delta T &= \delta T_0 + \frac{1}{\sqrt{\lambda}} \delta T_1 + ... \\
p &= p_0 + \frac{1}{\sqrt{\lambda}} p_1 + ... \ .
\end{aligned}
\end{equation}

Injecting (\ref{eq:powerexp_lamb-1}) in the momentum equation then yields the $O(1)$ and $O(1/\sqrt{\lambda})$ interior equations, that is

\begin{equation}
u_{\phi,0} = \frac{\partial p_0}{\partial s} - s \delta T_0, \quad  z \delta T_0=\frac{\partial p_0}{\partial z},  \quad {\rm and} \quad \left(\nabla^2 - \frac{1}{s^2}\right)u_{\phi,0}=0,  \ ,
\end{equation}
and

\begin{equation}
u_{\phi,1} = \frac{\partial p_1}{\partial s} - s \delta T_1, \quad z \delta T_1=\frac{\partial p_1}{\partial z}, \quad {\rm and} \quad \left(\nabla^2 - \frac{1}{s^2}\right)u_{\phi,1}=u_{s,1}  \ .
\end{equation}

Hence, for large values of $\lambda$, thermal stratification inhibits vertical motions in the interior, and the Ekman layer pumping/suction no longer controls the dynamics of the interior flow. The Taylor-Proudman is not verified and is replaced by the thermal wind equation

\begin{equation}
    (\be_z \cdot \bnabla) \buu = \br \times \bnabla \delta T  \ ,
\end{equation}
to $O(E  \buu)$, and the $O(1)$ azimuthal velocity can be obtained solving

\begin{equation}
\left( \nabla^2 -\frac{1}{s^2} \right) u_\phi = 0 \ .
\end{equation}

This yields, using boundary conditions (\ref{eq:BC_spect})

\begin{equation}\label{eq:omegar}
u_\phi \simeq \frac{K}{3} \left( \frac{1}{r^3} - \frac{1}{\eta^3}\right) r \sin \theta \ .
\end{equation}

\begin{figure}
\centering
     \includegraphics[width=0.6\textwidth]{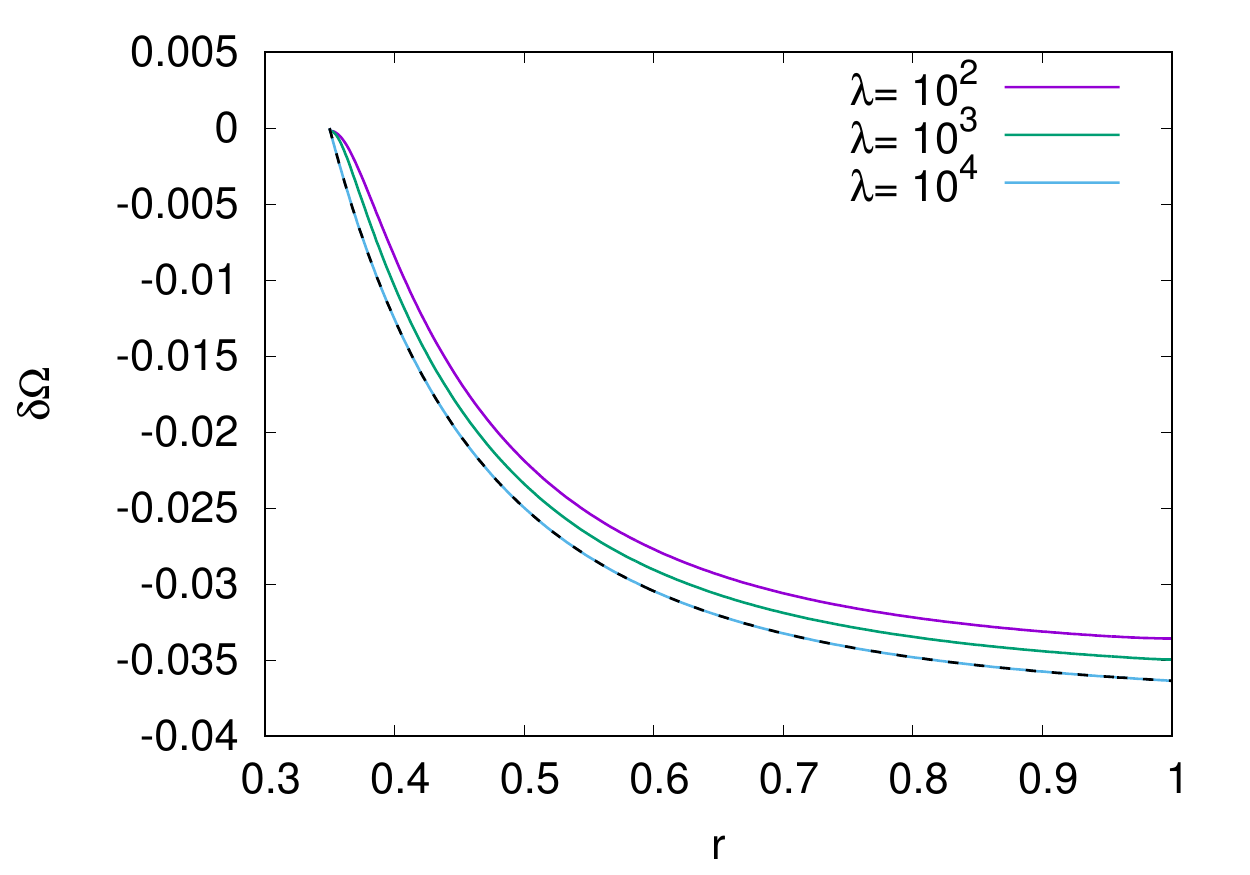}\hfill
          \caption{Angular velocity radial profiles in the rotating frame $\delta \Omega$, at $\theta=\pi/2$, $E=10^{-7}$, and for various $\lambda$. The black dashed line corresponds to the asymptotic solution (\ref{eq:omegar}).}
    \label{fig:omega_r}
\end{figure}

Hence, at the lowest order, $\delta\Omega=u_\phi/(r \sin \theta)$ is shellular in the asymptotic regime of large $\lambda$. Fig.~\ref{fig:omega_r} shows the angular velocity radial profiles at $\theta = \pi/2$, for $E=10^{-7}$ and for various $\lambda$. We see that the asymptotic limit where the  $O(1/\sqrt{\lambda})$ terms can be neglected corresponds to $\lambda \gtrsim 10^4$.

 Another interesting feature of the $\lambda \gg 1$ regime is that the circulation is concentrated in the boundary layers. This can be seen in Fig.~\ref{fig:diff_lambda} for $\lambda=10^2$. In the interior, the radial motion is so strongly inhibited by thermal stratification that it  prevents the existence of the Stewartson layer, thus undermining rotation-induced advection.
 
\begin{figure}
\centering
     \includegraphics[width=0.6\textwidth]{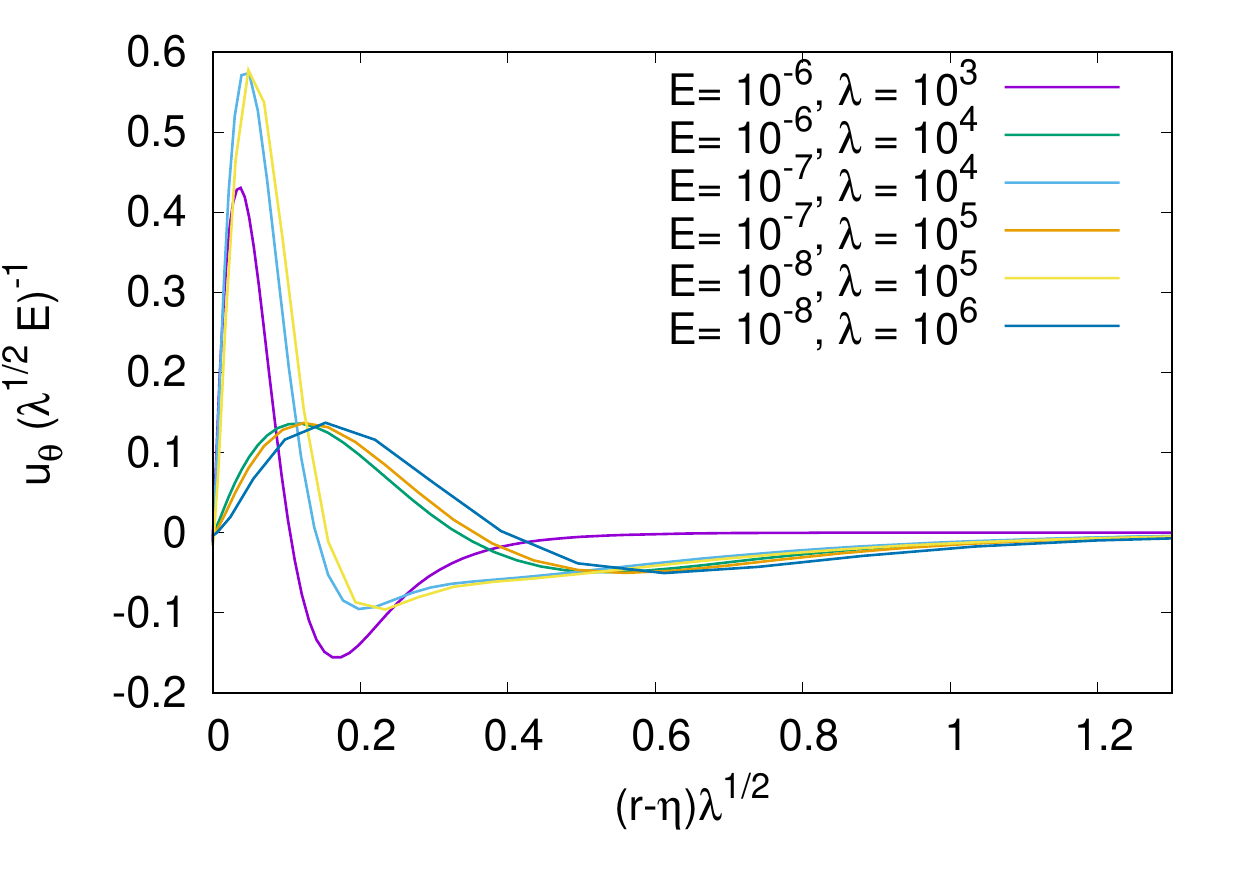}\hfill
          \caption{$u_\theta ( \sqrt{\lambda}E)^{-1} $ as a function of the stretched  radial coordinate $(r-\eta)\sqrt{\lambda}$, for various combinations of Ekman numbers and $\lambda$ parameters, $\eta=0.35$, $\theta=\pi/8$, and $A=0.01$.}
    \label{fig:uth_thermal}
\end{figure}

 In the thermal layers, radial gradients are increased by a factor $\sqrt{\lambda}$ which gives $\hat{u}_r= O(E)$, and from the continuity equation $\hat{u}_\theta= O(\sqrt{\lambda}E)$. In Fig.~\ref{fig:uth_thermal}, we plot $u_\theta/(E \sqrt{\lambda})$ as a function of the stretched radial coordinate $(r-\eta)\sqrt{\lambda}$, for various combinations of Ekman numbers and $\lambda$ parameters. We find that, indeed, the latitudinal velocity scales as $\sqrt{\lambda E}$ in the outer part of the layer of thickness $\delta_\lambda$, that is in the thermal boundary layer region, outside the Ekman layer (typically for $(r-\eta)\sqrt{\lambda} \gtrsim 0.5$). Additionally, we verify the $O(\sqrt{E})$ amplitude scaling of the latitudinal velocity in the Ekman boundary layer. Indeed, Fig.~\ref{fig:uth_thermal} shows that provided fixed values of $\delta_E / \delta_\lambda$, that is fixed values of $\sqrt{\lambda E}$, the latitudinal velocity scales as $\sqrt{\lambda}E$ in the Ekman boundary layer as well. However, taking $\sqrt{\lambda E} = D$, where $D$ is some constant, implies that $\sqrt{\lambda}E = D \sqrt{E}$, hence $\hat{u}_\theta \propto \tilde{u}_\theta = O(\sqrt{E})$.

\subsection{The transient flow}

We finally determine the scaling of the transient time with the Ekman number and the $\lambda$-parameter, in the $\lambda \gg 1$ limit. Assuming that in this regime, $\delta T$ remains $O(1)$ during the transient, (\ref{eq:eqstrat_time}) simplifies to
 
\begin{equation}\label{eq:syst_fin_time_strong}
\frac{\partial}{\partial t} \left(\cotan\theta \frac{\partial}{\partial \theta}\tan \theta u_\phi \right) \simeq  E \cotan \theta \frac{\partial }{\partial \theta} \tan \theta \left( \nabla^2 -\frac{1}{r^2 \sin^2 \theta}\right) u_\phi \ ,
\end{equation}
indicating that the (non-geostrophic) steady-state is reached on a $O(E^{-1})$ time scale as well. This is verified in Fig.~\ref{fig:tauss_lamb}. We note a weak $\lambda$ dependence of this time scale in the intermediate stratification regime.

\bibliographystyle{jfm}
% Note the spaces between the initials
\bibliography{bibnew}
\end{document}